\documentclass{gGAF2e}
\usepackage{color}

\newcommand{\be}{\begin{equation}}
\newcommand{\en}{\end{equation}}

\def\d{{\rm d}}
\def\uv{{\boldsymbol u}}
\def\av{{\boldsymbol a}}
\def\gv{{\boldsymbol g}}

\def\xv{{\boldsymbol x}}

\def\ev{{\boldsymbol e}}
\def\Bv{{\boldsymbol B}}

\def\cl{{\mathcal{L}^v}}
\def\clf{{\mathcal{L}^f}}
\def\clv{{\mathcal{L}^{vf}}}

\def\bm{\boldsymbol{\mathsfbi{B}}}

\def\grad{\boldsymbol\grad}

\def\wh{\mathcal{H}^v}
\def\whf{\mathcal{H}^f}

\def\grad{{\rm grad}\, }
\def\bnab{{\boldsymbol\nabla}}
\def\bdot{{\boldsymbol\cdot}}

\def\bsig{{\boldsymbol\sigma}}

\begin{document}

\jvol{00} \jnum{00} \jyear{2012} 

\markboth{Helicity and winding fluxes in flux emergence}{MacTaggart \& Prior}


\title{Helicity and winding fluxes as indicators of twisted flux emergence}

\author{{D. MACTAGGART${\dag}$}$^{\ast}$\thanks{$^\ast$Corresponding author. Email: david.mactaggart@glasgow.ac.uk
} and C. PRIOR${\ddag}$\\   ${\dag}$School of Mathematics and Statistics, University of Glasgow, Glasgow G12 8QQ, UK\\ ${\ddag}$Department of Mathematical Sciences, Durham University, Durham, DH1 3LE, UK}

\maketitle

\begin{abstract}
Evidence for the emergence of twisted flux tubes into the solar atmosphere has, so far, come from indirect signatures. In this work, we investigate the topological input of twisted flux tube emergence directly by studying helicity and winding fluxes. In magnetohydrodynamic simulations with domains spanning from the top of the convection zone to the lower corona, we simulate the emergence of twisted flux tubes with a range of different initial field strengths. One important feature of this work is the inclusion of a convectively-unstable layer beneath the photosphere. We find approximately self-similar behaviour in the helicity input for the different field strengths considered. As the tubes rise and reach the photosphere, there is a strong input of negative helicity since we consider left-handed twisted tubes. This phase is then followed by a reduction of the negative input and, for low initial field strengths, a net positive helicity input. This phase corresponds to the growing influence of convection on the field and the development of serpentine field structures during emergence. The winding flux can be used to detect when the twisted cores of the tubes reach the photosphere, {giving clear information about the input of topologically complex magnetic field into the solar atmosphere}. In short, the helicity and winding fluxes can provide much information about how a magnetic field emerges that is not directly available from other sources, such as magnetograms. {In evaulating the helicity content of these simualtions we test numerous means for creating synthetic magnetograms, including methods which acount for both the evolving geometry and the finite extent of the photosphere. Whilst the general qualitative behaviours are same in each case, the different forms of averaging do affect the helicity and winding inputs quantitatively.} 


\end{abstract}

\section{Introduction}

The idea of magnetic flux tubes rising through the convection zone to form bipolar active regions is long-standing \citep[e.g.][]{cowling1946sunspot,jensen1955tubes,parker1955}. The magnetic field lines in a rising flux tube are considered, generally, to be highly twisted, for the simple reason that a tube needs to survive the destructive effects of convection \citep[e.g.][]{fan2003convection,abbett2004convection,bushby2012convection}. Apart from this requirement, there is some observational evidence that suggests twisted flux tube emergence may be responsible for the formation of certain active regions \citep{poisson2015,poisson2016}. In this paper we will \emph{assume} that solar active regions are formed by twisted flux tubes {that reach the photosphere} and we will describe procedures that can be used to test this assumption. For an alternative theory on active region formation see, for example, \cite{brandenburg2010nempi,kemel2013} and \cite{mitra2014}.

Another appealing aspect of active regions being formed by twisted flux tubes is that twisted flux tubes provide an obvious source of magnetic helicity, which is important for the formation of eruptions. Since we cannot measure the magnetic field directly in the solar atmosphere, we require a robust method of indicating if strongly-twisted magnetic field emerges.  In this paper, we show that direct evidence of twisted flux tube emergence can be found by considering two quantities related to magnetic field topology: \emph{helicity flux} and \emph{winding flux}. The first of these quantities, the helicity flux, describes the amount of relative magnetic helicity passing through the photosphere and is used widely in both theoretical and observational studies \citep[e.g.][]{berger1984,berger1988,demoulin2003berger,prior2019interpreting,pariat2020book}. The second quantity, the winding flux, is a purely topological quantity that represents the average winding of magnetic field lines. It is identical in form to the helicity flux except that the magnetic field weighting is removed. The winding can be used to quantify reconnective activity in magnetic fields \citep{prior2018quantifying}. We will describe these quantities in more detail later.

 Simulations of twisted flux tube emergence have typically led to single signs for both the helicity rate and the total helicity input through the photosphere \citep[e.g.][]{yang2013evolution,moraitis2014validation,sturrock2016sunspot}. However, helicity input time series calculated from photospheric vector magnetograms have typically shown a far more mixed behaviour. For example, there are cases of bipolar emergence whose net helicity input has a consistent sign but whose input rate can vary in sign, as in \cite{kusano2002measurement} and \cite{chae2004determination}, or even the case where the total helicity input varies in sign, such as in \cite{chae2001observational} and \cite{vemareddy2017successive}. There is a clear need to attempt to explain this disparity through flux emergence simulations.

On this front, \cite{prior2019interpreting} simulated the emergence of both twisted and `mixed helicity' magnetic fields (fields with balanced but spatially varying negative and positve twisting). First, it was shown that in both cases mixed helicity input rates were possible. In the twisted tube case this was due to the core of the flux rope becoming trapped at the photosphere. In the mixed helicity case, it was due to a combination of the field's inherently mixed topology as well as emergence and subergence events.  A second important aspect of the results of \cite{prior2019interpreting} was to demonstrate that an examination of the winding, both the time series and density maps, in conjunction with the helicity, allows for a much more detailed interpretation of behaviour at the photosphere. One particular result was a clear jump in the winding time series indicating the emergence of the twisted core of the flux tube above the photospheric boundary. This result is important as it provides a clear signal that magnetic field of substantial twist has entered the solar atmosphere, something that is not clear from only examining the helicity data or other sources, such as magnetograms.

In this paper, we extend the work of  \cite{prior2019interpreting} by including a convectively-unstable solar interior. Since our focus is on the emergence of twisted flux tubes, we will be able to test if this field line topology, in conjunction with the effects of convection, is enough to lead to mixed-sign helicity signatures. We also aim to describe what clear signatures (if any) arise in the helicity and winding data that can give a clear indication that significant twist has entered the solar atmosphere.

The outline of the paper is as follows: first, the main equations and simulation setup are described.  This section is followed by a description of the helicity and winding fluxes. Then the analyses of the simulations are presented in detail and the paper concludes with a summary and a discussion.

\section{Simulation and initial condition setup}
\subsection{Main equations}
To model the large-scale behaviour of an emerging active region, we solve the compressible and ideal MHD equations using a Lagrangian remap scheme \citep{arber2001staggered}. In dimensionless form, the MHD equations are
\be\label{mass_con}
\left(\frac{\upartial}{\upartial t} + \uv\bdot{\bnab}\right)\rho = -\rho{\bnab}{\bdot}\uv,
\en

\be\label{mom_con}
\rho\left(\frac{\upartial}{\upartial t} + \uv\bdot{\bnab}\right)\uv = -\bnab p + (\bnab\times\Bv)\times\Bv + \bnab\bdot\bsig + \rho\gv,
\en

\be\label{flux_con}
\left(\frac{\upartial}{\upartial t} + \uv\bdot{\bnab}\right)\Bv = (\Bv\bdot\bnab)\uv - (\bnab\bdot\uv)\Bv,
\en

\be\label{energy_con}
\rho\left(\frac{\upartial}{\upartial t} + \uv\bdot{\bnab}\right)\varepsilon = -p\bnab\bdot\uv + Q_{\rm visc} - \frac{\varepsilon-\varepsilon_0}{\tau},
\en
with specific energy density
\be\label{en_den}
\varepsilon = \frac{p}{(\gamma-1)\rho}.
\en
The basic variables are the density $\rho$, the pressure $p$, the magnetic induction $\Bv$ (referred to as the magnetic field) and the velocity $\uv$. The gravitational acceleration is denoted $\boldsymbol{g}$ (uniform in the $z$-direction) and $\gamma =5/3$ is the ratio of specific heats. The dimensionless temperature $T$ can be found from
\be\label{temp}
T = (\gamma-1)\varepsilon.
\en
The variables are made dimensionless against photospheric values, namely, pressure $p_{\rm ph} = 1.4\times 10^4$ Pa; density $\rho_{\rm ph} = 2\times 10^{-4}$ kg~m$^{-3}$; scale height $H_{\rm ph}=170$ km;  surface gravity $g_{\rm ph} = 2.7\times 10^2$ m~s$^{-2}$; speed $u_{\rm ph} = 6.8$ km~s$^{-1}$; time $t_{\rm ph} = 25$ s; magnetic field strength $B_{\rm ph} = 1.3\times 10^3$ G and temperature $T_{\rm ph} = 5.6\times 10^3$ K. In the non-dimensionalization of the temperature we use a gas constant $\mathcal{R}=8.3\times 10^{3}$ m$^2$~s$^{-2}$~K$^{-1}$ and a mean molecular weight $\tilde{\mu}=1$. {This last quantity and the use of an ideal gas law throughout the domain are simplifying approximations in our model, whose main focus is on fundamental magnetic field dynamics and not on reproducing thermodynamic observables.}   

 The fluid viscosity tensor and the viscous contribution to the energy equation are respectively
\be\label{sigma}
\boldsymbol\sigma = 2\mu\left[\boldsymbol{D}-\frac13({\rm tr}\boldsymbol{D})\boldsymbol{I}\right] \quad {\rm and} \quad Q_{\rm visc} = \boldsymbol\sigma:\bnab\boldsymbol{u},
\en
where
\be
\boldsymbol{D} = \frac12\left(\bnab\boldsymbol{u} + \bnab\boldsymbol{u}^{\rm T}\right)
\en
is the symmetric part of the rate of strain tensor and $\boldsymbol{I}$ is the identity tensor. We take $\mu = 10^{-4}$ and use this form of viscosity primarily to aid stability. The code accurately resolves shocks by using a combination of shock viscosity \citep{wilkins1980use} and Van Leer flux limiters \citep{van1979towards}, which add heating terms to the energy equation. From now on, values will be expressed in non-dimensional form unless explicitly stated otherwise. 

The last term in equation (\ref{energy_con}) represents the non-adiabatic contribution to the energy of the plasma. Following \cite{leake2006}, we model this complex collection of physical phenomena using a simple Newton cooling term throughout the domain above the photosphere (see below for how the photospheric boundary is defined). The purpose of this term is to preserve the photosphere from being destroyed by the solar interior, which is convectively-unstable (we will return to this point shortly). At $t=0$ of the simulation,  $\varepsilon=\varepsilon_0$ and the Newton cooling term forces the energy profile above the photosphere to return to this initial state on a fast time scale of $\tau=0.5$. 

\subsection{Initial conditions}

\subsubsection{Background atmosphere and convective perturbation}

At the start of the simulations, the computational domain is filled by a plane-parallel atmosphere that represents the solar atmosphere ranging from the top of the solar interior (convection zone) to the corona. The temperature profile above the lower boundary of the photosphere ($z=0$) is given by

\be\label{initial_temp}
{T(z) = \left\{\begin{array}{cc}
1, & 0 \le z \le 10,  \\
150^{[(z-10)/10]}, & 10 < z < 20,  \\
150, & z \ge 20,
\end{array}\right.}
\en
where the three rows in equation (\ref{initial_temp}) correspond to the photosphere/chromosphere, the transition region and the corona respectively.  The other state variables, pressure and density, are found by solving the hydrostatic equation in conjunction with the ideal equation of state
\begin{equation}\label{EOS}
\frac{{\rm d}p}{{\rm d}z} = -\rho g, \quad p = \rho T.
\end{equation}
The initial atmosphere described above is common to many idealized flux emergence simulations \citep{hood2012review}. In this study, we also consider a solar interior that is convectively-unstable. To achieve this in our idealized setup, the Schwarzchild condition \citep[e.g.][]{archontis2009bomb,priest2014book} states that a convective instability will set in for an adiabatic fluid if the following criterion (written in our notation) is satisfied,
\be\label{schwarz}
-\frac{{\rm d}T}{{\rm d}z}>\frac{\gamma-1}{\gamma}.
\en  
In the solar interior ($z<0$) we consider an initial temperature profile of
\be
T_{\rm si}(z) = 1 - \delta z\frac{\gamma-1}{\gamma},
\en
where $\delta$ is a non-dimensional parameter used to satisfy condition (\ref{schwarz}). In this paper we take $\delta=1.3$ as this value leads to convection that has velocity magnitudes that are close to those of solar convection.

Our purpose here is to investigate, incrementally, the effect of convection on topological signatures and compare the results to previous work in \cite{prior2019interpreting} where convection is not included. We excite a stable mode of convection with hexagonal cells based on the solution in \cite{christopherson1940} (see also \cite{chandrasekhar1961book}). At $t=0$, the velocity is perturbed as
\be\label{christop}
u_z = \frac{3}{10}\exp(-(z+5)^2)\left(\frac{1}{54\sqrt{3}}\cos(2\pi x)\cos(2\pi y) + \frac{1}{18}\cos(4\pi y)\right).
\en
This perturbation acts below the photospheric boundary at $z=-5$ and the hexagonal cells of Christopherson's solution develop quickly {to fill} the solar interior {(an example of this will be displayed shortly in section \ref{initial_rise})}. The numerical parameters in equation (\ref{christop}) are chosen to produce convection cells with typical solar values, i.e. speeds of $O(1)$ km~s$^{-1}$ and sizes of $O(1)$ Mm.

\subsubsection{Magnetic field}
To model a magnetic flux tube, we adopt the standard profile for a cylindrical flux tube with uniform twist \citep[e.g.][]{fan2001emergence}. In cylindrical coordinates $(r,\theta,y)$, with $r^2=x^2+(z-z_{\rm axis})^2$, the magnetic field components are
\be\label{initial_b}
B_r=0, \quad B_y = B_0\exp(-r^2/d^2), \quad B_{\theta}=\alpha r B_y,
\en 
where $B_0$ is the axial field strength, $d$ is the flux tube radius and $\alpha$ is the field line twist. In this paper, we fix $d=3$ and $\alpha=-0.4$ (left-handed twist). We consider a range of field strengths $B_0=3$, 5 and 7.

The magnetic field in (\ref{initial_b}) is not force-free and is balanced by a pressure gradient (when there is no flow), i.e.
\be
\bnab p_m = (\bnab\times\Bv)\times\Bv,
\en
where $p_m$ is an additional pressure compared to the background hydrostatic pressure $p_b$ such that $p=p_b+p_m$. For the magnetic field in (\ref{initial_b}), the additional pressure has the form
\be
p_m = \frac14B_0^2\exp(-2r^2/d^2)(\alpha^2d^2-2-2\alpha^2r^2).
\en
In order to make the flux tube rise up buoyantly towards the photosphere, we introduce a density deficit within the tube of the form
\be
\frac{\rho_m}{\rho_b} = \frac{p_m}{p_b}\exp(-y^2/\lambda^2),
\en
where the $\rho_m$ is the density deficit, $\rho_b$ is the background hydrostatic density and $\lambda$ is a parameter that allows the flux tube to rise in the shape of an $\Omega$-loop. Here we fix $\lambda=20$.

 The flux tube is placed initially at $z_{\rm axis}=-15$. Since the density deficit is proportional to $B_0^2$, the rise times will vary for the values that we consider. The initial position of the tube axis is deep enough, however, for the flux tube to encounter fully developed convection before reaching the photosphere.
 
{There are clear modelling reasons for setting up the initial conditions as described above. First, as stated in the Introduction, we are assuming that a strongly-twisted flux tube reaches the photosphere - an assumption that is common in solar physics studies and worthy of testing. Therefore, we do not insert a flux tube deep into a well-developed convection zone as the subsequent deformation of the magnetic field would complicate our analysis of the topological signatures of twisted tube emergence. Secondly, we are interested in how convective motions affect the helicity and winding inputs. In order to identify the direct effects of convection on these signatures we need to compare closely to the setup of non-convective simulations \citep[e.g.][]{sturrock2016sunspot,prior2019interpreting} by introducing convection incrementally in the numerical model.}

 \subsubsection{Simulation parameters}
  All the simulations we present have a computational domain of size $(x,y,z)\in[-80,80]\times[-80,80]\times[-30,80]$ with a {uniform} resolution of 432$^3$. These values are suitable for studying the initial emergence of the flux tubes and their growth into the corona. The simulations are stopped before the expanding magnetic field reaches the side boundaries. The boundary conditions are periodic on the side boundaries and closed on the top and bottom boundaries. A damping layer is placed near the top boundary to help prevent the reflection of waves generated by convection at the photosphere. 
 
 \section{Helicity and winding fluxes}
 \subsection{Basic definitions}
{
The winding flux $\d L/ \d t$ measures the input of magnetic field line entanglement through a planar boundary,
 \be\label{windflux}
 \frac{\d L}{\d t} = -\frac{1}{2\pi}\int_{P\times P} \sigma_z(\av_1)\sigma_z(\av_2)\frac{\d\theta}{\d t}\, \d^2 a_1\, \d^2 a_2,
 \en
 where
 \be\label{sigma1}
 \sigma_z(\xv) = \left\{\begin{array}{cc}
 1 & {\rm if}\, B_z(\xv) > 0, \\
 -1 &  {\rm if}\, B_z(\xv) < 0, \\
  0 & {\rm if}\, B_z(\xv) = 0,\end{array}\right.
 \en
and the angle $\theta$ between points $\av_1$, $\av_2\in P$ is
\be
\theta = \arctan\left[\frac{(\av_2-\av_1)\bdot\ev_y}{(\av_2-\av_1)\bdot\ev_x}\right],
\en
 and $\{\ev_x,\ev_y,\ev_z\}$ is a basis of Cartesian space with $\ev_z$ normal to $P$.  The position $\av$ of a magnetic field line intersecting $P$ and satisfying ideal MHD is given by
 \be\label{dadt}
 \frac{\d\av}{\d t} = \uv_{\|}-\frac{u_z}{B_z}\Bv_{\|} =  \uv_{\|} - u_z\left(\frac{\d\av}{\d z}\right)_{\|},
 \en
 where $\uv_{\|}$ and $\Bv_{\|}$ are the projections of $\uv$ and $\Bv$ onto $P$. The angle $\theta$ measures the extent to which points in the plane $P$ of a pair of field lines  intertwine as the field undergoes ideal motion. 

Analogous to results in \cite{prior2014helicity,prior2018quantifying}, the rate of change of magnetic helicity $H$ through a planar boundary $P$ can be defined as the flux-weighted winding, 
\be\label{helflux}
\frac{\d H}{\d t} = -\frac{1}{2\pi}\int_{P\times P} B_z(\av_1)B_z(\av_2)\frac{\d\theta}{\d t}\, \d^2 a_1\, \d^2 a_2.
\en
We prefer this topologically motivated definition of helicity, as opposed to the more standard definition of relative helicity with arbitrary vector potentials and reference fields, since it highlights that helicity can be derived from a more fundamental quantity $L$ whose definition follows from purely geometric considerations {of the field line curves} \citep{berger1986,berger2006writhe}. In particular, it can be shown that $L$ can be used to uniquely categorize classes of magnetic field topology which $H$ cannot. Therefore, deriving $H$ from this more mathematically fundamental quantity is a logical step. {Equation (\ref{helflux}) is equivalent to the rate of change of relative helicity through $P$ when making use of a specific choice of gauge, namely the winding gauge discussed by \cite{prior2014helicity}, and has been used in observational studies \citep[e.g.][]{vemareddy2017successive}.}

To understand the difference between $L$ and $H$, consider a section of magnetic field with a complex field line topology but very weak field strength. The helicity contribution from this magnetic field could be low but its winding contribution high since the winding does not `read' the field strength.
}

Although the winding can pick up more topological detail than the helicity, there is the danger it that picks up weak field with complex topology that does not play a significant dynamical role (such field would be essentially ignored by helicity due to the weak field strength). In practice, it is important to test winding results against cut-off values in order to determine if results are dynamically important. Therefore, equation (\ref{sigma1}) is modified to
\be
 \sigma_z(\xv) = \left\{\begin{array}{cc}
 1 & {\rm if}\, B_z(\xv) > 0\,\, {\rm and}\,\, |\Bv|>\epsilon, \\
 -1 & {\rm if}\, B_z(\xv) < 0\,\, {\rm and}\,\, |\Bv|>\epsilon, \\
 0 & {\rm if}\, B_z(\xv) = 0\,\, {\rm or}\,\, |\Bv|\le\epsilon,\end{array}\right.
\en    
where $\epsilon$ is a chosen cut-off. In this paper, we will consider a cut-off of $\epsilon=0.01$. 
 
 \subsection{Related quantities}
 The helicity and winding rates of equations (\ref{helflux}) and (\ref{windflux}) respectively provide time series data. Time series of the actual helicity and winding input can be produced from the time integration of the rates, i.e.
 \be
\label{totalinputs}
  L(t) = \int_{t_0}^t\frac{\d L}{\d t'}\,\d t' \quad  H(t) = \int_{t_0}^t\frac{\d H}{\d t'}\,\d t',
 \en 
Spatial information about the distribution of helicity and winding can be found by determining the contribution of one point $\av_0\in P$ to the rate. Since $\av_0$ represents the intersection of a field line with $P$, this quantity represents the contribution of one field line to the helicity or winding input rate.   Repeating this calculation for a large set of $\av_0$ produces a map of how the helicity or widing rate is distributed spatially. For helicity, the field line helicity input rate is
 \be
\label{helinputdist}
 \frac{\d\mathcal{H}}{\d t}(\av_0)= -\frac{1}{2\pi}B_z(\av_0)\int_P B_z(\av)\frac{\d\theta}{\d t}(\av_0,\av)\,\d^2a.
 \en
 {A similar quantity was first introduced by \cite{berger1988}. In the solar physics literature,  ${\d\mathcal{H}}(\av_0)/{\d t}$ is often labelled as $G_{\theta}$ \citep[e.g.][]{dalmasse2014}}. For the winding rate, the analogous quantity is
  \be
\label{windinputdist}
 \frac{\d\mathcal{L}}{\d t}(\av_0)= -\frac{1}{2\pi}\sigma_z(\av_0)\int_P \sigma_z(\av)\frac{\d\theta}{\d t}(\av_0,\av)\,\d^2a.
 \en
{There are a number of additional analyses that we could potentially perform on the distributions (\ref{helinputdist}) and (\ref{windinputdist}), such as combining components through field line connectivity. However, they do not feature in our analysis here for reasons discussed in detail in \cite{prior2019interpreting}.}
 
 \subsection{Accounting for a changing photosphere}
  Following \cite{prior2019interpreting}, we calculate approximate helicity and winding fluxes by projecting values from a non-planar and evolving photosphere onto $P$. This process is intended to mimic what observations are recording, {assuming that there is some variation in height in the locations where the magnetic field is recorded}. In \cite{prior2019interpreting}, the photospheric boundary was taken to be the surface where $\rho=1$, which represents the photosphere in the initial condition. As the simulation evolved and the flux tube emergence led to the rise and churn of dense plasma from below the location of the initial $\rho=1$ surface, the surface became deformed. The field was sampled on this evolving surface and then projected onto $P$ to represent the production of planar vector magnetograms used in observational helicity studies.

 Due to the inclusion of convection in this study, this simple approach is no longer applicable as the $\rho=1$ surface can become multi-valued. To account for this multi-valued structure and the fact that the photosphere is really a thin volume instead of a surface, with a depth of approximately $100$ km \cite[e.g.][]{muller1975model}, we calculate `photospherically averaged'  magnetic field values as
\be
\label{averagefield1}
\widehat{{\Bv}}(x,y) = \frac{\int_{z_{\rm min}}^{z_{\rm max}}\exp[-2 (\rho-1)^6]{\Bv}(x,y,z)\,\d{z}}{\int_{z_{\rm min}}^{z_{\rm max}}\exp[-2 (\rho-1)^6]\,\d{z}},
\en
and similarly for the velocity field.  The idea is that the above average is calculated for a range of values whose density is suitably close to $\rho=1$. The exponential weighting is chosen so that the  density values that affect the average significantly are those that, with respect to the initial condition (determined by equation (\ref{initial_temp}) in conjunction with equations (\ref{EOS})), correspond (approximately) to a vertical extent of $100\mathrm{km}$ which contains $\rho=1$ at is centre. The domain $[z_{\rm min}, z_{\rm max}]$ is chosen here to be $[-7.5,7.5]$, i.e. suitably large to always contain the required density range.  Whilst the photospheric domain deforms (and its vertical width varies) it corresponds to a consistent range of densities. {When calculating the helicity and winding rates, the averaged magnetic field $\widehat{\Bv}$ and velocity field $\widehat{\uv}$ are projected onto a horizontal plane (as indicated in equation (\ref{averagefield1}) where the averaged quantity is only a function of  $x$ and $y$).} 

{We should be clear that using averaged fields means that the winding rates $\d{\cal L}/\d t$ and $\d {\cal H}/\d t$ do not measure the exact winding and helicity input of the field, since $\widehat{\Bv}$ and $\widehat{\uv}$ will not (generally) be tangent to the actual fields' integral curves. {This procedure allows us to study how recording the magnetic field at different heights, and then projecting these values onto a plane, affects the topological signatures. As this procedure is likely to be present in actual observations, it is important to consider it in our calculations.} To demonstrate the efficacy of our averaged measure, we perform two checks: 
\begin{enumerate}
\item{We directly link features of the distributions and time series of ${\cal L}$ and ${\cal H}$ (and hence $L$ and $H$) to features of the magnetic field's development.}
{
\item{We compare calculations with a moving photosphere to those assuming a fixed and horizontal photosphere.}
}

\end{enumerate}
{A similar comparison between  moving and fixed vector magnetogram production proecedures} has already been utilized succesfully in \cite{prior2019interpreting} to validate the moving boundary helicity calculations. A final point to make on this matter is that some progress has recently been made with regard to defining helicity inputs on moving boundaries \citep{schuck2019determining}. We do  not adopt this approach here as it does not currently match observational practice and our aim is to use flux emergence simulations to provide insight to observational findings. 
}

In what follows we drop the hat notation from equation (\ref{averagefield1}). {Any quantities obtained using the moving-average fields calculated using (\ref{averagefield1}), {i.e} the winding and helicity rates (\ref{helflux}), (\ref{windflux}), (\ref{totalinputs}) and the distributions (\ref{helinputdist}) and (\ref{windinputdist}), are labelled with a superscript $v$ for varying (e.g.  $H^v$). {For example, the varying (photosphere) helicity rate would be
\be\label{helflux2}
\frac{\d H^v}{\d t} = -\frac{1}{2\pi}\int_{P\times P} B^v_z(\av_1)B^v_z(\av_2)\frac{\d\theta^v}{\d t}\, \d^2 a_1\, \d^2 a_2,
\en
where $\Bv^v$ and $\uv^v$ are determined from equation (\ref{averagefield1}) and are used to find $B_z^v$ and $\d\theta^v/\d t$ (where all of these quantities are coplanar due to projection onto $P$).} The same quantities calculated using the fixed plane fields ${\Bv}(x,y,0)$ and $\uv(x,y,0)$ are labelled with a superscript $f$ for fixed (e.g. $H^f$).}

\section{Simulation analysis}  

We now present a detailed analysis of the simulations, considering each initial field strength in turn. Since there are similarities between the cases, we will describe the $B_0=7$ case in detail and only focus on specific highlights and differences for the $B_0=3$, 5 cases. For the basic properties of flux emergence, we refer the reader to the reviews by \cite{hood2012review} and \cite{cheung2014review}. Our focus here is on how to interpret the topological signatures associated with emergence.

\subsection{$B_0=7$, $\alpha=-0.4$}\label{sub_b07}

{
\subsubsection{Initial rise}\label{initial_rise}

\begin{figure}
\begin{center}
\subfigure[]{\includegraphics[scale=0.2115]{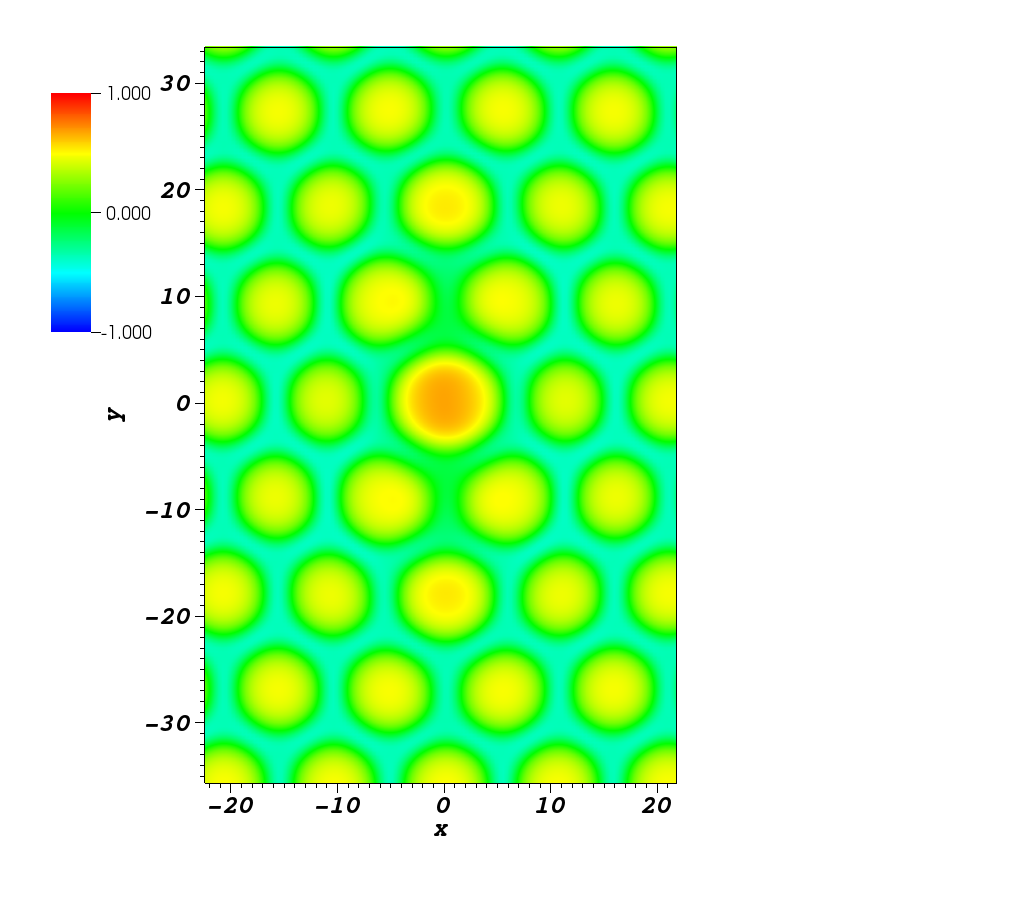}} \subfigure[]{\includegraphics[scale=0.2115]{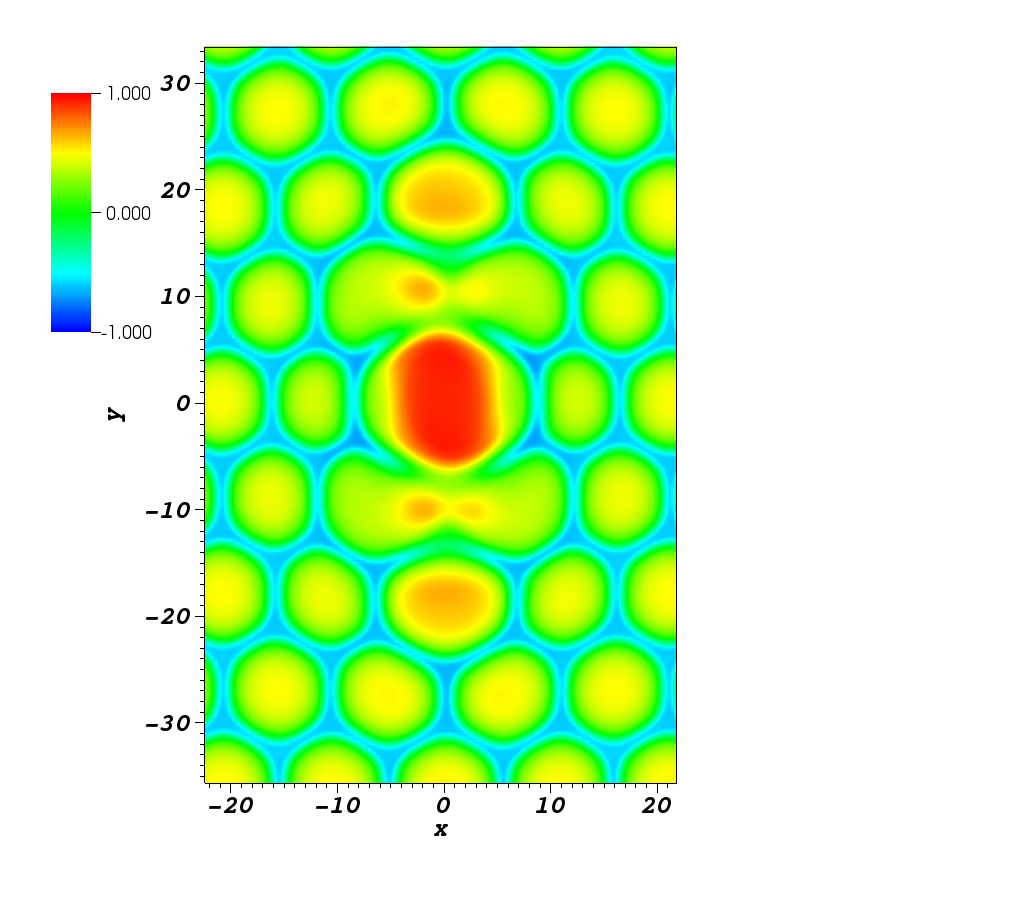}}\quad 
\caption{{ Maps of $u_z$ at $z=0$ at times (a) $t=15$ and (b) $t=20$. }\label{convection}}
\end{center}
\end{figure}
As we will be comparing the results of this work to those of non-convective simulations, we briefly describe how the tube  interacts with convection initially. Figure \ref{convection} shows maps of $u_z$ at $z=0$ for times (a) $t=15$ and (b) $t=20$, which are before the flux tube reaches the photospheric boundary. The flux tube rises at the centre of the domain, where the largest positive values of $u_z$ are present. Even before emergence, the tube is interacting with the convection cells. The tube is being buffeted by the cells (the plasma $\beta =2p/|\Bv|^2>1$ in this region) and is causing the cells directly above it to expand. The effect of the former feature will be made clear in figures that we present later. The latter feature is also found in simulations of convection that include radiative transfer \citep[e.g.][]{cheung2008emergence,tortosa2009emergence}. Later, it will be shown that as the flux tubes weaken and convection can deform them more easily, convection will be responsible for pulling mangetic field back down into the convection zone. Again, this this a feature that is found in convection simulations with radiative transfer \citep[e.g.][]{tortosa2009emergence,cheung2010emergence}. The effects of convection on emergence into the atmosphere will be described throughout the paper.}

\subsubsection{Helicity and winding - general features}\label{gen_features}
\begin{figure}
\begin{center}
\subfigure[]{\includegraphics[width=7.5cm]{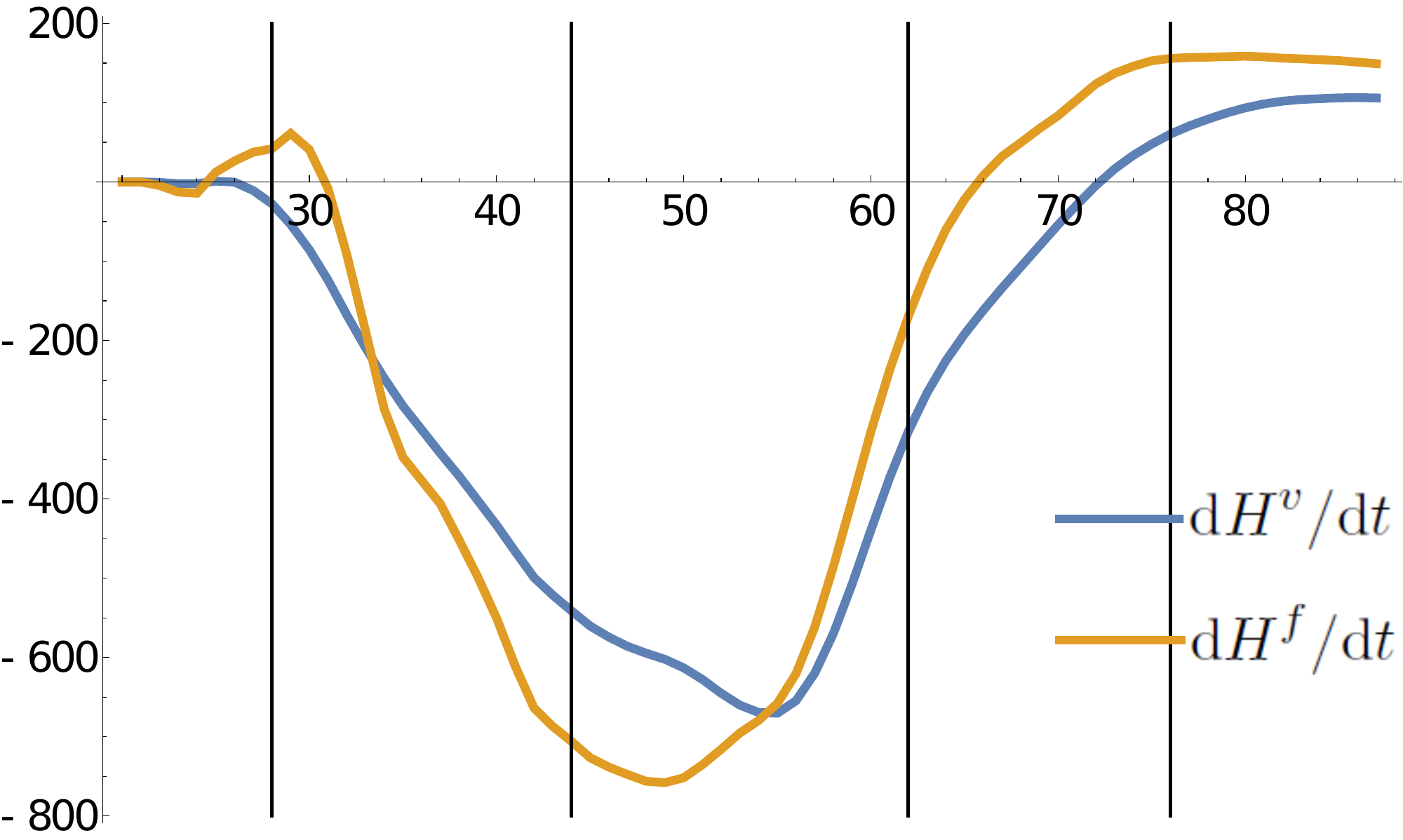}}\quad \subfigure[]{\includegraphics[width=7.5cm]{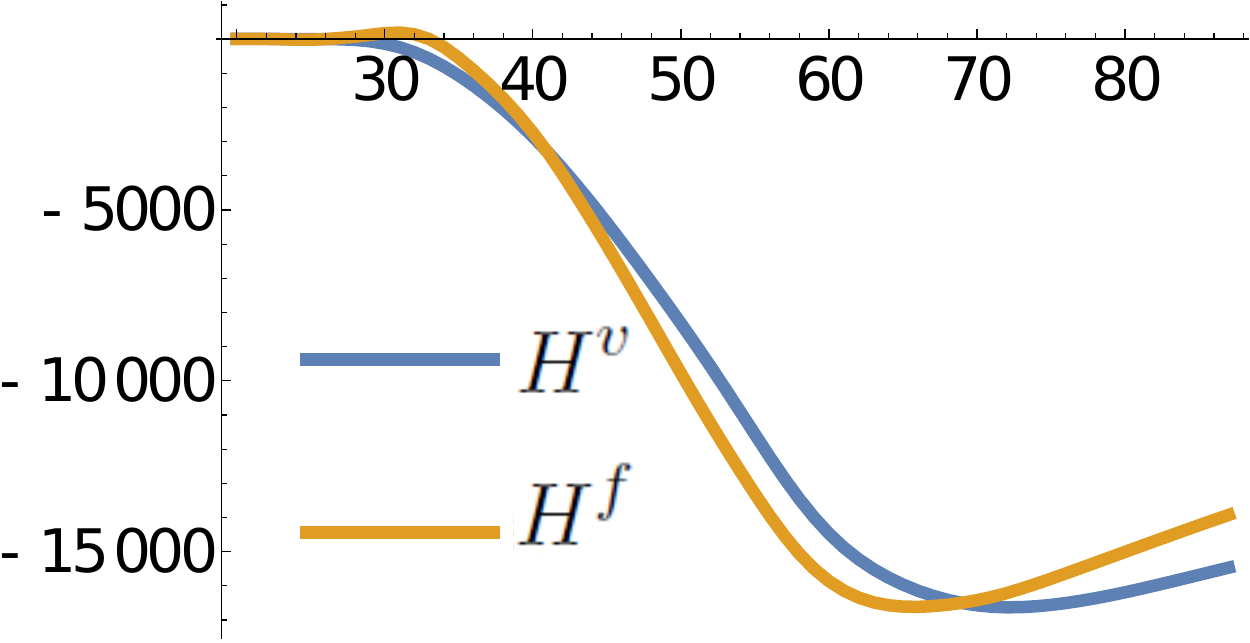}}\quad 
\subfigure[]{\includegraphics[width=7.5cm]{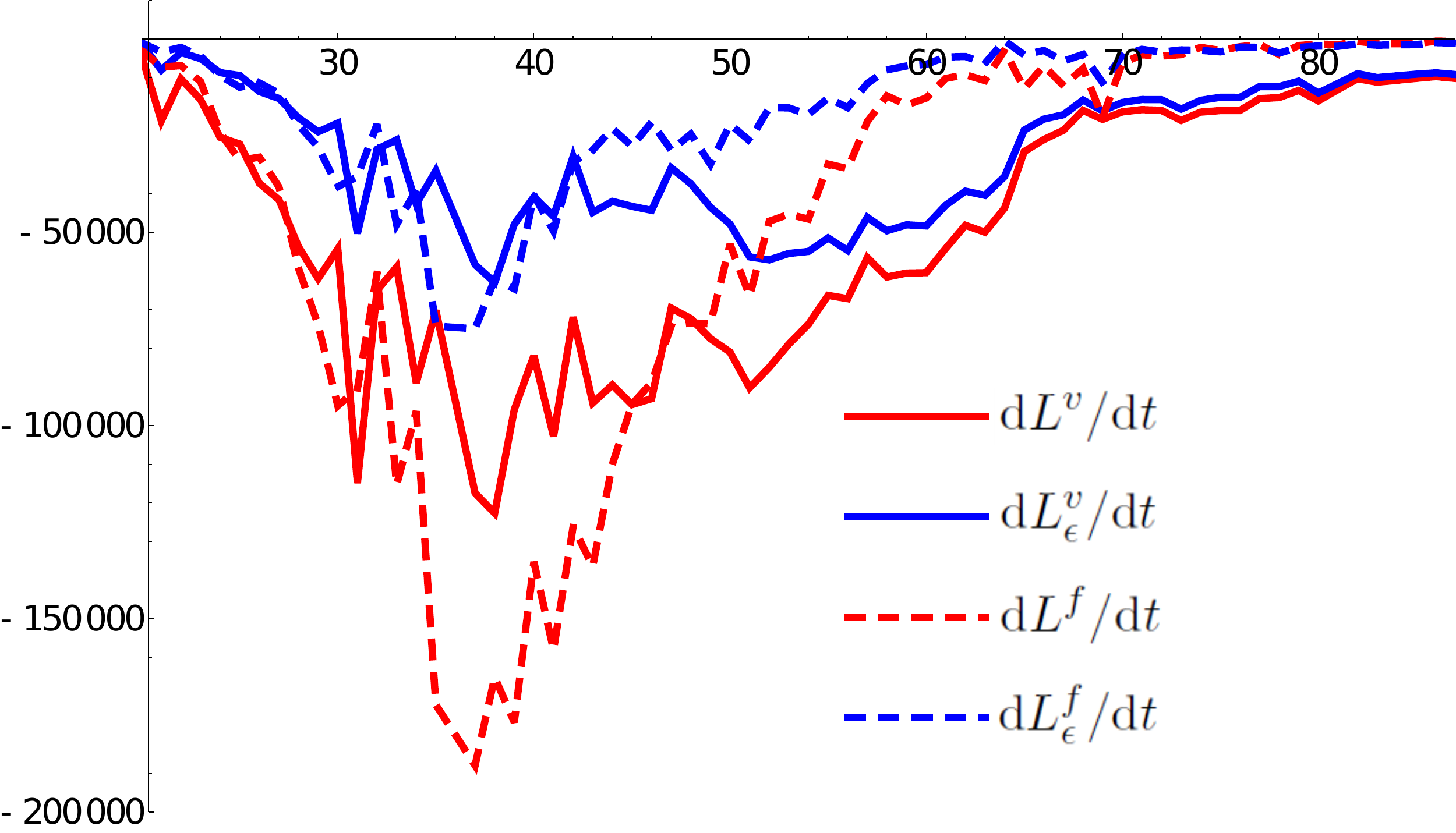}}\quad \subfigure[]{\includegraphics[width=7.5cm]{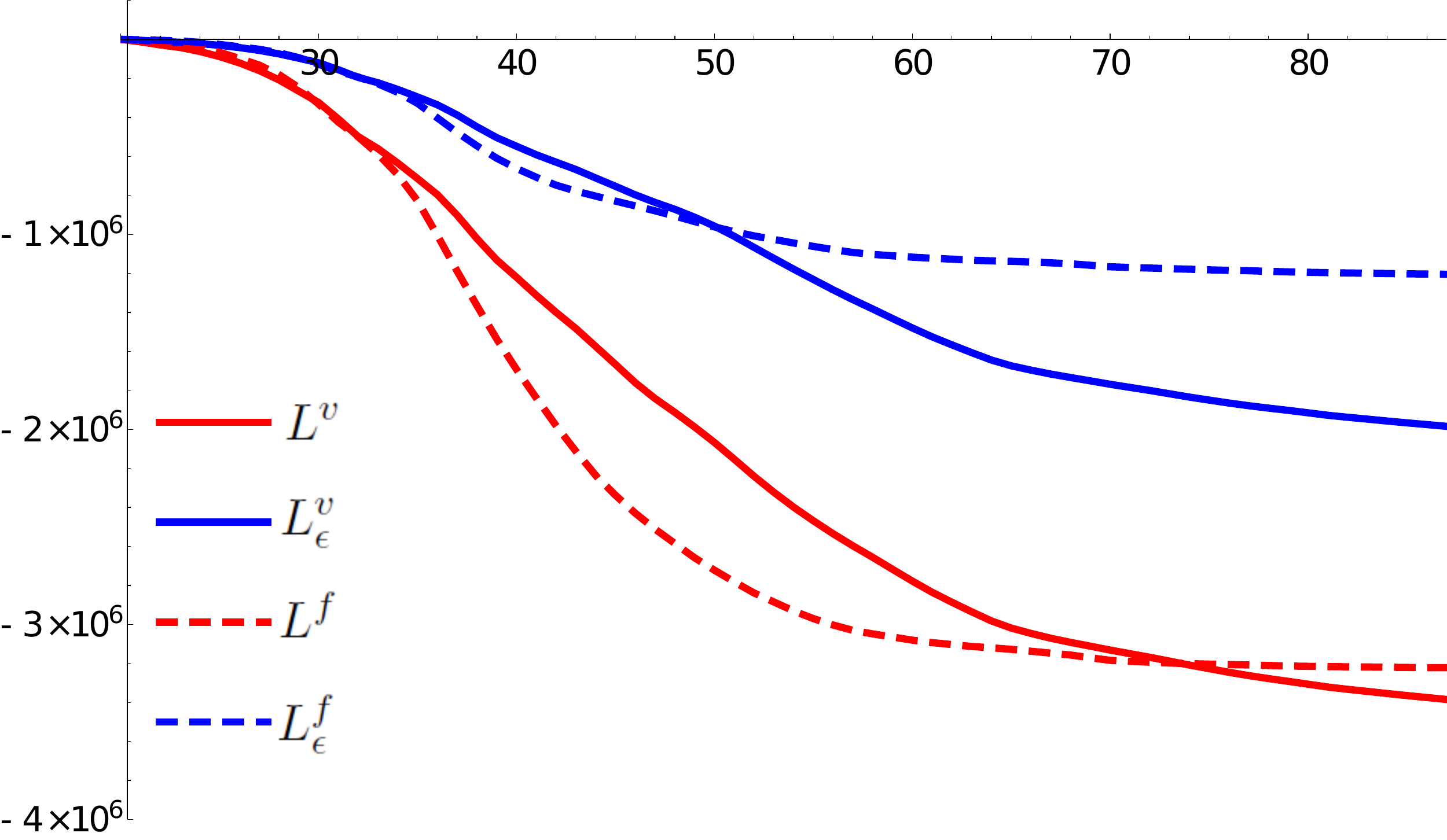}}
\caption{{ Helicity and winding time series for the $B_0=7$, $\alpha=-0.4$ case. Panel (a) shows the helicity input rates $\d{H}^v/\d{t}$ and $\d{H}^f/\d{t}$. The vertical lines indicate times for which helicity rate distributions are plotted in other figures. Panel (b) shows the integrated helicities $H^v(t)$ and $H^f(t)$ over the same period. Panel (c) displays the winding input rates $\d L^v/\d t$ (solid lines) and $\d L^f/\d t$ (dashed lines)  with no cut-off $\d L^{v/f}/\d t$ and with the cut-off $\varepsilon$ , $\d L^{v/f}_{\epsilon}/\d t$. Panel (d)  displays the total winidng inputs $L^v(t)$ and $L^v(t)$ with the same format of (c). }\label{helicityvaryb07}}
\end{center}
\end{figure}
The helicity input {rates  $\d H^v/\d t$  and $\d H^f/\d t$ are} shown in figure \ref{helicityvaryb07}(a).  {In both cases,} for most of the emergence period shown, the input is negative. This result is to be expected for a left-handed twisted flux rope. After $t=65$, however, the {rates} are overall positive. This input pattern has the obvious effect of injecting overall negative helicity into the atmosphere as indicated in figure \ref{helicityvaryb07}(b). { The peak in the negative input rate for the flat measure occurs before the peak of the varying measure. As we will see shortly, this is due to the fact that the $\rho=1$ surface is (generally) pushed upwards above $z=0$ due to the flux rope bringing up dense plasma. }

The winding input {rates $\d L^v/\d t$  and $\d L^f/\d t$  are shown in figure \ref{helicityvaryb07}(c) and are consistently negative in all cases}. The result is a net input of negative winding consistent with the field's structure (see the time series of $L(t)$ in figure \ref{helicityvaryb07}(d)). { In both cases the  winding time series} with and without and the cut-off  $\varepsilon =0.01$, are shown. {The cut-off series are qualitatively similar to those without the cut-off and, in both cases,  the magnitudes of values in the series with the cut-off are smaller. }

Before investigating what causes the particular features in the time series in figure \ref{helicityvaryb07}, we will compare these results to those of \cite{prior2019interpreting} where twisted flux tube emergence was studied in an atmosphere with a convectively-stable solar interior. Making this comparison will help us assess the effects of convection on the topological signatures.  

First, we point out that the magnetic field model used in this paper is different to that used in \cite{prior2019interpreting}. Here, a cylindrical tube is used rather than a toroidal tube. Although the emergence behaviours of these two magnetic field models are different \citep{mactaggart2009emergence}, we will show that convection plays a fundamental role in determining the shape of the time series. For the benefit of the reader, let us recap briefly the main results for twisted tube emergence from \cite{prior2019interpreting} (see their figure 11 in particular). First, the helicity rate exhibited an oscillatory character that was due to the flux rope core oscillating at the photosphere. The helicity input continued to become more negative in time. The winding rate was generally quite smooth, containing small oscillations and one significant spike. The winding input had a `step function' profile meaning that once most the twisted core of the flux tube emerged, there was little new topological complexity input into the atmosphere after this event.  With the above picture in mind, we will now investigate how the time series in figure \ref{helicityvaryb07} differ from this picture.

\begin{figure}
\begin{center}
\subfigure[$t=30$]{\includegraphics[width=6cm]{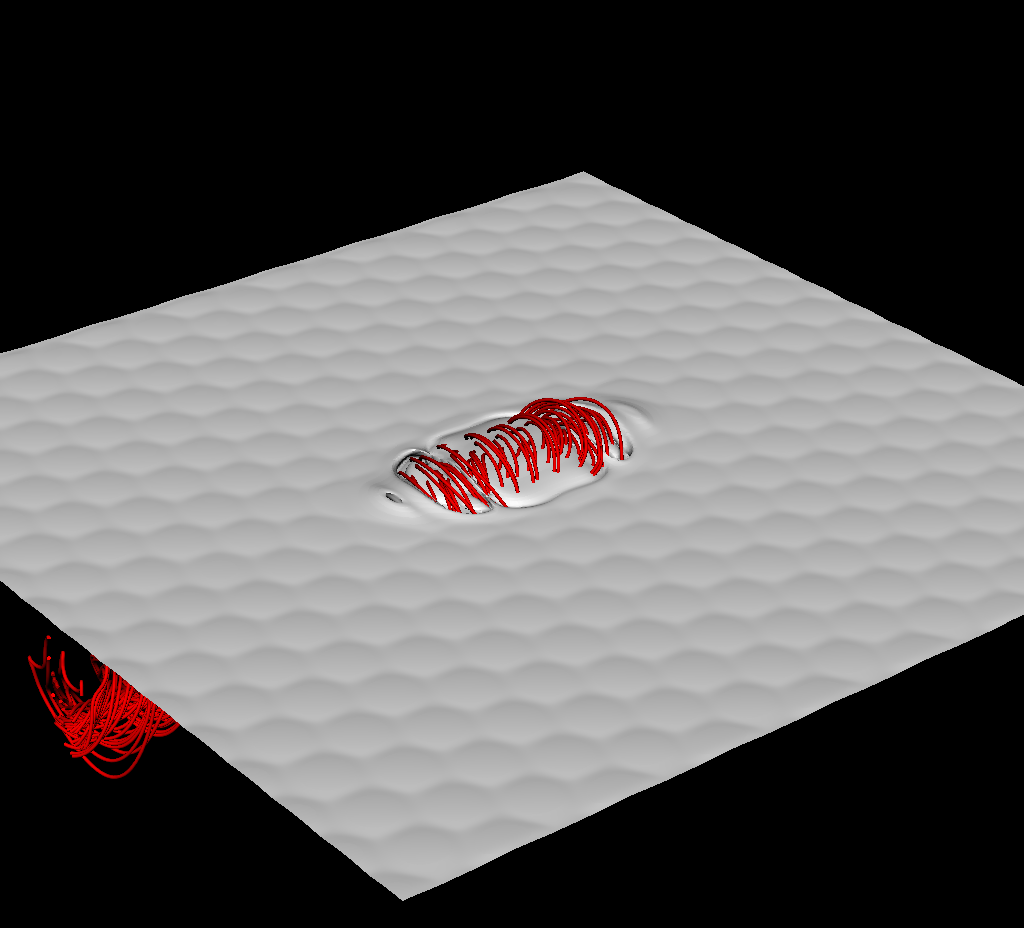}}\quad\subfigure[$t=30$]{\includegraphics[width=6cm]{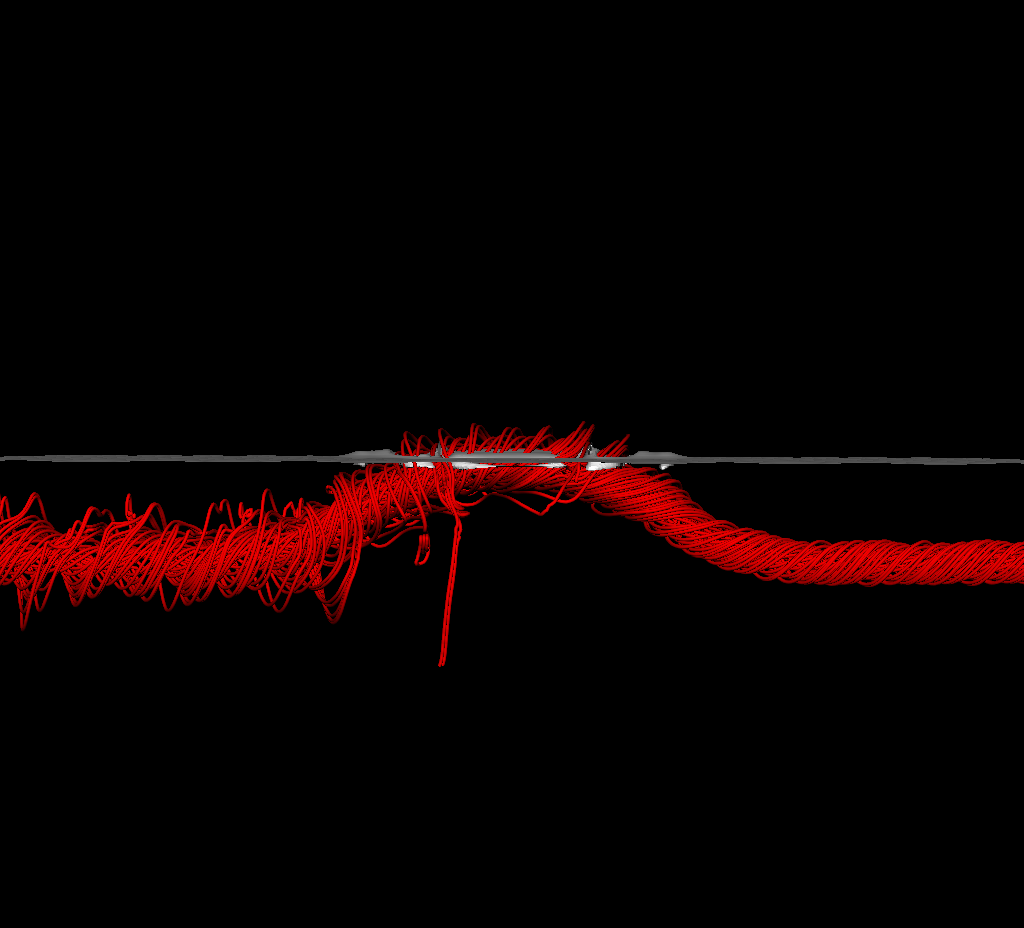}}\quad\subfigure[$t=50$]{\includegraphics[width=6cm]{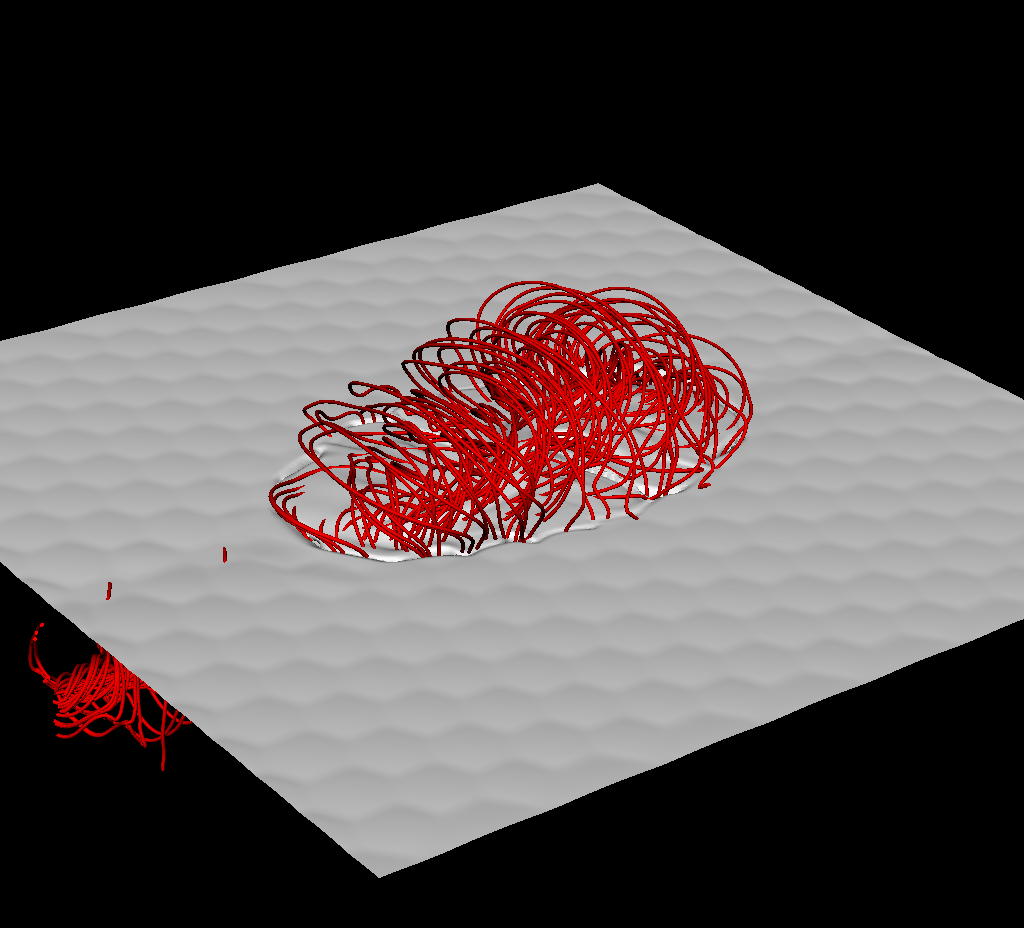}}\quad\subfigure[$t=50$]{\includegraphics[width=6cm]{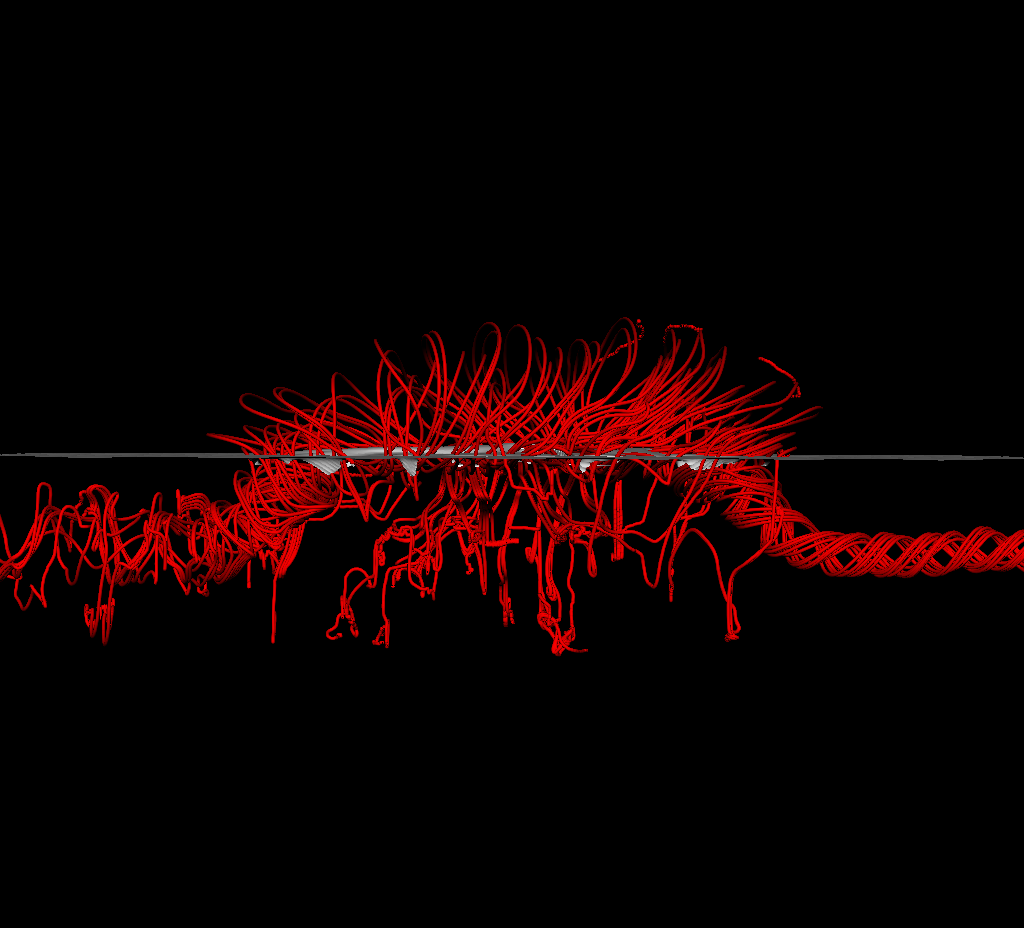}}\quad\subfigure[$t=80$]{\includegraphics[width=6cm]{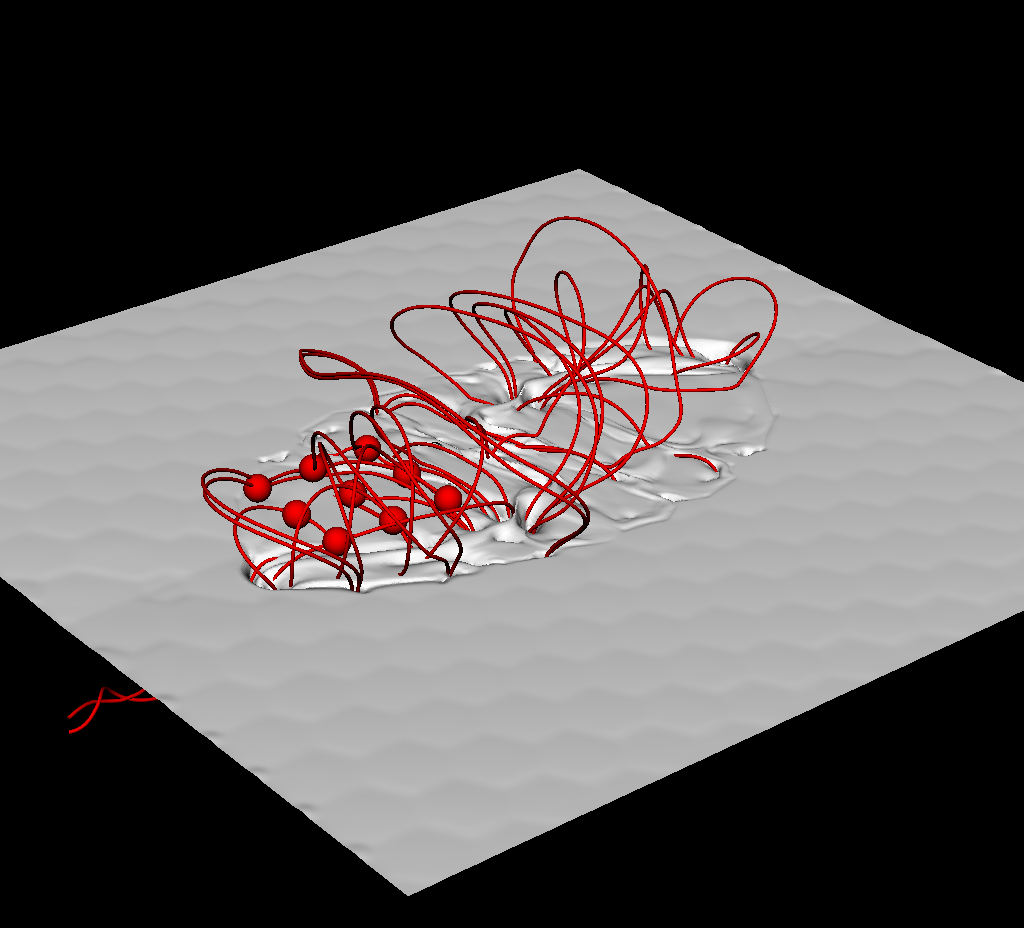}}\quad\subfigure[$t=80$]{\includegraphics[width=6cm]{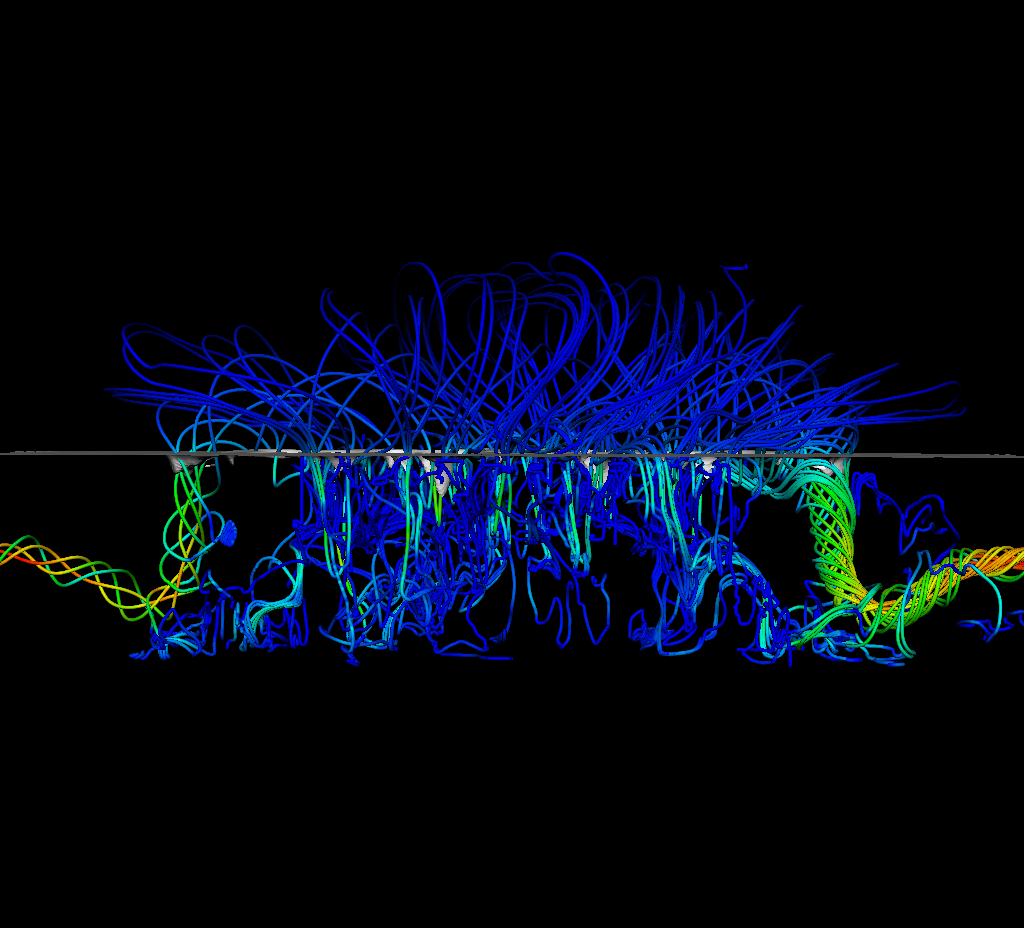}}
\caption{\label{b07fieldlines1} Field line renderings for the $B_0=7$, $\alpha=-0.4$ case at various points in its evolution. Panels (a) and (b) depict the field at $t=30$. The surface shown is that of the plasma density $\rho=1$. Panels (c) and (d) depict the field at $t=50$. Panel (e) depicts field lines in the emerged field at $t=80$. The spheres shown are the starting points for the field line integration. Panel (f) depicts field lines side-on, with lighter colours indicating increased field strength. }
\end{center}
\end{figure}

Snapshots of the field at various times during its emergence are shown in figure \ref{b07fieldlines1}. At $t=30$ (figures \ref{b07fieldlines1}(a) and (b)) the field is seen to have only just emerged above the $\rho=1$ surface. By $t=50$, the field has emerged much further into the atmosphere. Figure \ref{b07fieldlines1}(c) shows a sheared arcade and figure \ref{b07fieldlines1}(d) indicates that convection has begun to affect the flux tube field lines significantly. At $t=80$, in figures \ref{b07fieldlines1}(e) and (f), we see that the field has developed a strong `serpentine' structure  with field lines submerging and re-emerging at several points. This behaviour is commonly observed in nonlinear force-free coronal field extrapolations \citep[e.g.][]{pariat2004serpentine,pariat2009current,harra2010response,valori2012nonlinear} and is highlighted in figure \ref{b07fieldlines1}(e) which shows seed points at one side of the region and field lines, emanating from these point, submerging and emerging throughout the active region. The serpentine behaviour is further indicated in figure \ref{b07fieldlines1}(f) where the field lines are coloured by field strength and several loops of significantly strong field strength are present below the photosphere. Such behaviour is not found in non-convective simulations.

\begin{figure}
\begin{center}
\subfigure[$t=30$]{\includegraphics[width=6.5cm]{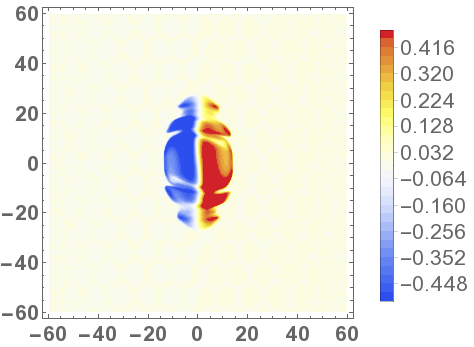}}\quad \subfigure[$t=50$]{\includegraphics[width=6.5cm]{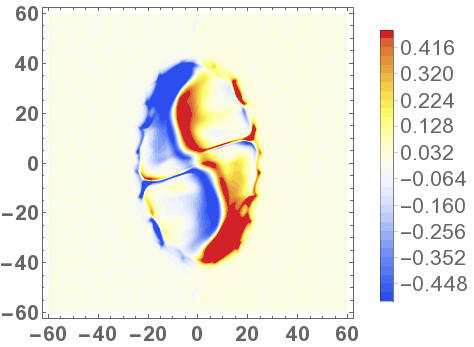}}\quad\subfigure[$t=80$]{\includegraphics[width=6.5cm]{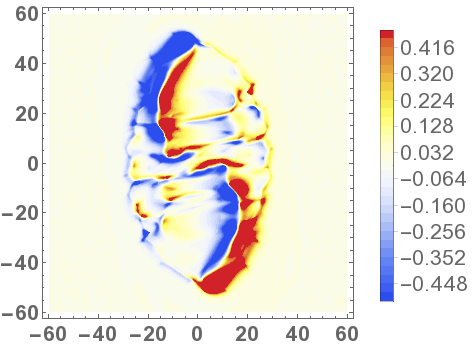}}
\caption{\label{b07magnetograms} { Varying field ($B_z^v$)}  magnetograms for the times shown in figure \ref{b07fieldlines1}. }
\end{center}
\end{figure}
{Magnetograms (of $B_z^v$) }of the emerging field, at the times shown in figure \ref{b07fieldlines1}, are displayed in figure \ref{b07magnetograms}. These magnetograms exhibit many features that are similar to those from non-convective simulations. There are, however, some important differences due to convection. At $t=30$ in figure \ref{b07magnetograms}(a), the field has just reached above the photosphere and there is no clear signal of twisted field. There is some disruption of the field due to convection cells. By $t=50$ in figure \ref{b07magnetograms}(b), the sigmoidal pattern  \citep[e.g.][]{fan2004numerical,archontis2009sigmoid} of the twisted field has emerged and some disruption due to convection is evident (particularly around the edge of the region). At $t=80$ in figure \ref{b07magnetograms}(c), we see many more smaller-scale positive and negative flux features surrounding the sigmoid, indicative of the serpentine behaviour displayed in figure \ref{b07fieldlines1}. 

To summarize the results so far, as the field emerges through the photosphere and expands into the atmosphere, it becomes more susceptible to deformation by convection (since the field strength is, overall, becoming weaker). The action of convection is to create a serpentine structure by pulling loops of magnetic field beneath the photosphere. Evidence for this behaviour is shown in field line plots (figure \ref{b07fieldlines1}) and magnetograms (figure \ref{b07magnetograms}). The question now is how to relate this behaviour to the topological information given in figure \ref{helicityvaryb07}? To proceed, we will focus on the helicity first and then the winding.

\subsubsection{Helicity evolution}\label{sec_hel_evo}
The {varying} helicity time series in figures \ref{helicityvaryb07}(a) and (b) differ substantially from those for non-convective simulations \citep[e.g.][]{sturrock2016sunspot,prior2019interpreting}. Whereas for the non-convective case helicity continues to become more negative, despite an oscillation about this mean trend, the helicity input in the convective case eventually stops becoming more negative and changes direction, becoming more positive. The time scale of this turning point in the helicity flux coincides with the development of serpentine field lines driven by convection. By examining specific features in detail, we will now show that convection is ultimately responsible for the turning point in the helicity time series.

\begin{figure}
\begin{center}
\subfigure[$t=27$]{\includegraphics[width=6.5cm]{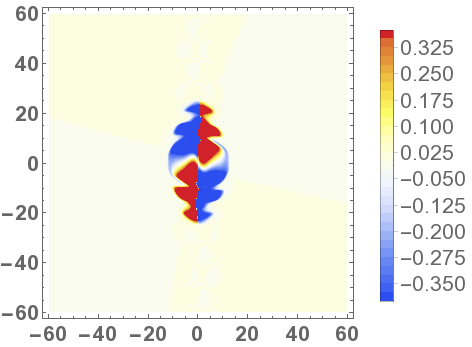}}\quad \subfigure[$t=44$]{\includegraphics[width=6.5cm]{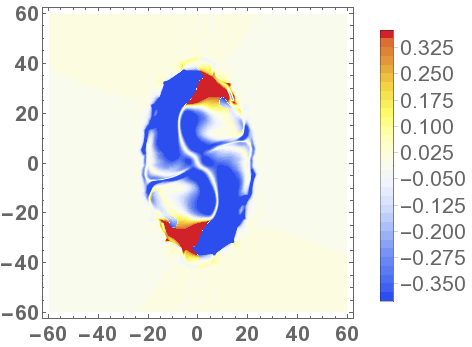}}\quad\subfigure[$t=62$]{\includegraphics[width=6.5cm]{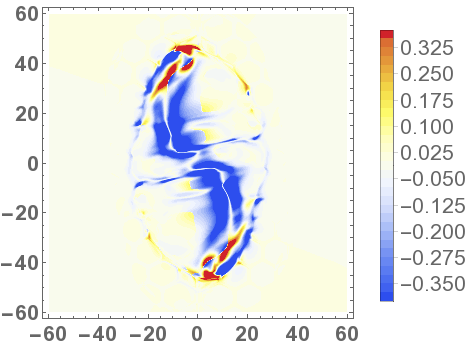}}\quad\subfigure[$t=76$]{\includegraphics[width=6.5cm]{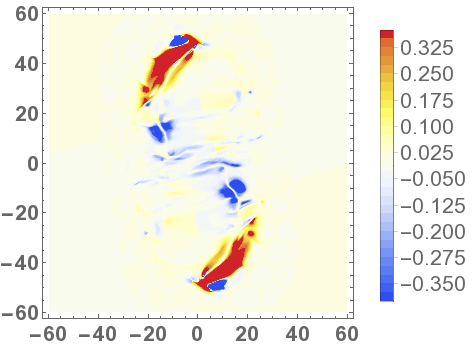}}
\caption{\label{helicitydistsb07} {Varying} helicity input rate distributions $\d \wh(\av_0)/\d t$ at the times indicated by vertical lines in the helicity rate time series shown in figure \ref{helicityvaryb07}(a), i.e. (a) $t=27$, (b) $t=44$, (c) $t=62$ and (d) $t=76$. }
\end{center}
\end{figure}

\begin{figure}
\begin{center}
\subfigure[$t=27$]{\includegraphics[width=6.5cm]{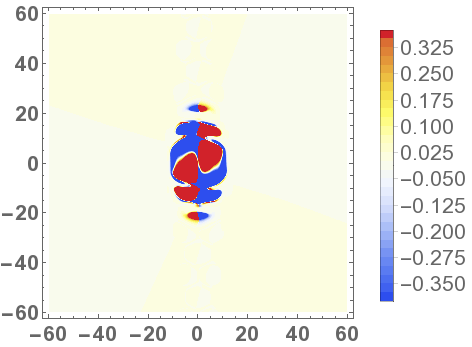}}\quad \subfigure[$t=44$]{\includegraphics[width=6.5cm]{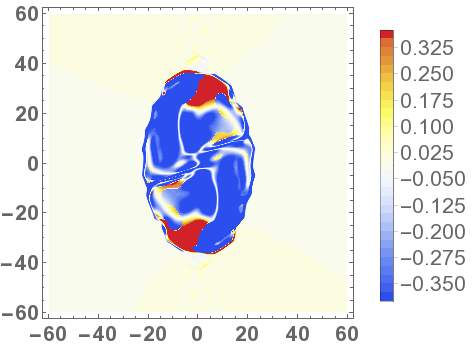}}\quad\subfigure[$t=62$]{\includegraphics[width=6.5cm]{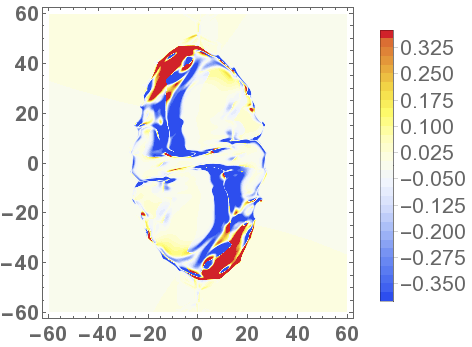}}\quad\subfigure[$t=76$]{\includegraphics[width=6.5cm]{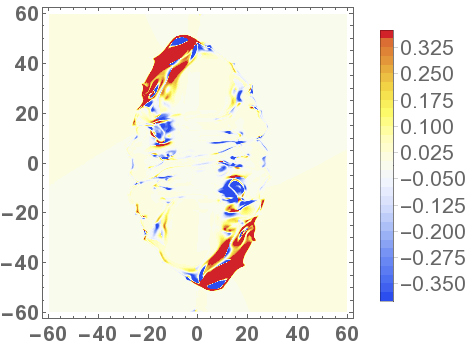}}
\caption{\label{helicitydistsb07flat} { Flat} helicity input rate distributions $\d \whf(\av_0)/\d t$ at the times indicated by vertical lines in the helicity rate time series shown in figure \ref{helicityvaryb07}(a), i.e. (a) $t=27$, (b) $t=44$, (c) $t=62$ and (d) $t=76$. }
\end{center}
\end{figure}

Figure \ref{helicitydistsb07} shows the field line helicity rate distributions $\d\wh(\av_0)/\d t$ for times representing key phases in the helicity evolution. Figures \ref{helicitydistsb07}(a)-(c) reveal an evolution that is similar qualitatively to that found in non-convective simulations. As the magnetic field first reaches the photosphere (figure \ref{helicitydistsb07}(a)), there is an approximate balance of positive and negative helicity.  Later (figure \ref{helicitydistsb07}(b)), as emergence of the tube with left-handed twist proceeds, the helicity rate distribution is dominated, naturally, by negative helicity. Later still (figure \ref{helicitydistsb07}(c)), the twisted core of the flux tube reaches the photosphere and this feature dominates the negative helicity rate signature. Although the band of negative helicity rate density, representing the twisted core, follows the sigmoidal polarity inversion line clearly, it also shows locations of disruption, particularly near (0,0).

 At $t=76$ (figure \ref{helicitydistsb07}(d)) the band of negative helicity rate density has been distorted significantly, although the regions of positive helicity rate density at $y=\pm 40$ persist. To investigate what causes the features in the helicity rate map in figure \ref{helicitydistsb07}(d), we need to investigate the behaviour of other quantities in the locations of particular helicity features.  { Before doing so, however, we briefly consider the equivalent distributions in the flat case $\d\whf(\av_0)/\d t$ in figure \ref{helicitydistsb07flat}. At each stage there is significant qualitative similarity between the two sets of distributions. In particular, at $t=76$ the regions of strong positive helicity rate present in the varying case are also present in the flat case. The distribution in the middle of the domain differs significantly with much small-scale helicity rate structure present in the flat case which is not there in in the varying case. It is the large positive regions which ensure that the helicity time series in figures \ref{helicityvaryb07}(a) and (b) have very similar magnitudes (if slightly offset temporally).}

\begin{figure}
\begin{center}
\subfigure[]{\includegraphics[width=7cm]{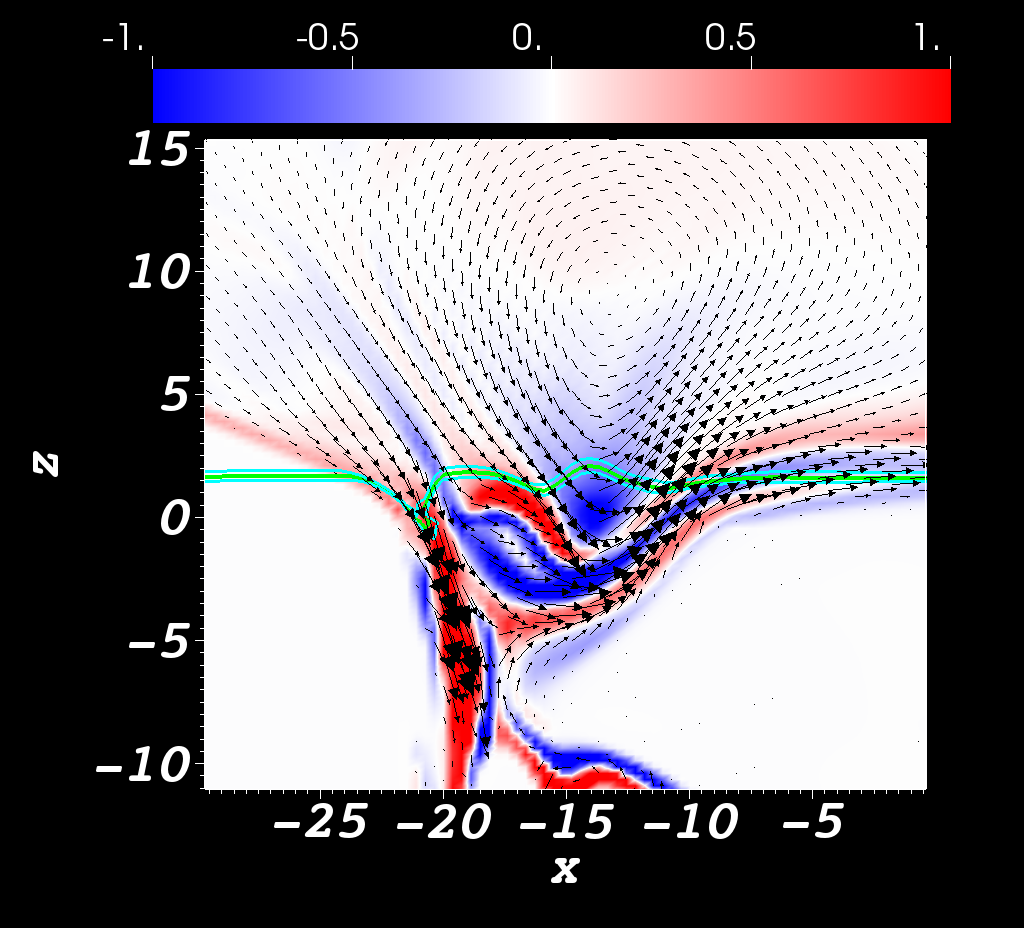}}\quad\subfigure[]{\includegraphics[width=7cm]{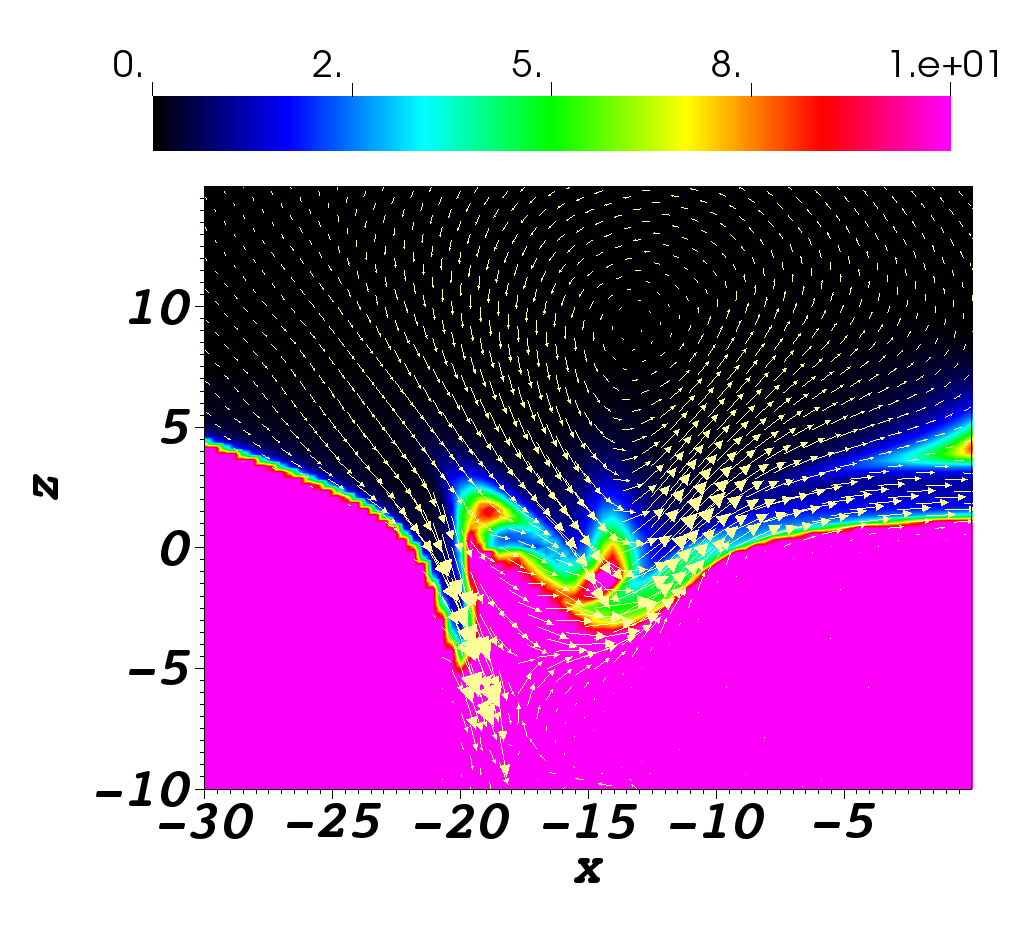}}
\quad\subfigure[]{\includegraphics[width=7cm]{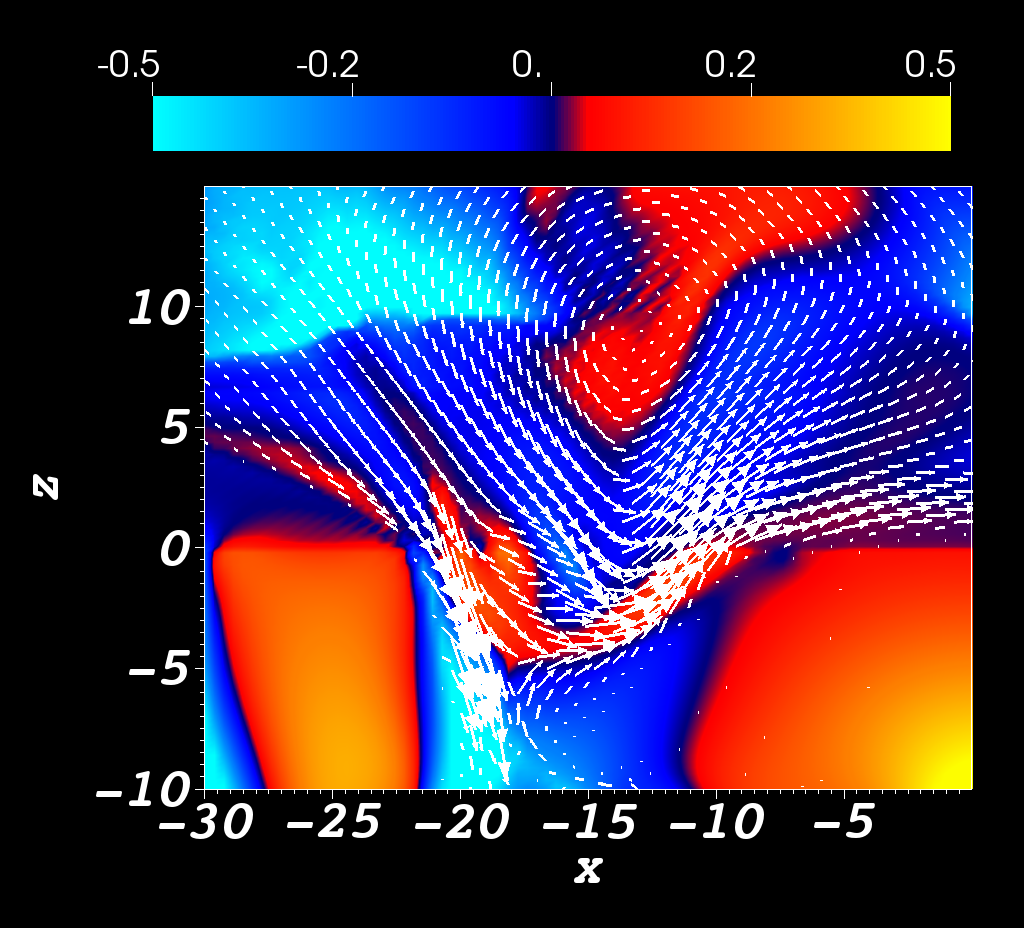}}
\caption{\label{slicesb07_1}Slices at $t=76$ taken in the $x$-$z$ plane at $y=30$ which intersects a region of strong positive helicity input rate $\d \wh(\av_0)/\d t$ in figure \ref{helicitydistsb07}(d).  In (a) a map of the $y$-component of $\bnab\times\Bv$ is shown and the vector field arrows represent the projection of the $\Bv$-field onto the plane. The green curve represents the intersection of the $\rho=1$ surface and the plane and the cyan curves are the same but for $\rho=0.8,1.2$. In panel (b) a map of the plasma $\beta$ is displayed with the $\Bv$-field vector arrows superimposed on top. In panel (c) a map of $u_z$ is displayed, again with the vector arrows of the $\Bv$-field superimposed on top. }
\end{center}
\end{figure}

Figure \ref{slicesb07_1} displays slices (the $x$-$z$ plane) of various quantities at $y=30$ and $t=76$, corresponding to a region of strong positive helicity rate (as shown in figure \ref{helicitydistsb07}(d)). Figure  \ref{slicesb07_1}(a) displays a map of $\ev_y\bdot\bnab\times\Bv$, vector arrows of the in-plane magnetic field and photospheric boundaries corresponding to $\rho=1$ (green) and $\rho=$0.8, 1.2 (cyan). The twisted core of the emerged flux tube can be seen far above the photospheric boundary in figure \ref{slicesb07_1}(a). Just below the photospheric boundary at $x\approx -15$, there are strong currents of both signs. Further left at $x\approx -20$, there is a thin region of strong downward-pointing magnetic field, indicative of the serpentine behaviour visualized in figure \ref{b07fieldlines1}(f).

Figure \ref{slicesb07_1}(b) shows a map of the plasma $\beta$ and vector arrows of the in-plane magnetic field.  In the region just below the photosphere at $x\approx -15$, the plasma $\beta$ varies between 2 and 8, with a complex mixing pattern that suggests a competition for dominance between the plasma and magnetic forces. The plasma $\beta$ is significantly low for part of the region in which there is a strong downward-pointing field at $x\approx -20$. This value, however, quickly shifts to a much higher plasma $\beta$ even where the magnetic field is still strong. This suggests that plasma motions (due to convection) are pulling down magnetic loops. 

This conclusion is backed by the behaviour displayed in figure \ref{slicesb07_1}(c), which displays a map of $u_z$ and the in-plane $\Bv$-vectors.  In the thin region of downward-pointing field at $x\approx-20$, there are strong downward flows coincident with the location where the plasma $\beta$ rises sharply. In the mixed current region at $x\approx-15$, there is a strong correlation between the vertical component of the magnetic field and the sign of $u_z$, where $\beta\approx 5$. This evidence points strongly to plasma motions in the convectively-unstable layer generating the structural complexity of the later-stage magnetic field evolution.

\begin{figure}
\begin{center}
\subfigure[]{\includegraphics[width=7cm]{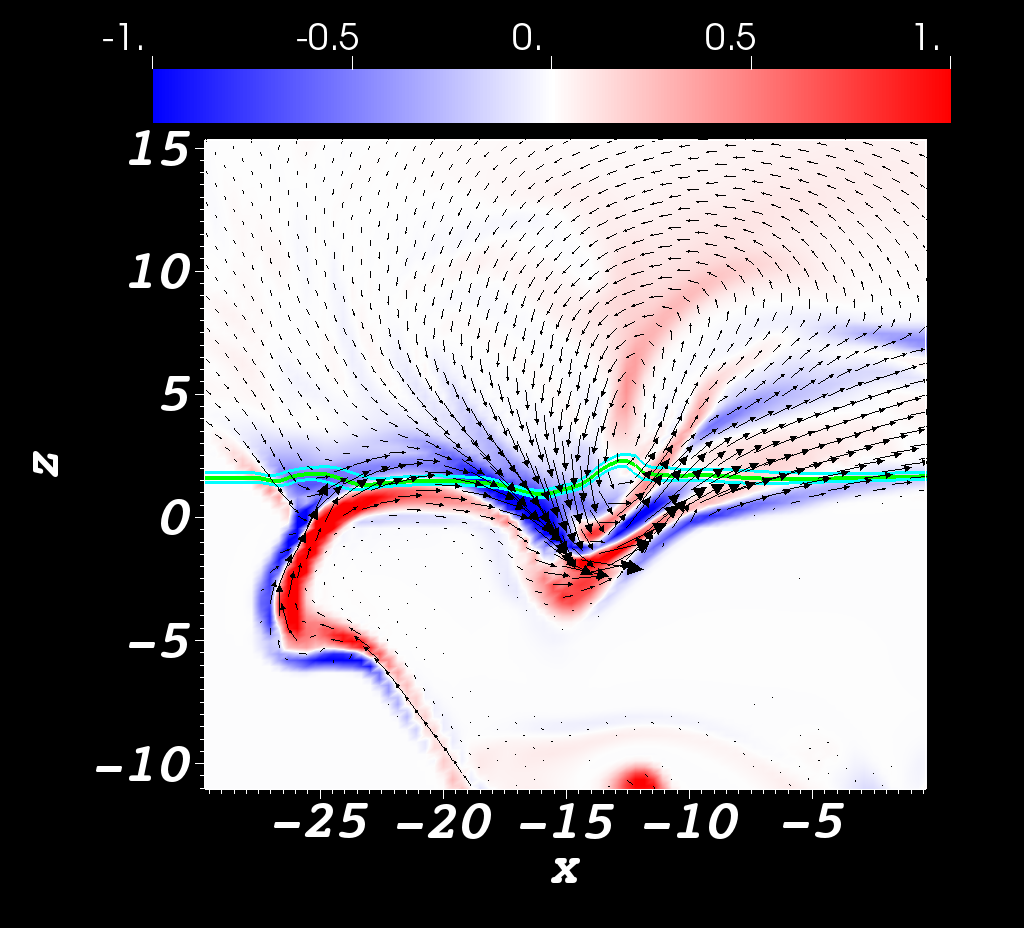}}\quad\subfigure[]{\includegraphics[width=7cm]{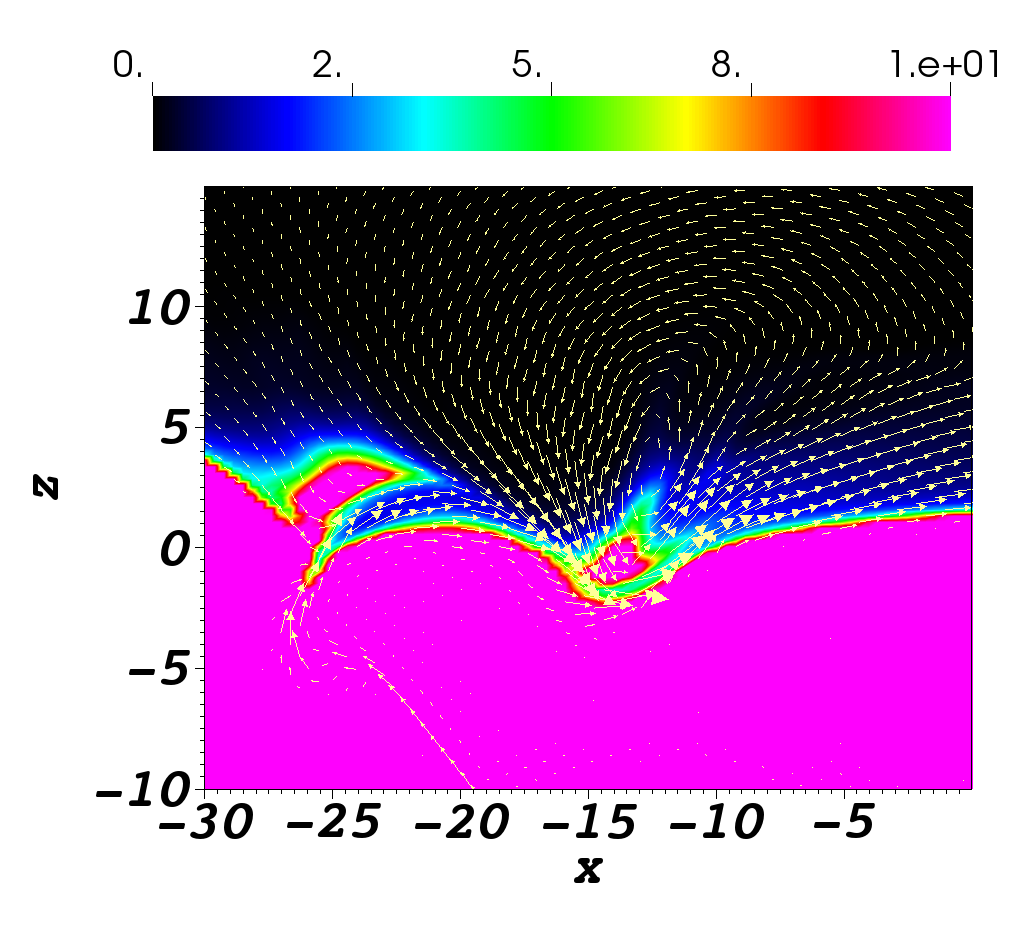}}
\quad\subfigure[]{\includegraphics[width=7cm]{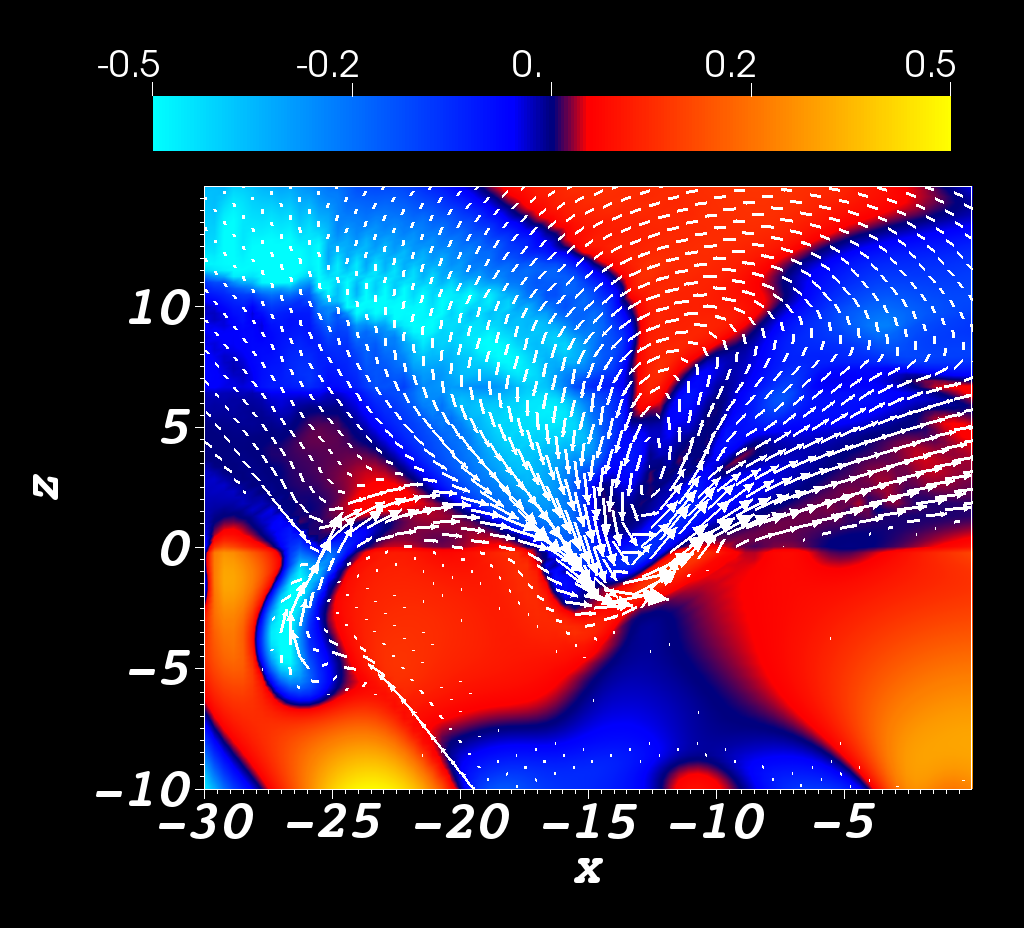}}
\caption{\label{slicesb07_2}Slices at $t=76$ taken in the $x$-$z$ plane at $y=10$ which intersects a region of strong negative helicity input rate $\d \wh(\av_0)/\d t$ in figure \ref{helicitydistsb07}(d).  In (a) a map of the $y$-component of $\bnab\times\Bv$ is shown and the vector field arrows represent the projection of the in-plane $\Bv$-field. The green curve represents the intersection of the $\rho=1$ surface and the plane and the cyan curves are the same but for $\rho=0.8,1.2$. In panel (b) a map of the plasma $\beta$ is displayed with the $\Bv$-field vector arrows superimposed on top. In panel (c) a map of $u_z$ is displayed, again with the vector arrows of the $\Bv$-field superimposed on top. }
\end{center}
\end{figure}

Figure \ref{slicesb07_2} shows the same quantities (in the same format) as figure \ref{slicesb07_1} for $y=10$. This slice corresponds to a region of strong negative helicity rate density, as shown in figure \ref{helicitydistsb07}(d). In figure  \ref{slicesb07_2}(a), the flux rope core is found at the photospheric boundary at $x\approx-13$. Below the photospheric boundary, there are strong currents of both signs, similar to the situation at $y=30$ in figure  \ref{slicesb07_1}. At $x\approx-25$ and below the flux tube core there is a strong upward-pointing magnetic field (more distorted and with more complex current structure than in the $y=30$ case) that is another example of the serpentine loop structure. In figure \ref{slicesb07_2}(b), there is a `pocket' of very strong plasma $\beta$ at the base of the flux tube core at the photosphere. In figure \ref{slicesb07_2}(c), the vertical direction of the magnetic field correlates strongly with the direction of $u_z$. Also, to the left of the twisted core, magnetic field is being pulled down in a similar way to the case at $y=30$. These results add weight to our claim that convection is primarily responsible for the increased (serpentine) structural complexity with time. Further, the flux rope core itself behaves in a serpentine manner, being both fully emerged and submerged at various locations.

\begin{figure}
\begin{center}
\quad\subfigure[]{\includegraphics[width=8cm]{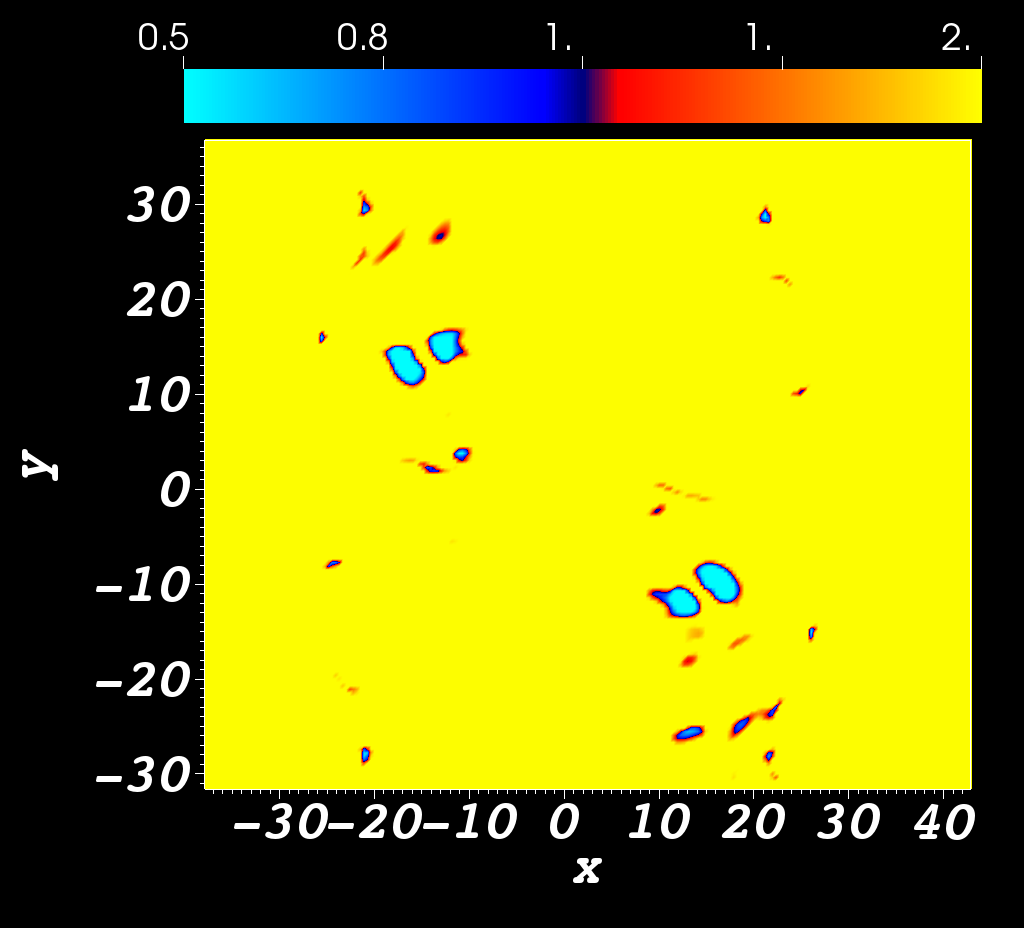}}
\caption{\label{dendip}A distribution of density in the plane $z=0$ at $t=76$. Regions of low density correspond to downward spikes in the surface of $\rho=1$. }
\end{center}
\end{figure}

{
\subsubsection{Significant convective action}\label{sec_conv}
We see in figure \ref{slicesb07_1} that the $\rho=1$ surface is pulled down in the region where the field is being dragged down. As shown in figure \ref{dendip}, there are regions where this effect is extreme with two regions of much lower density (on the $z=0$ plane) compared to the surrounding area (at $y=14$)}. These two regions indicate downward spikes in the photospheric boundary exactly where the serpentine field enters and leaves the photosphere. 
\begin{figure}
\begin{center}
\subfigure[]{\includegraphics[width=7cm]{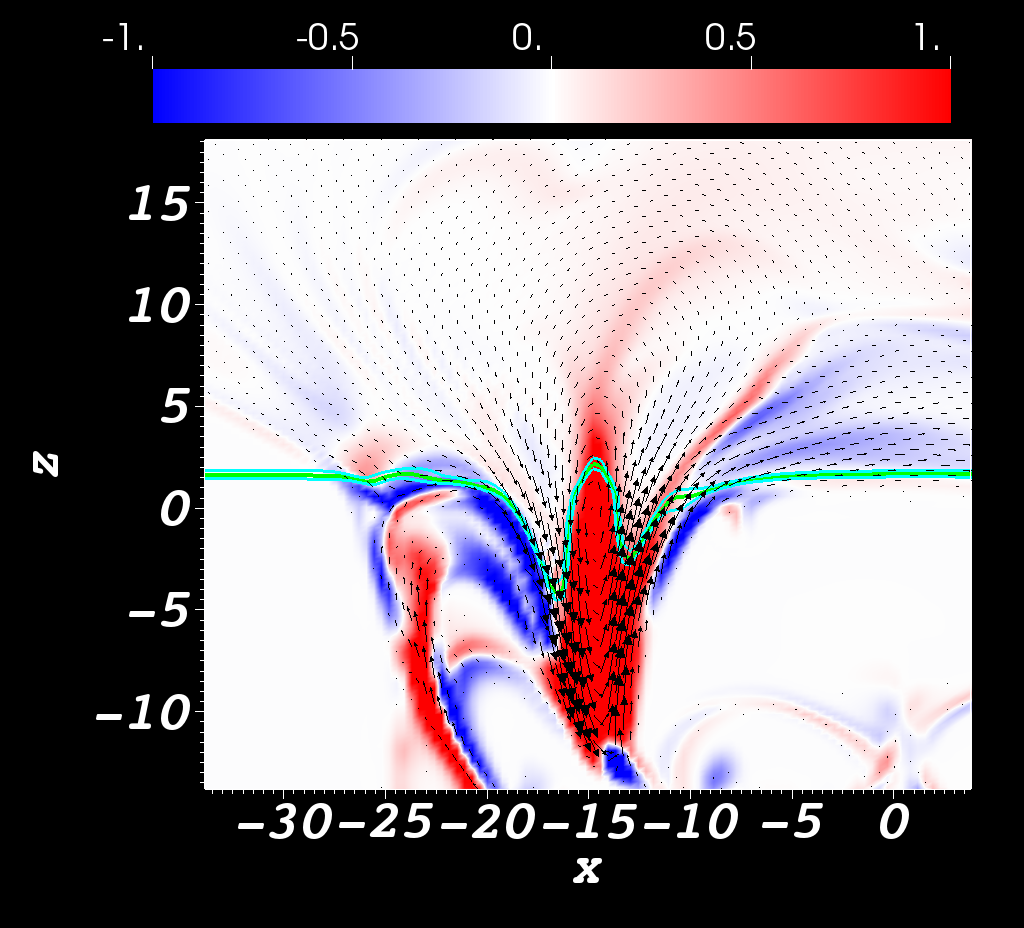}}\quad\subfigure[]{\includegraphics[width=7cm]{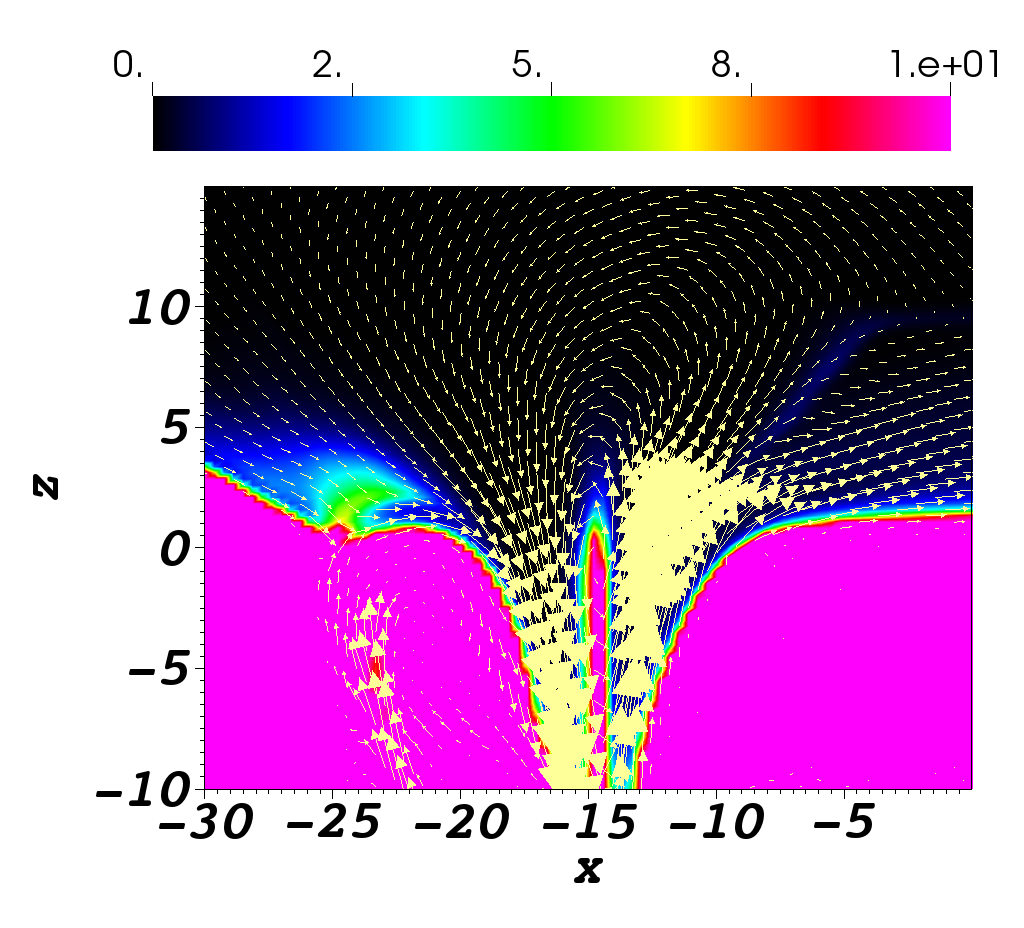}}
\quad\subfigure[]{\includegraphics[width=7cm]{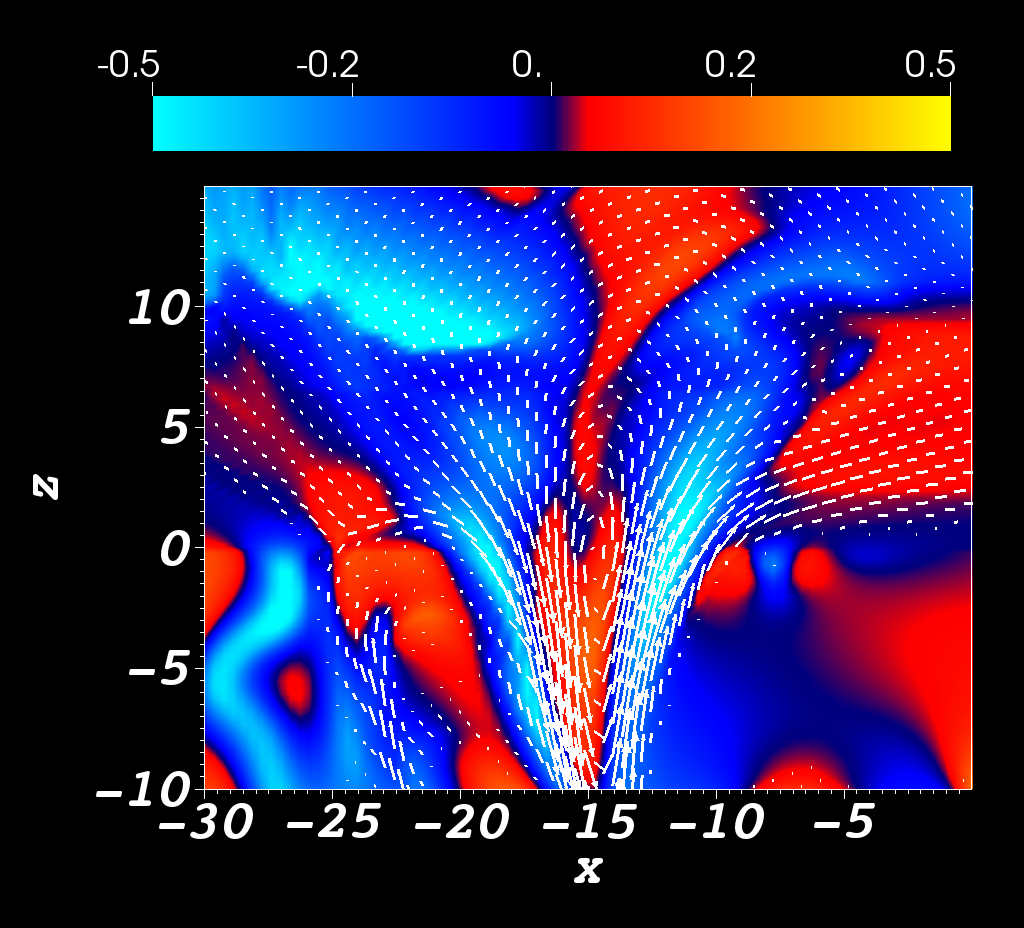}}
\caption{\label{slicesb07_3}Slices at $t=76$ taken in the $x$-$z$ plane at $y=14$ which intersects a region of bipolar helicity input rate $\d \wh(\av_0)/\d t$ in figure \ref{helicitydistsb07}(d).  In (a) a map of the $y$-component of $\bnab\times\Bv$ is shown and the vector field arrows represent the projection of the in-plane $\Bv$-field. The green curve represents the intersection of the $\rho=1$ surface and the plane and the cyan curves are the same but for $\rho=0.8,1.2$. In panel (b) a map of the plasma $\beta$ is displayed with the $\Bv$-field vector arrows superimposed on top. In panel (c) a map of $u_z$ is displayed, again with the vector arrows of the $\Bv$-field superimposed on top. }
\end{center}
\end{figure}

Figure \ref{slicesb07_3} displays slices at $y=14$ and has exactly the same format as figures \ref{slicesb07_1} and \ref{slicesb07_2}. In figure \ref{slicesb07_3}(a), the photospheric boundary dips significantly around the central twisted core, which is almost completely below the photospheric boundary. {The surface $\rho=1$ is also far below the $z=0$ plane indicating this effect features strongly in both measures of the helicity input}. From figure \ref{slicesb07_3}(b), the plasma $\beta$ at the twisted core is high compared to the two dips either side of it, where there is very low plasma $\beta$ and relatively strong upward and downward-pointing magnetic field. Figure \ref{slicesb07_3}(c) confirms that the dips on either side of the core are being pulled down. {In figure \ref{helicitydistsb07flat}(d) (the flat distribution at $t=76$), we see this leads to a} bipolar signature in the helicity rate maps {which can} be understood with reference to the field line emergence velocity in equation (\ref{dadt}), namely $-u_z\Bv_{\|}/B_z$. With $|\Bv_{\|}|>0$ and $u_z<0$ for both dips, the changing sign of $B_z$ in the dips results in the observed bipolar structure. 

All of the features that we have investigated in this section point strongly towards convection deforming the magnetic field and changing the behaviour of the helicity input. At the start of emergence, the tube is affected by convection but is still strong enough to resist significant deformation. Later, as the field expands into the atmosphere, it becomes weaker and thus more prone to deformation by convection (in combination with other common effects associated with emergence, such a plasma drainage).

{ 
\subsubsection{Winding series and distributions}\label{winddistsb07}
\begin{figure}
\subfigure[]{\includegraphics[width=7.5cm]{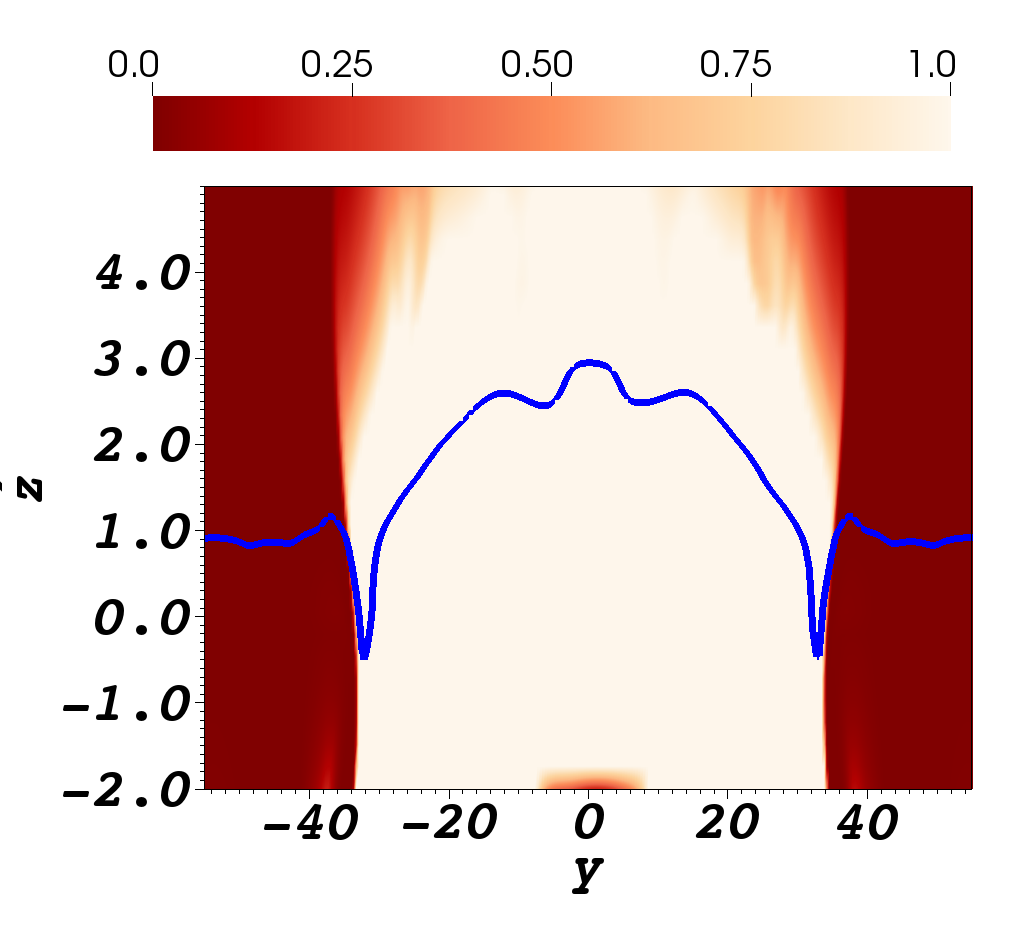}}\quad\subfigure[]{\includegraphics[width=7.5cm]{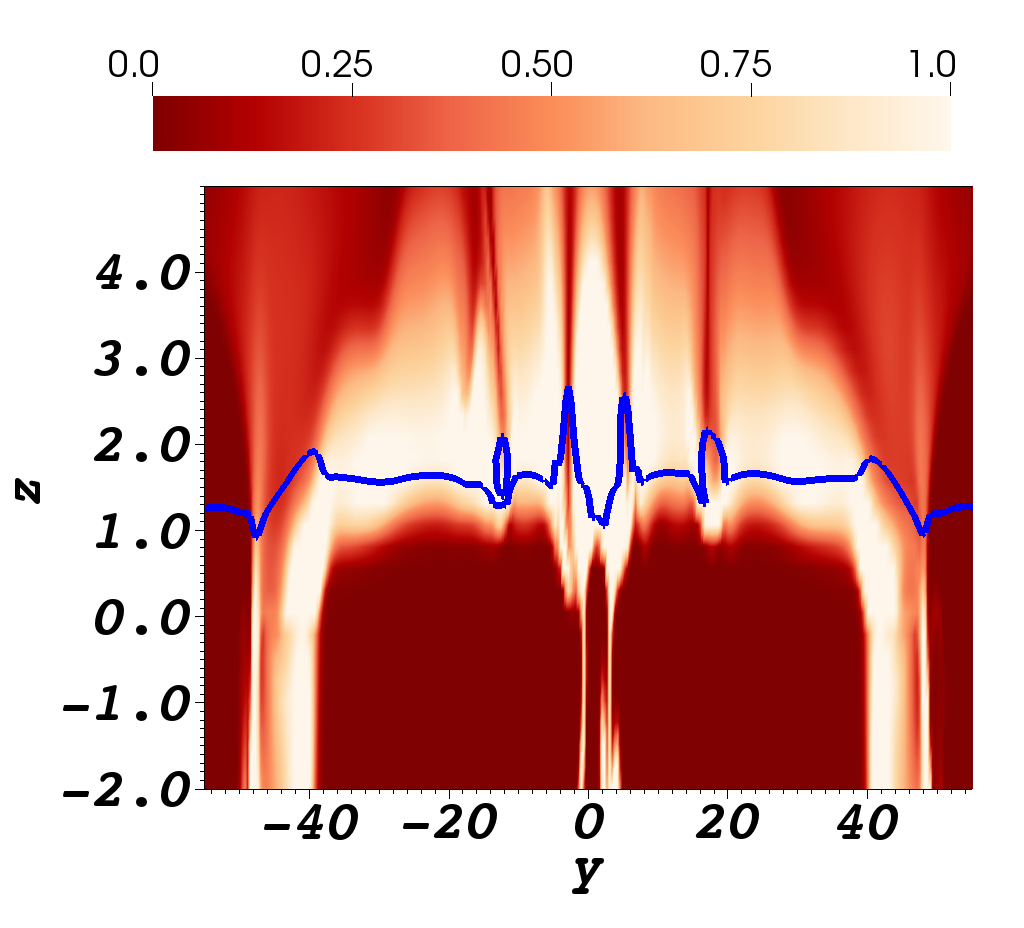}}
\quad \subfigure[]{\includegraphics[width=7.5cm]{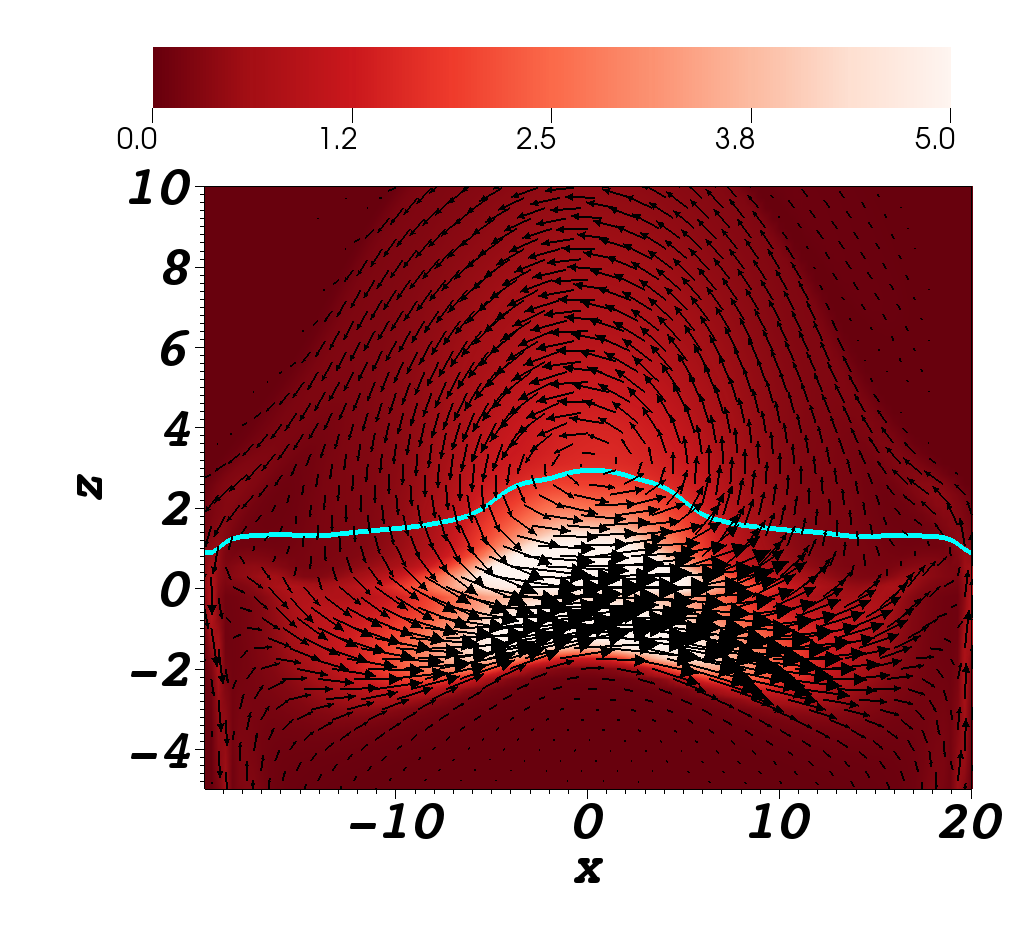}}\quad \subfigure[]{\includegraphics[width=7.5cm]{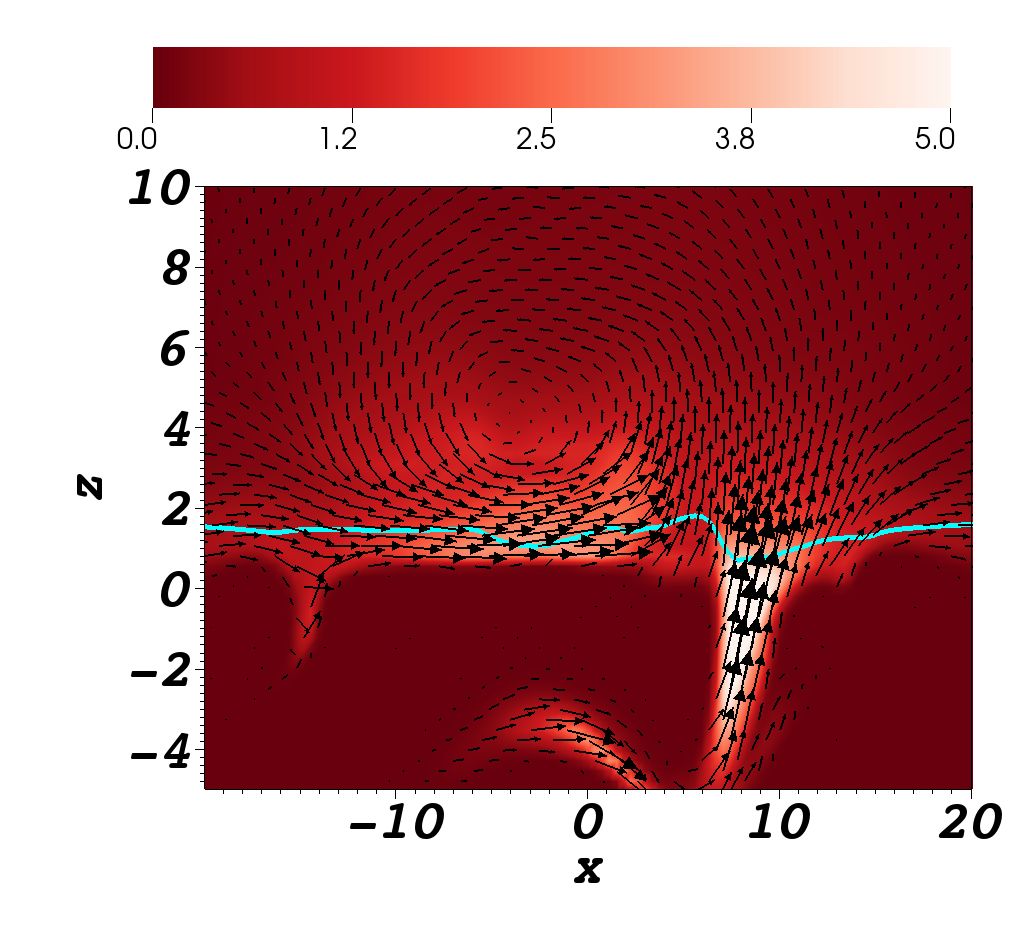}}
\caption{\label{photoslices}{Planar slices of the distribution of $|\Bv|^2$ for the $B_0=7$ case at (a) $t=40$ and (b) $t=70$. Also shown is the curve intersection of the surface $\rho=1$ with these planes. In (a) and (b) the slices are taken in the $y$-$z$ plane at $x=0$. The $\rho=1$ surface is generally significantly above the $z=0$ line In (c), $t=40$ and (d), $t=70$, the slices are taken in the $x$-$z$ plane at $y=0$ (cutting across the flux rope). Once again the $\rho=1$ curve is generally well above the $z=0$ line. The vector field shown is the projection in the plane of the magnetic field, clearly revealing the magnetic field's core to be at or above the $\rho=1$ surface.} }
\end{figure}
In figure \ref{helicityvaryb07}(c), the flat winding input series $\d L^f/\d t$ has a much larger peak of negative input than that of $\d L^v/\d t$. This results in an initially larger net input of $L^f(t)$ in comparison to $L^v(t)$. A second substantial difference is that, in the flat case, the input rate drops to almost zero whilst it does not for the varying case. To see how this difference occurs, we observe in figure \ref{photoslices} that the surface $\rho=1$ is, generally, significantly above the $z=0$ plane. Also we see that the core of the flux rope is at or about the $\rho=1$ surface, which is to be expected as this region of the photosphere is where the tube cannot continue to rise by buoyancy alone \citep[e.g.][]{hood2012review}. Thus the flat measure views the flux rope as `fully emerged' as its core moves (approximately) with the $\rho=1$ surface which rises above the $z=0$ plane. The varying measure, however, tracks this varying density structure and its input instead reflects the motion of the  flux rope \emph{relative} to the surface. We might expect that this would lead to a larger net winding input in the flat case, as the majority of the flux rope has emerged in the flat case, whilst it only partially emerges in the varying case. However, we see in figure \ref{helicityvaryb07}(d) that over the whole simulation there is actually higher total input for the varying case.}

\begin{figure}
\subfigure[]{\includegraphics[width=8cm]{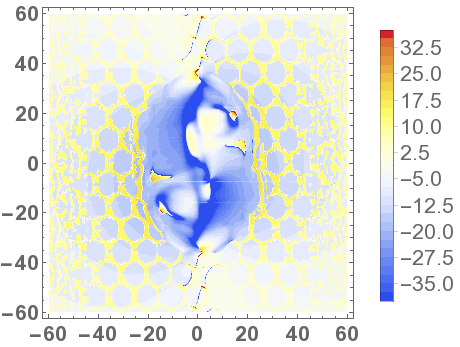}}\quad \subfigure[]{\includegraphics[width=8cm]{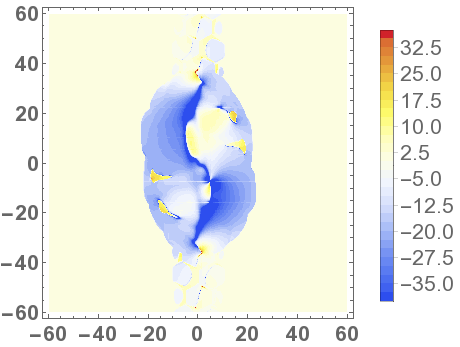}}
\caption{\label{lcutoff} Distributions of  $\d\cl(\av_0)/\d t$ at $t=40$: (a) without the cut-off and (b)  with the cut-off. The significant difference between (a) and (b) is the removal of structure associated with the convective motion. }
\end{figure}

{ 
In figures \ref{helicityvaryb07}(c) and (d) it is clear that including the cut-off in the calculations changes the winding inputs quantitatively but not qualitatively. We note here that the effect of utilising the cut-off is simply to exclude the winding input of regions outside the main emergence domain where convection dominates the input pattern. This effect is indicated in figure \ref{lcutoff} for the varying case, but it is found to be true of the flat distributions also. The distinction will not be critical in what follows and the distributions we present are those with the cut-off (for the sake of clarity).

\begin{figure}
\subfigure[]{\includegraphics[width=8cm]{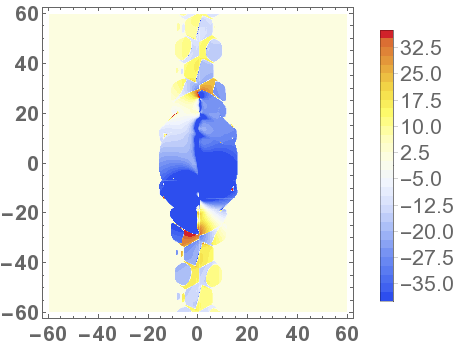}}\quad\subfigure[]{\includegraphics[width=8cm]{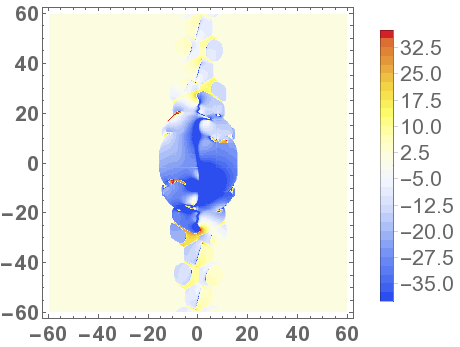}}\quad 
\subfigure[]{\includegraphics[width=8cm]{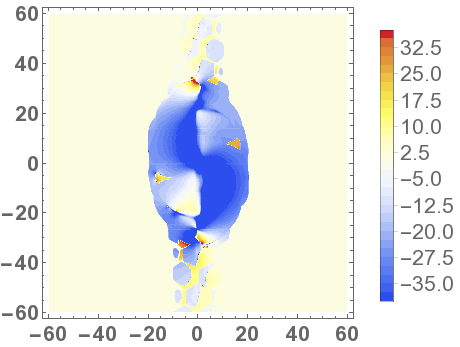}}\quad\subfigure[]{\includegraphics[width=8cm]{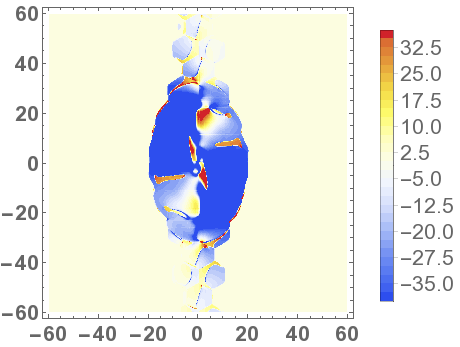}}
\caption{\label{windningdistb07set1} Distributions of  $\d\cl(\av_0)/\d t$ and $\d\clf(\av_0)/\d t$ at $t=31$, (a) and (b) respectively, and $t=37$, (c) and (d) respectively. }
\end{figure}

\begin{figure}
\subfigure[]{\includegraphics[width=8cm]{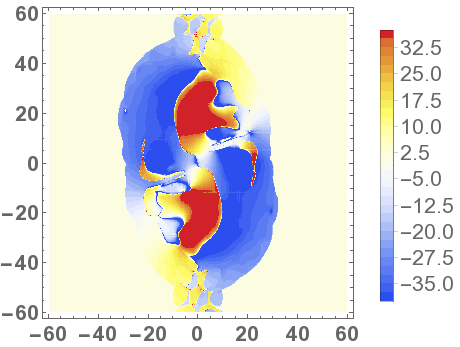}}\quad\subfigure[]{\includegraphics[width=8cm]{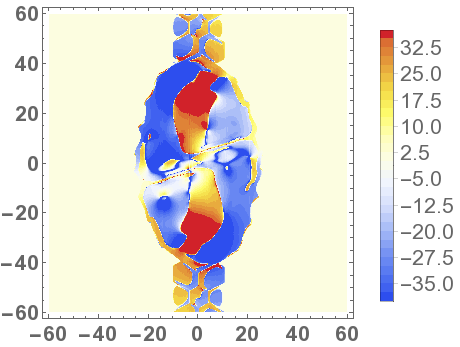}}\quad 
\subfigure[]{\includegraphics[width=8cm]{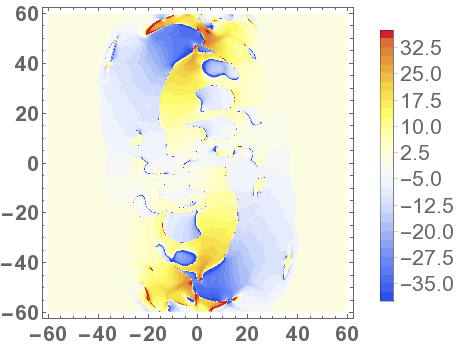}}\quad\subfigure[]{\includegraphics[width=8cm]{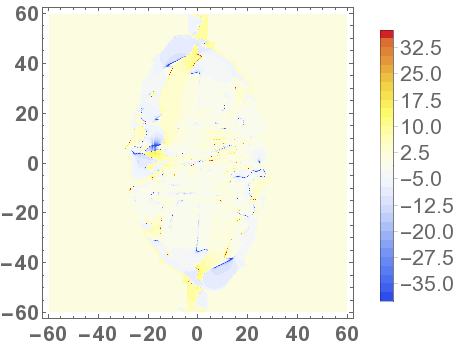}}
\caption{\label{windningdistb07set2} Distributions of  $\d\cl(\av_0)/\d t$ and $\d\clf(\av_0)/\d t$ at $t=50$, (a) and (b) respectively, and $t=75$, (c) and (d) respectively. }
\end{figure}

In figures \ref{windningdistb07set1}(a) and (b) we see comparative snapshots of the distributions $\d\cl(\av_0)/\d t$ and $\d\clf(\av_0)/\d t$ at $t=31$, when the net input rates are almost identical (see figure \ref{helicityvaryb07}(c)). In both cases the central region is dominated by negative winding as a result of the flux rope structure protruding into each surface. There is also a contribution from the convective cells which are directly above the emerging flux rope. The averaged varying distribution is somewhat smoother than the flat counterpart. A final observation is that a number of small islands of positive winding, present in the main flux rope regions  for the flat distribution (b), are absent in the varying case (a). In figures \ref{windningdistb07set1}(c) and (d) we see the distributions $\d\cl(\av_0)/\d t$ and $\d\clf(\av_0)/\d t$ at $t=37$, approximately the time when the flat input rate shows its peak and the net input rate $\d L^f/\d t$ significantly exceeds that of the varying case $\d L^v/\d t$ (see figure \ref{helicityvaryb07}(c)). Both  $\d\cl(\av_0)/\d t$ and $\d\clf(\av_0)/\d t$ still have very similar distributions qualitatively, however, the relative strength of the negative input regions is consistently larger for the flat distribution. It was confirmed that the centre of the flux rope core is at (or around) the $z=0$ plane at this point. Once again the averaged varying distribution is somewhat smoother and there are also more regions of positive density in the flat case. 

In figures \ref{windningdistb07set2}(a) and (b) we see comparative snapshots of the distributions $\d\cl(\av_0)/\d t$ and $\d\clf(\av_0)/\d t$ at $t=50$, when the net input rate of the flat measure $\d L^f/\d t$ is significantly smaller than at its peak, whilst the net varying input rate $\d L^f/\d t$  is still similar to its peak value (see figure \ref{helicityvaryb07}(c)). We now see a much larger area for the winding input in the varying case. That said, the qualitative nature of the distributions (varying and flat) are similar, with both exhibiting the development of two regions of significant positive winding input surrounded by larger negative input. It is noticeable that the relative sizes of the negative and positive input regions are larger in the case of the varying distribution. 

In figures \ref{windningdistb07set2}(c) and (d) we see comparative snapshots of the distributions $\d\cl(\av_0)/\d t$ and $\d\clf(\av_0)/\d t$ at $t=75$, when the net input rate of the flat measure $\d L^f/\d t$  has dropped close to zero but the varying distribution is still significant. The distribution in the flat case (d) is negligible, coherent with the evidence in figure \ref{photoslices} that the flux rope structure has largely risen above the $z=0$ plane. The varying distribution (c) is also relatively weak compared to earlier times, but there are still regions of significant winding input. There are now multiple interspersed sub-regions of positive and negative winding input, consistent with the development of the magnetic field's serpentine structure.

\begin{figure}
 \subfigure[]{\includegraphics[width=8cm]{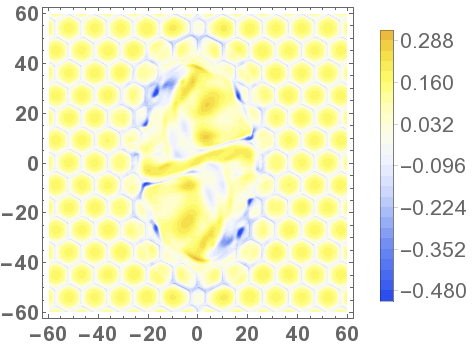}}\quad \subfigure[]{\includegraphics[width=8cm]{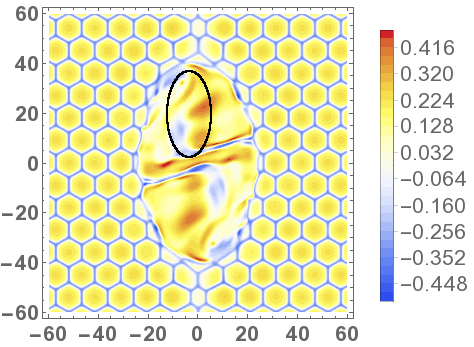}}
\caption{\label{veldistsb07} Vertical velocity distributions at $t=50$. Panel (a) shows $u_z^v$ and panel (b) $u_z^f$. In (a), the varying case, the convective cell velocities are weaker than in (b),  the flat case. In (a) the strongest regions of negative velocity are around the edge of the emerged structure. In (b), there are significant down flows in the core of the emerging region. The highlighted region contains both (relatively) strong up and down flows. }
\end{figure}
{ 
The mixture of positive and negative winding rate regions can be understood with reference to the behaviour of upward and downward flows. Figure \ref{veldistsb07} shows the distributions of $u_z^v$ and $u_z^f$ at $t=50$, which are cotemporal to the winding rate distributions in figures \ref{windningdistb07set2}(a) and (b). By comparing figure \ref{windningdistb07set2} to figure \ref{veldistsb07}, it is clear that the positive winding input rates occur in regions which contain (relatively) strong upward and downward flows, an example of which is highlighted on the velocity distribution $u_z^f$. We can see that the flows is this region are significantly weaker in the varying case, as is generally true for the convective motions also. This feature is consistent with the evidence of convective down flows reversing the sign of the helicity input rates discussed above. The difference in the winding case is that the winding measure is not weighted by magnetic field strength and the regions described above have relatively strong magnetic field (see figures \ref{slicesb07_1}-\ref{slicesb07_3}). Therefore, these regions affect the helicity inputs far more than the winding rate inputs. This is why the winding distribution correctly identifies the underlying initial topological nature of the emerging field, emphasising the benefit of calculating both winding \emph{and} helicity inputs. 
\subsubsection{Differing net winding inputs}\label{differingnetinputs}
The last significant issue to address is the fact that the total winding input $L(t)$ is systematically less in magnitude for the flat case (see figure \ref{helicityvaryb07}(d)), despite the apparent total emergence of the majority of the flux rope above this plane (which is not the case for the varying surface as we have seen above). There are two key factors that influence this result:
\begin{itemize}
\item{As indicated in the analysis of the helicity time series, the field line plots (figure \ref{b07fieldlines1}) and above in the winding rate analysis, convective motion is acting to pull down regions of the magnetic field. Convective motion is stronger at the lower $z=0$ plane and hence we should expect the effect to be more significant there. This is evidenced by the relatively large proportion of positive winding rate in the flat input rate distributions in figures \ref{windningdistb07set1} and \ref{windningdistb07set2}, by comparison to the varying distributions.}
\item{If there is significant variation in both the $\Bv$ and $\uv$ fields in height, the averaging procedure that we have adopted can capture more complex winding patterns, not found on only one plane. In this application of flux emergence, the result is an enhanced (varied) winding input compared to the flat case.}
\end{itemize}
In Section \ref{b03case} we consider a third field-sampling method that tracks the varying $\rho=1$ surface but averages over a much smaller domain. It will be shown that the effect of spatial averaging can be significant for affecting the magnitudes of quantities whilst the qualitative behaviours of helicity and winding found with different averages are similar.
}

\subsection{$B_0=5$, $\alpha=-0.4$}

\begin{figure}
\begin{center}
\subfigure[]{\includegraphics[width=7.5cm]{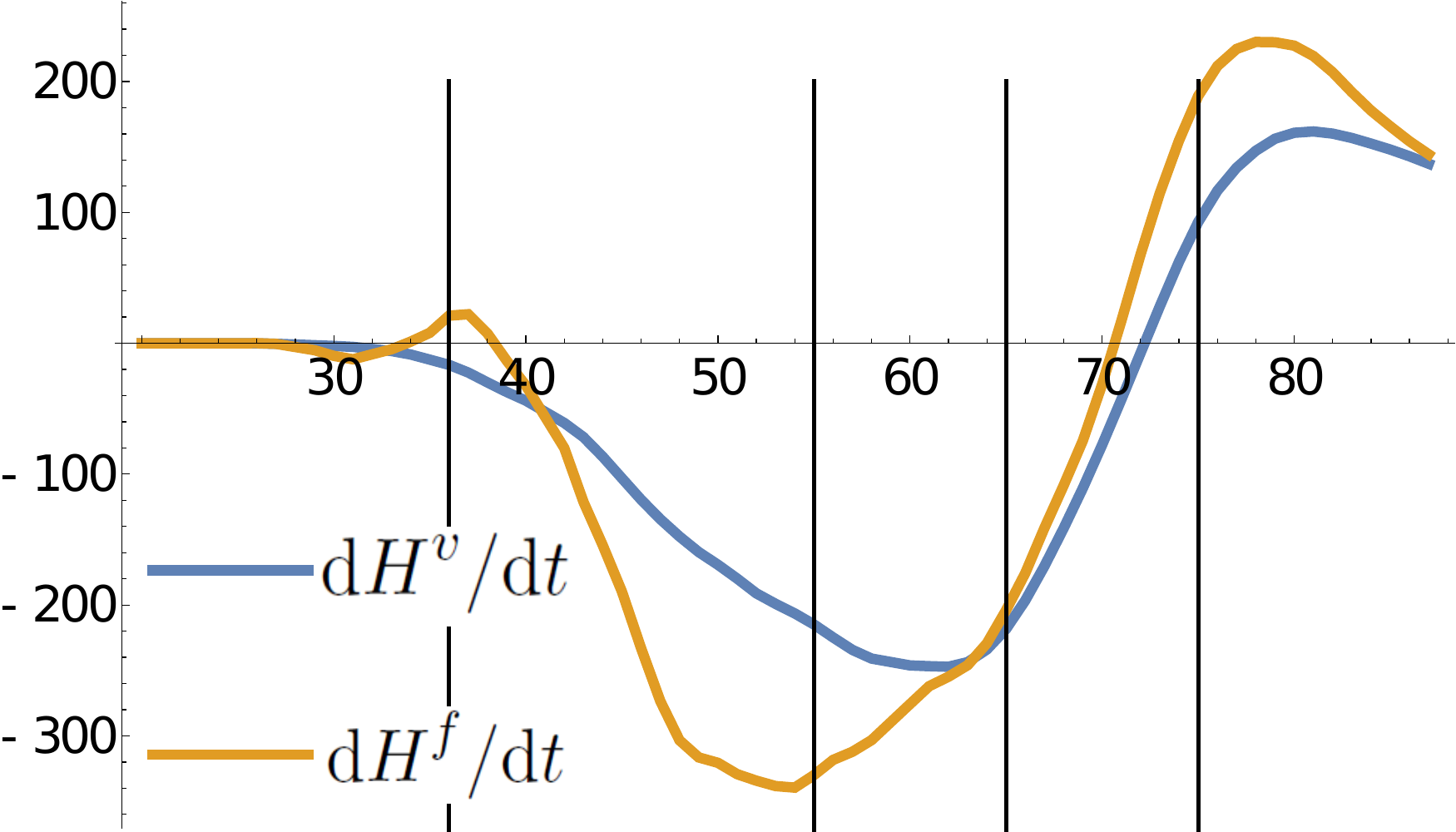}}\quad \subfigure[]{\includegraphics[width=7.5cm]{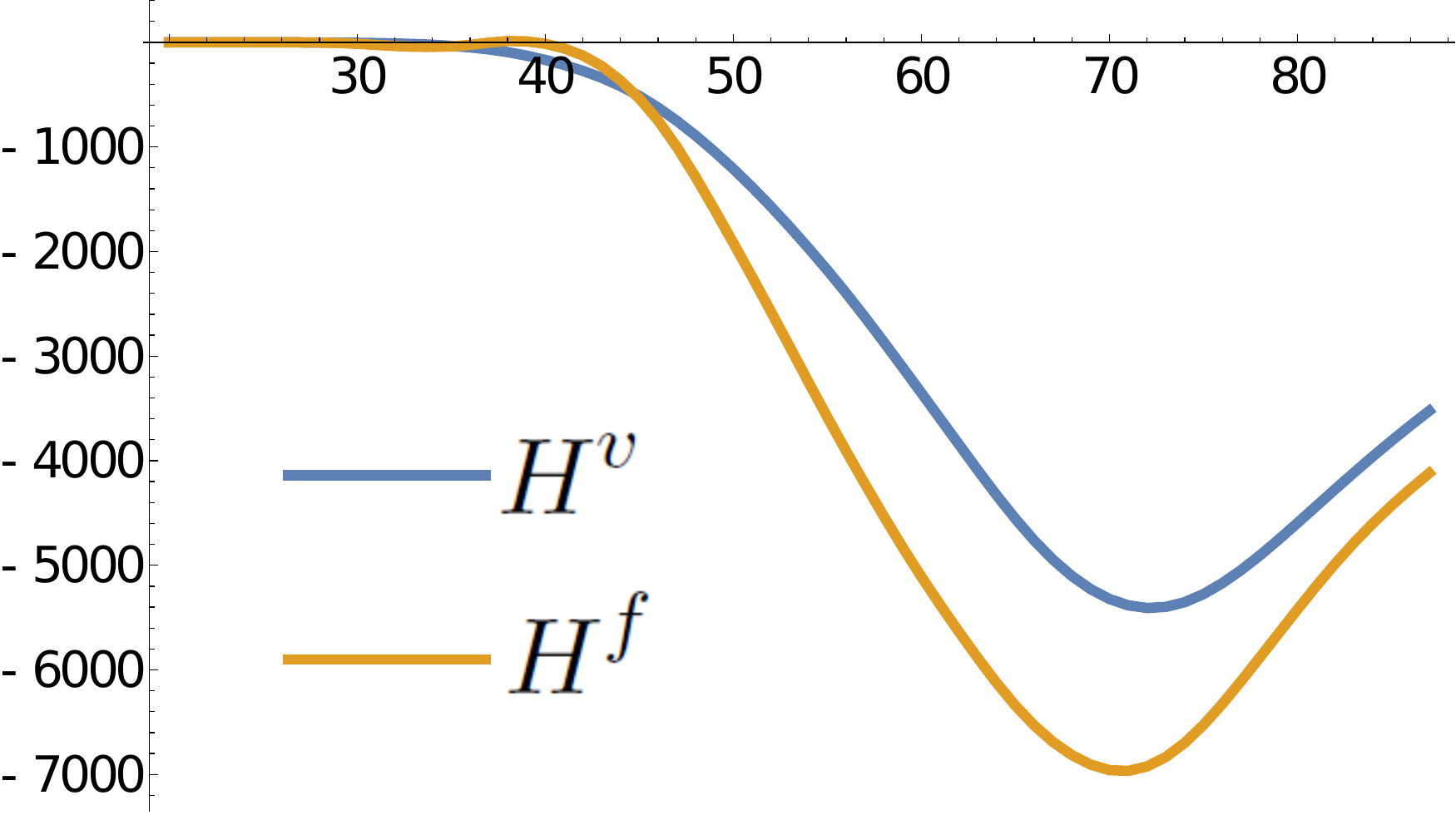}}\quad 
\subfigure[]{\includegraphics[width=7.5cm]{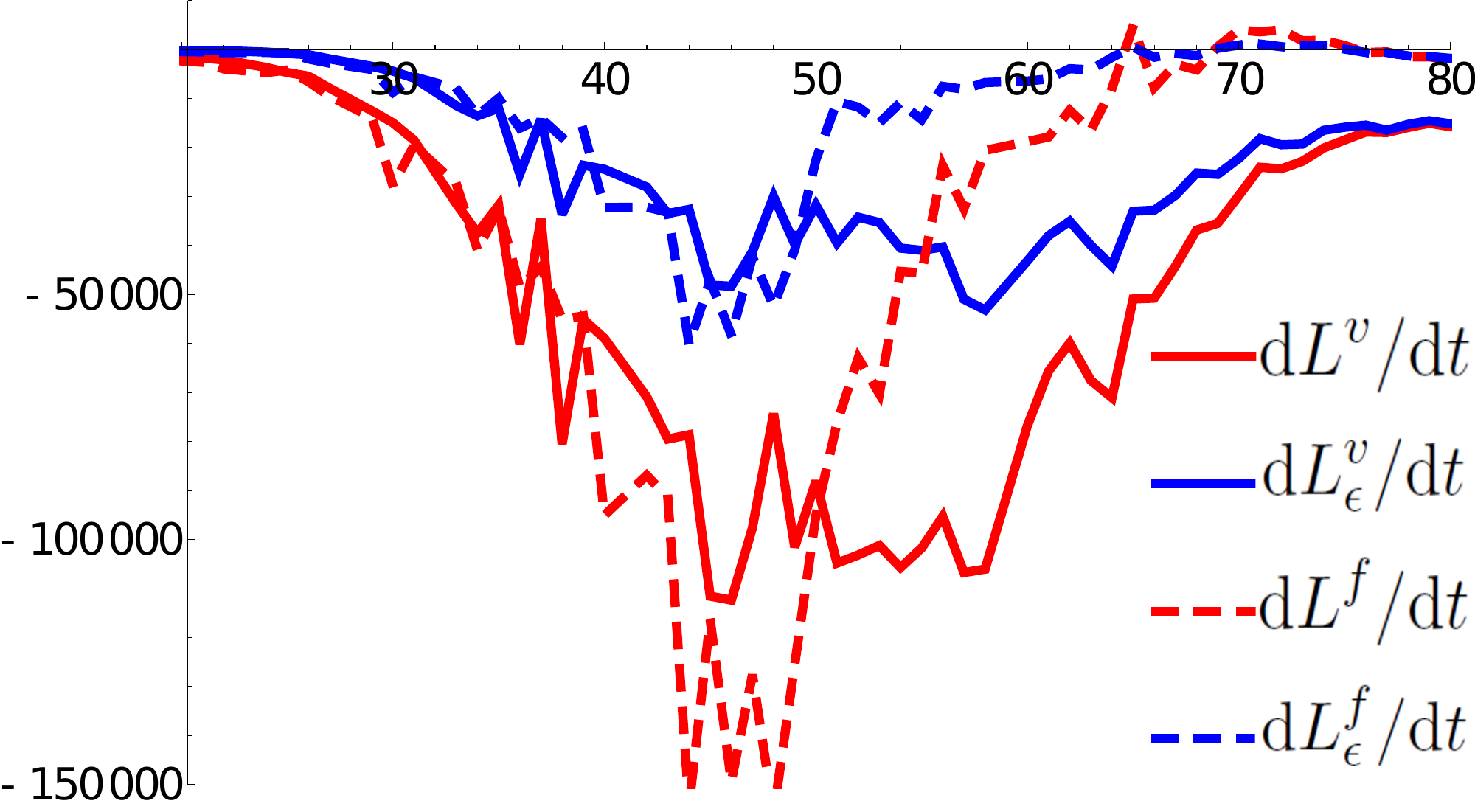}}\quad \subfigure[]{\includegraphics[width=7.5cm]{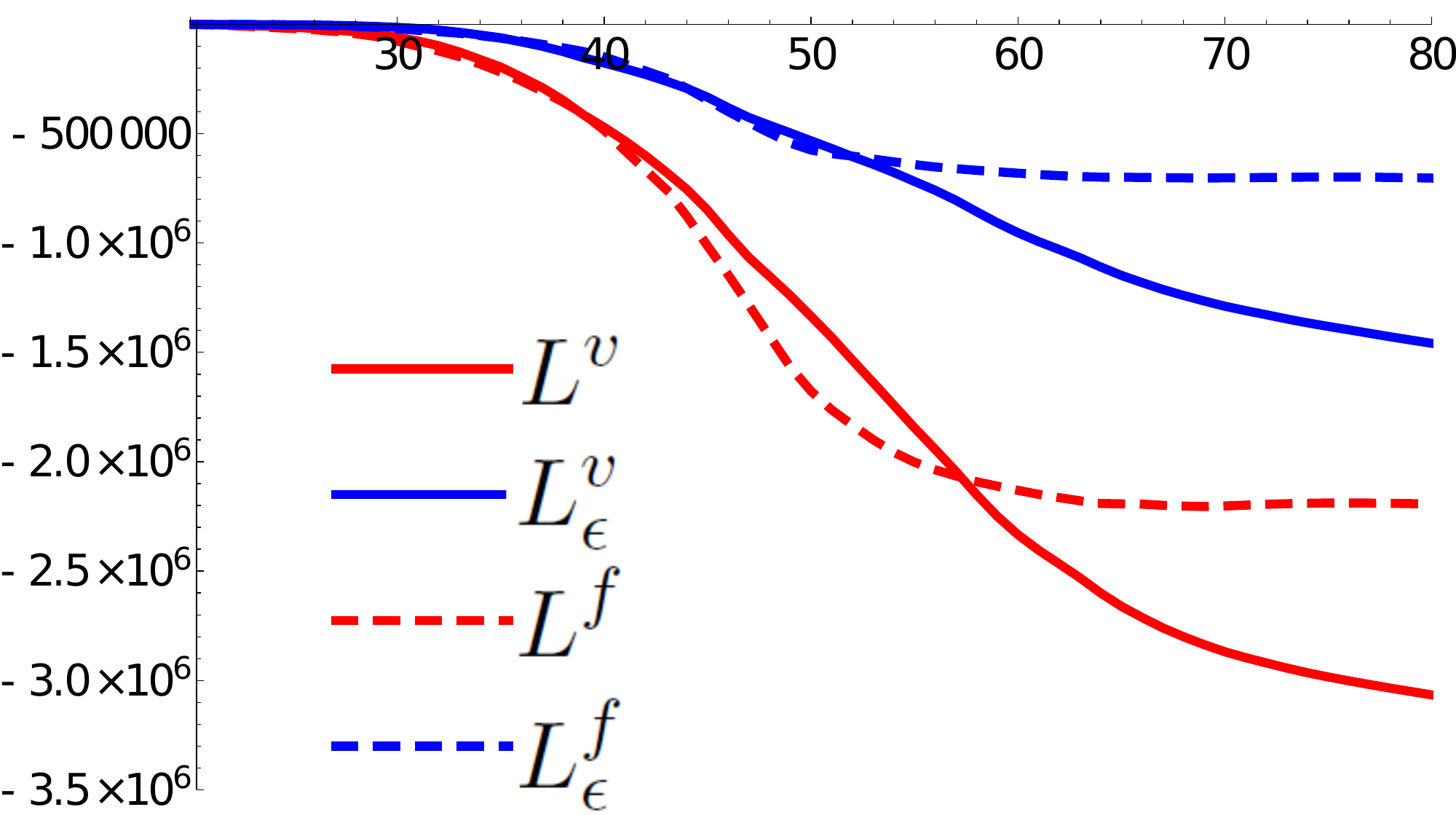}}
\caption{\label{helicityvaryb05}{ Helicity and winding time series for the $B_0=5$, $\alpha=-0.4$ case. Panel (a) shows the helicity input rates $\d{H}^v/\d{t}$ and $\d{H}^f/\d{t}$. The vertical lines indicate times at which maps of $\d\wh(\av_0)/\d t$ are plotted in figure \ref{helicitydistsb05}. Panel (b) shows the integrated helicities $H^v(t)$ and $H^f(t)$ over the same period. Panel (c) displays the winding input rates $\d L^v/\d t$ and $\d L^f/\d t$ both with the cut-off (blue) and without it (red). Panel (d)  displays the total winding inputs $L^v(t)$ and $L^v(t)$ and has the same format as (c).} }
\end{center}
\end{figure}
The helicity time series in figures \ref{helicityvaryb05}(a) and (b) are qualitatively similar to those of the $B_0=7$ case. The bulk of the input is negative, corresponding to the left-handed twisted flux tube, {with a switch to positive input around $t=70$. Also, as in the $B_0=7$ case, we see the flat input series has slightly larger and earlier peaks than the varying time series. This can again be traced to the fact that the $\rho=1$ surface rises significantly above the $z=0$ plane and the flux rope's core tracks it}. We will show that the turning point in the sign of  {both} the helicity input series is due to the effects of convection. 

{ The winding time series, in figures \ref{helicityvaryb05}(c) and (d), are qualitatively similar to those for the $B_{0}=7$ case. The input is generally negative for both cases. As with the $B_0=7$ case, the flat input rate series show a much larger peak input than that of the varying case (figure \ref{helicityvaryb05}(c)).  Also, its input rate drops to nearly zero whilst that of the varying case remains non-zero for the time span of the simulation.  Again, this feature leads to more net input in the varying series, in contrast to the helicity input where the flat series has a larger total input.}

\begin{figure}
\begin{center}
\subfigure[$t=35$]{\includegraphics[width=6cm]{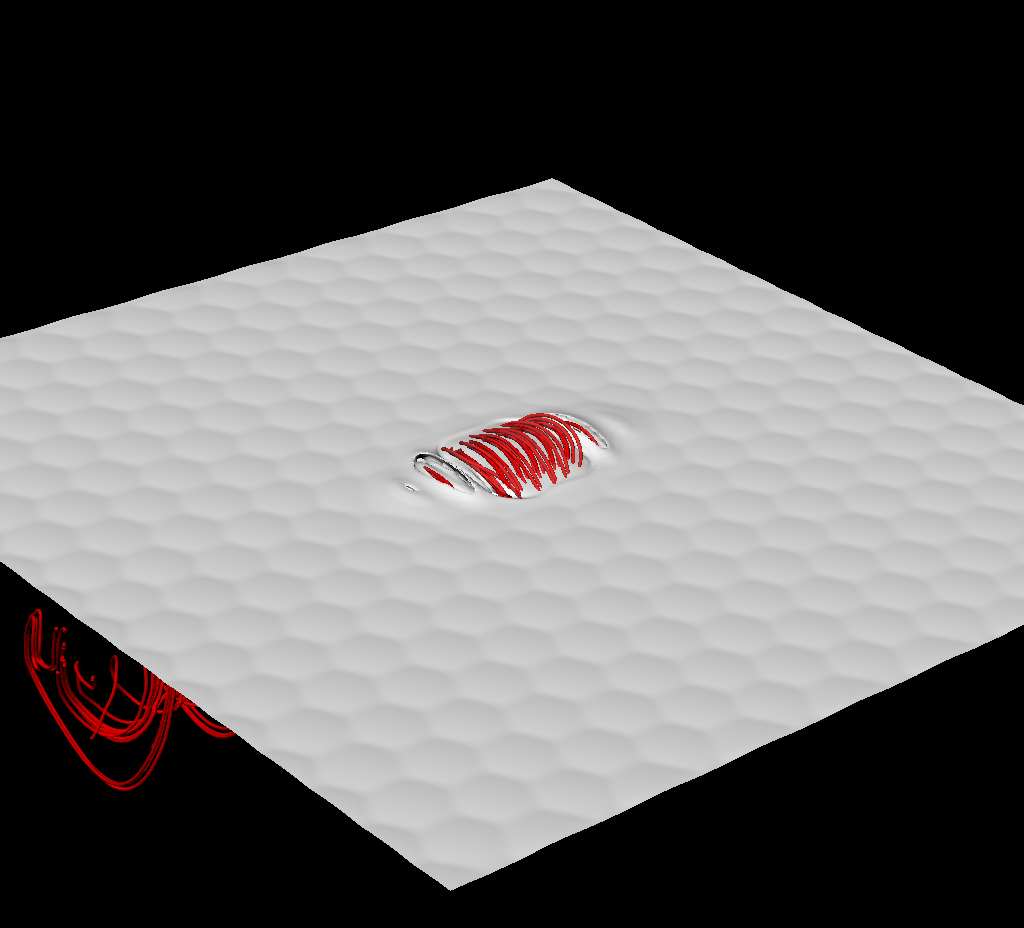}}\quad\subfigure[$t=35$]{\includegraphics[width=6cm]{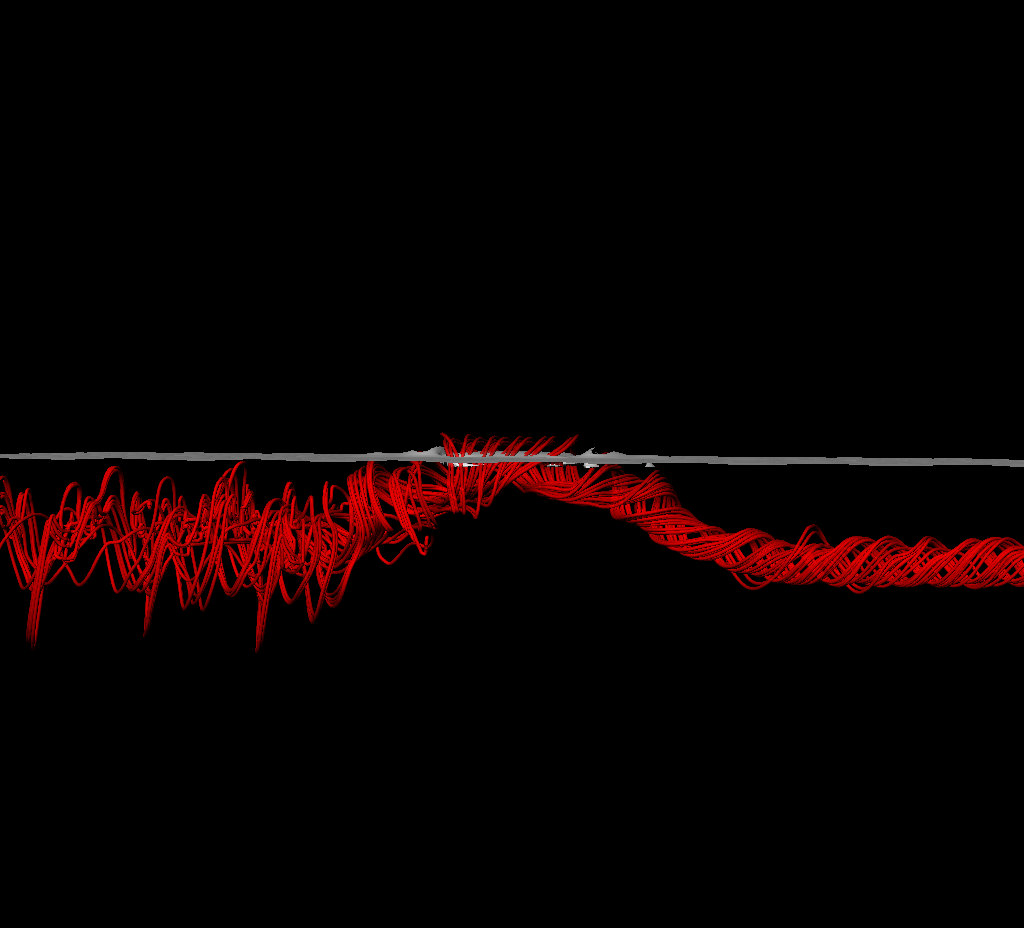}}\quad\subfigure[$t=55$]{\includegraphics[width=6cm]{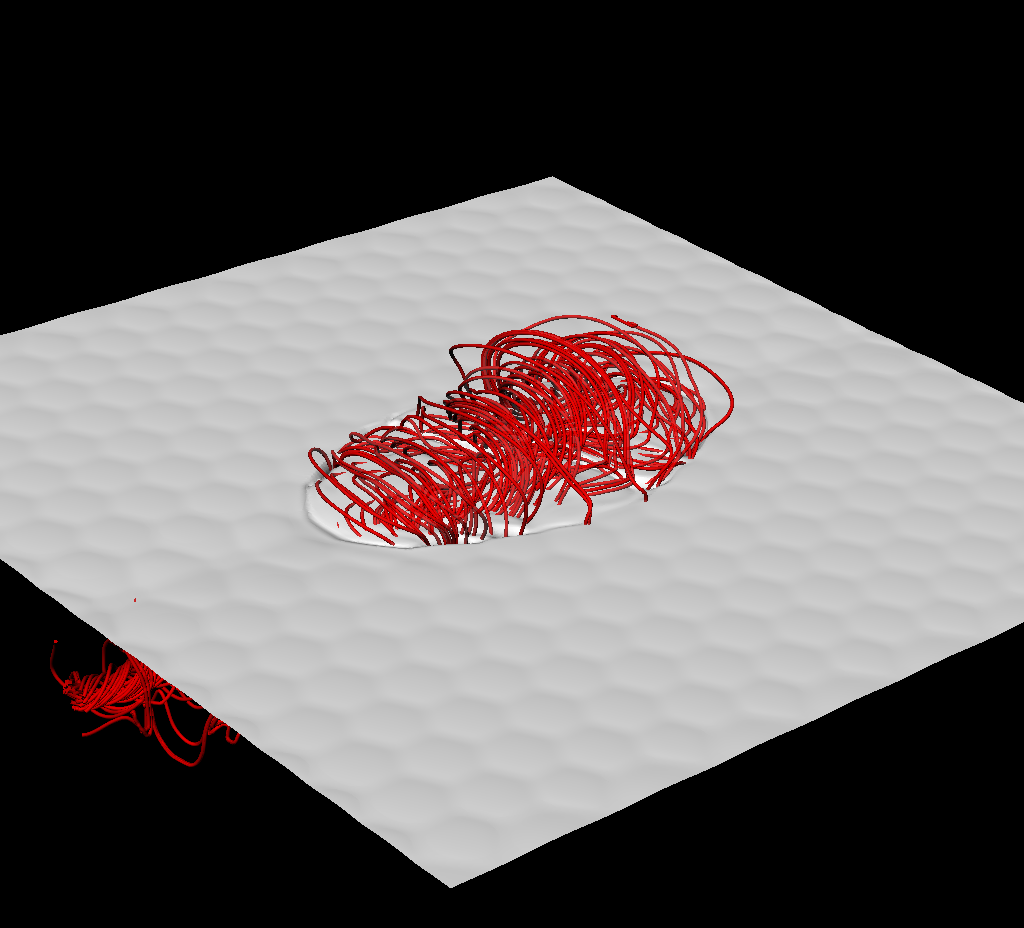}}\quad\subfigure[$t=55$]{\includegraphics[width=6cm]{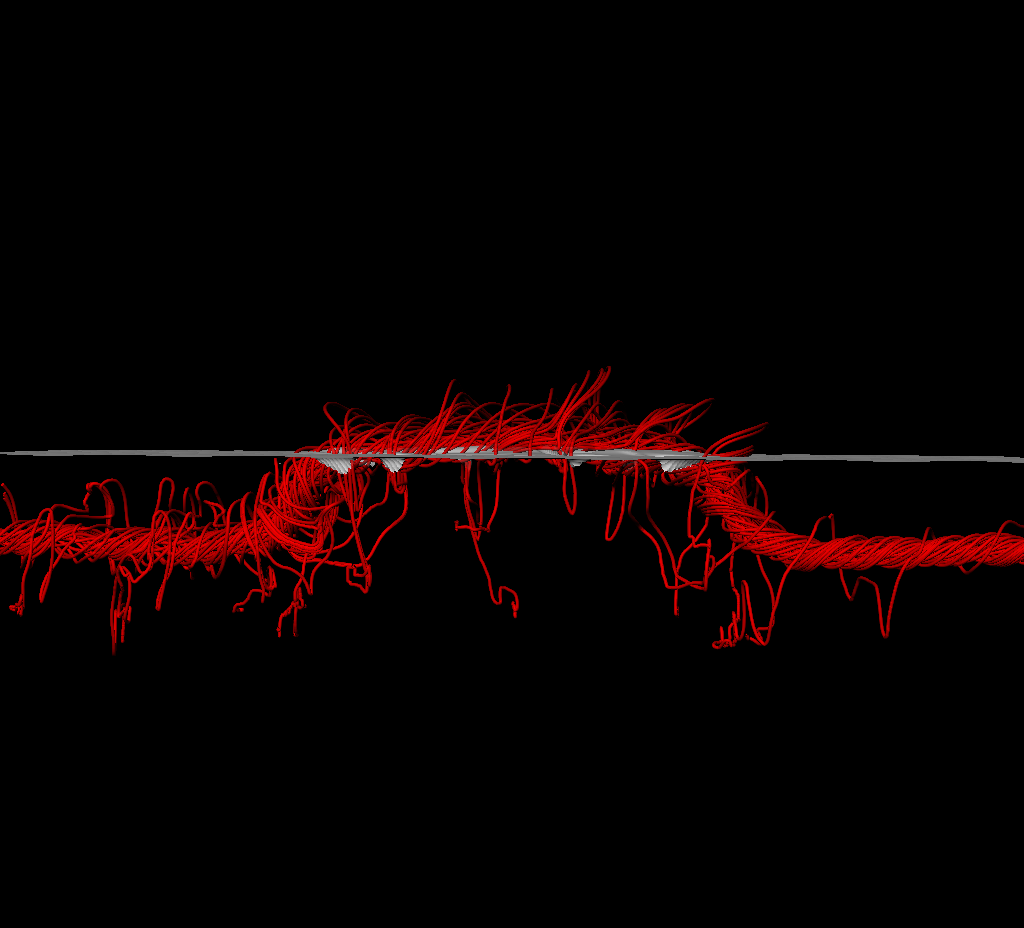}}\quad\subfigure[$t=80$]{\includegraphics[width=6cm]{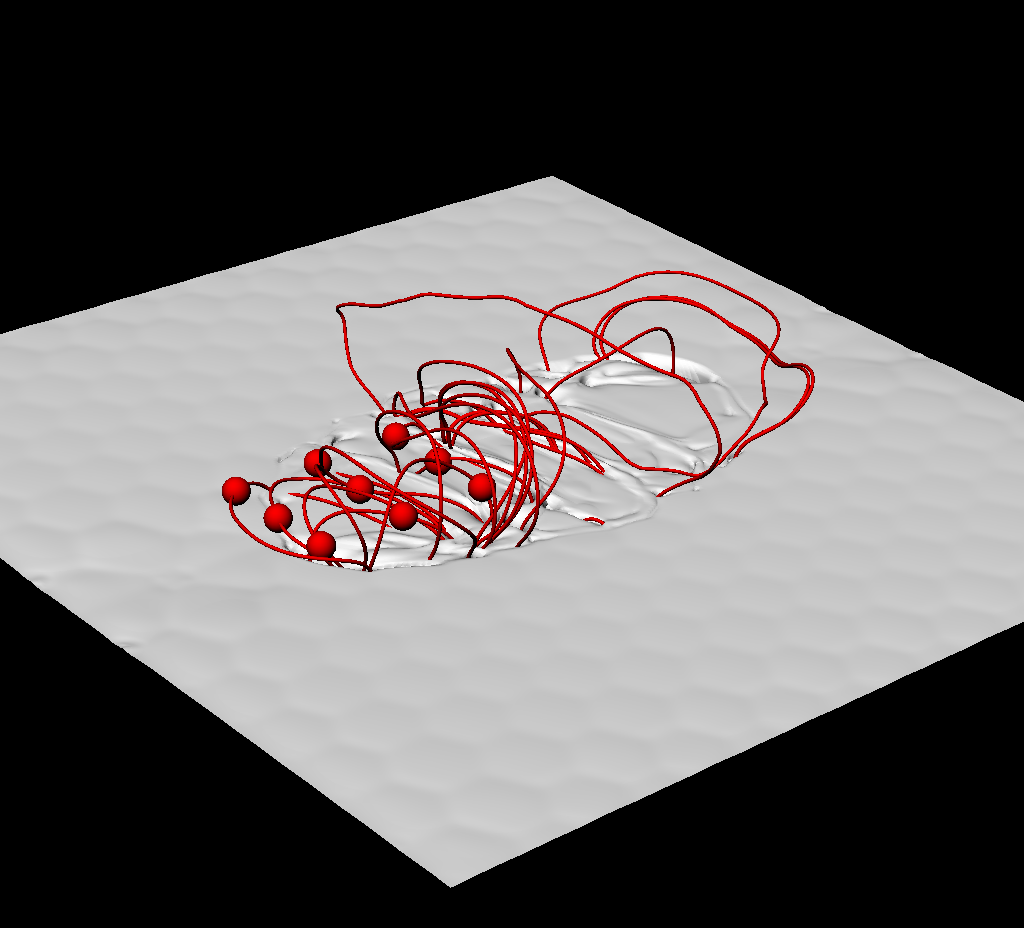}}\quad\subfigure[$t=80$]{\includegraphics[width=6cm]{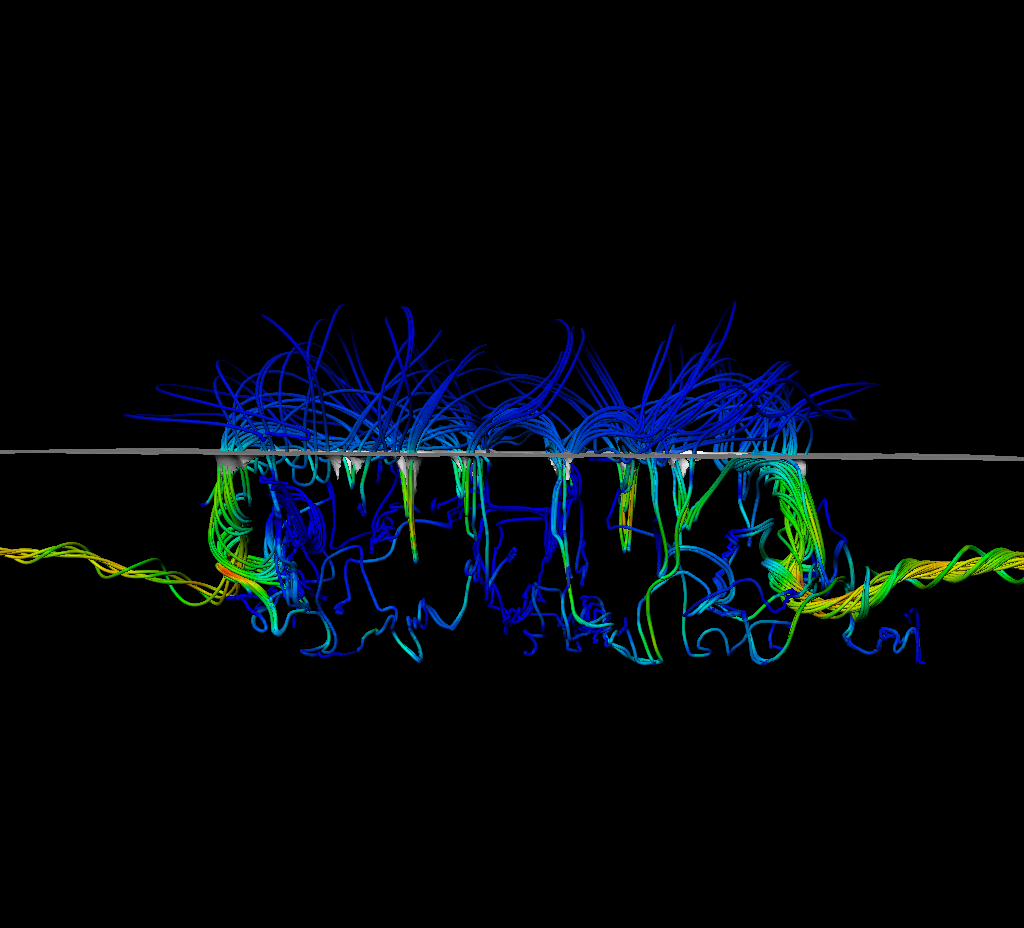}}
\caption{\label{b05fieldlines}Field line renderings for the $B_0=5$ $\alpha=-0.4$ case at various points in its evolution. Panels (a) and (b) depict the field at $t=35$. In (a) the surface shown is that of the plasma density $\rho=1$. Panels (c) and (d) depict the field at $t=55$. Panel (e) depicts field lines in the emerged field at $t=80$. The spheres shown are the starting points for the field line reconstruction. Panel (f) depicts field lines side-on, with lighter colours indicating increased field strength. }
\end{center}
\end{figure}

Figure \ref{b05fieldlines} displays snaphots of magnetic field lines during various stages of emergence (the format is the same as figure \ref{b07fieldlines1}). The general evolution, at first glance, is very similar to that of the $B_0=7$ case. The development of serpentine field lines is particularly clear in  figure \ref{b05fieldlines}(f).

\begin{figure}
\begin{center}
\subfigure[$t=36$]{\includegraphics[width=6.5cm]{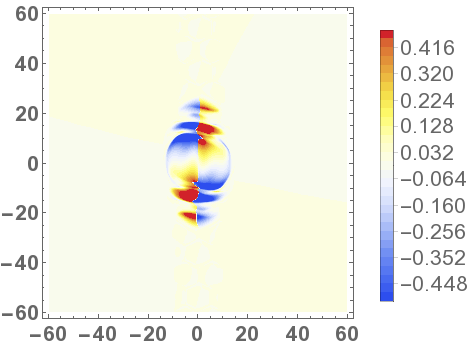}}\quad\subfigure[$t=50$]{\includegraphics[width=6.5cm]{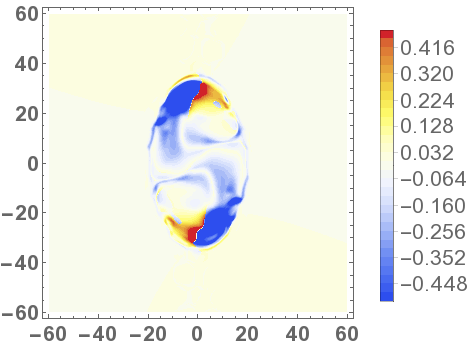}}\quad\subfigure[$t=65$]{\includegraphics[width=6.5cm]{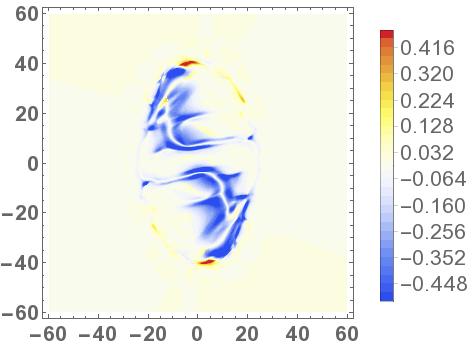}}\quad\subfigure[$t=75$]{\includegraphics[width=6.5cm]{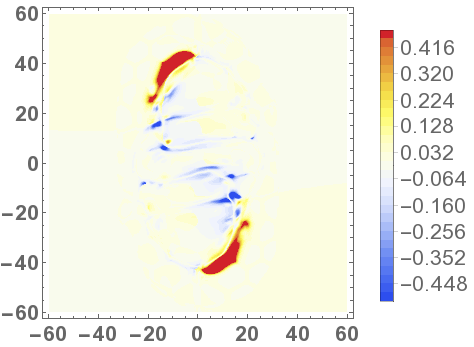}}
\caption{\label{helicitydistsb05} Varying helicity rate distributions at the times indicated by vertical lines in the helicity rate time series shown in figure \ref{helicityvaryb05}(a), i.e. (a) $t=36$, (b) $t=55$, (c) $t=65$ and (d) $t=75$. }
\end{center}
\end{figure}

Figure \ref{helicitydistsb05} displays maps of the field line helicity input rate $\d\wh(\av_0)/\d t$ at various times during emergence. Again, the overall evolution is very similar to the $B_0=7$ case, as shown in figure \ref{helicitydistsb07}. Compared to the $B_0=7$ case, the flux tube has weaker field strength and takes longer to reach the photosphere. Both these facts allow convection to deform the magnetic field to a greater extent than the $B_0=7$ case. 

{ \subsubsection{Winding inputs}
Further analysis of the winding input time series, similar to that performed in Section \ref{winddistsb07} for the $B_0=7$ case, leads to similar conclusions and so we do not detail these results here for the sake of brevity. Our conclusion as to how the net winding input $L^v$ is larger than that of $L^{f}$ is the same as before.
}

\subsubsection{Partial emergence of the Serpentine core }
\begin{figure}
\begin{center}
\subfigure[]{\includegraphics[width=8cm]{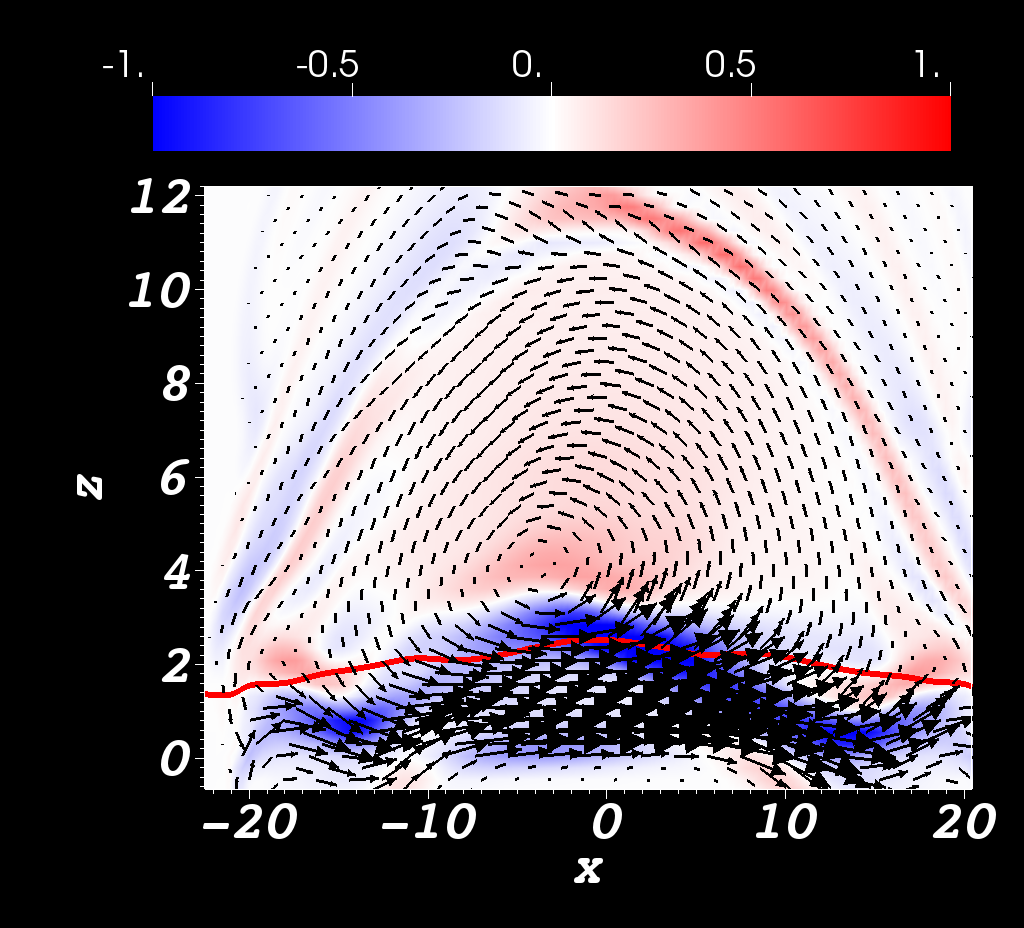}}\quad\subfigure[]{\includegraphics[width=8cm]{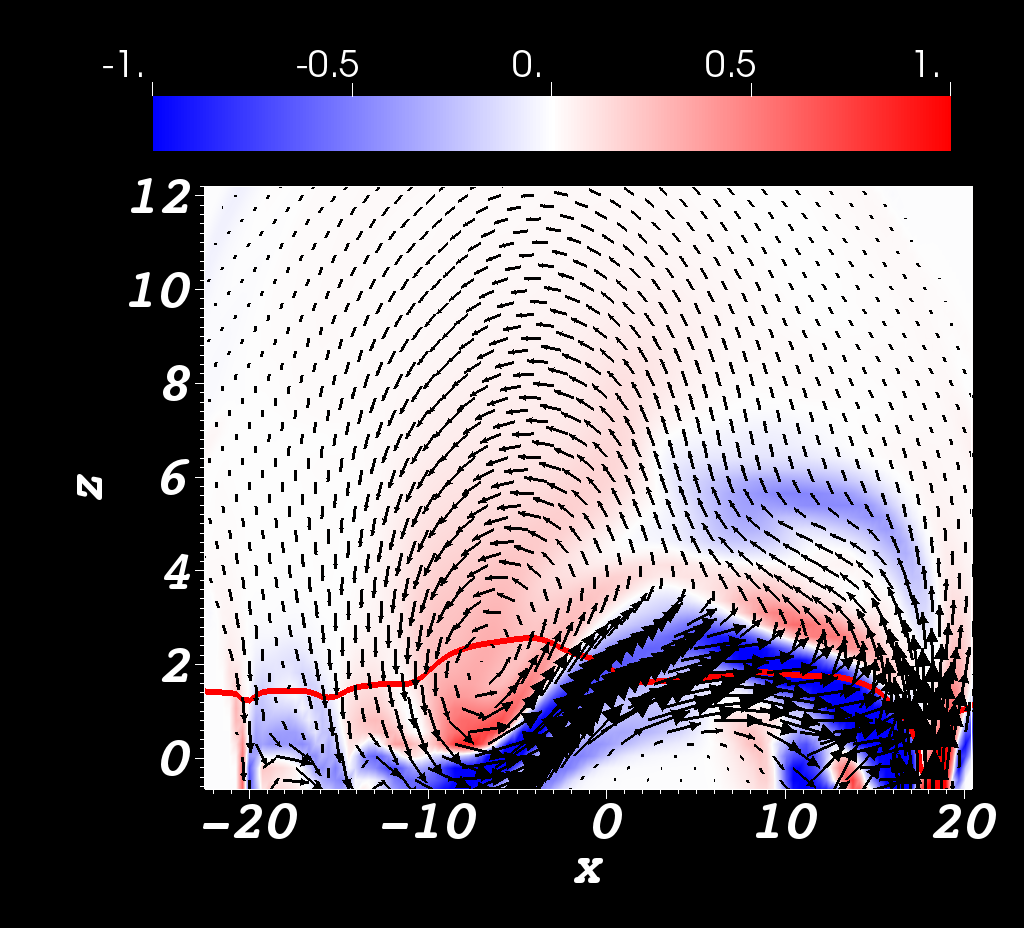}}
\caption{\label{sigmoidgapb05}Slices of the $B_0=5$ case at $t=55$ in the $x$-$z$ plane. The density plotted is the ${y}$-component of $\bnab\times {\Bv}$. The red curves represent the intersection of the slices with the $\rho=1$ surface. In (a) the intercept with the $y$-axis is at $y=0$ and the tube's core is above the $\rho=1$ line. In (b) the intercept with the $y$-axis is at $y=10$ and the tube's core is centered at the $\rho=1$ line. }
\end{center}
\end{figure}

As for the $B_0=7$ case, the core of the $B_0=5$ case flux tube also develops a serpentine structure. In figures \ref{helicitydistsb05}(c) and (d), the negative sigmoidal band corresponding to the twisted core appears much weaker near the centres of these maps at $(0,0)$. The reason can be seen to be due to serpentine emergence by examining slices through the domain. Figure \ref{sigmoidgapb05} shows two such slices. In figure \ref{sigmoidgapb05}(a) a slice at $y=0$ displays a map of $\ev_y\bdot\bnab\times\Bv$ and a projection of $\Bv$-field vectors onto the plane. The main part of the twisted core is seen clearly above the photospheric surface (indicated by the red line). In figure \ref{sigmoidgapb05}(b), the same variables are shown but now at $y=10$. Here, the twisted core is held at the photospheric boundary, as suggested by figures \ref{helicitydistsb05}(b) and (c).

 
\subsection{$B_0=3$, $\alpha=-0.4$}\label{b03case}
{ In this section we include an additional test for the effect of choosing a particular form of field sampling. We introduce the following new magnetic field average, 
\be
\label{averagefield}
\widehat{{\Bv}}(x,y) = \frac{\int_{z_{\rm min}}^{z_{\rm max}}\exp[-2000 (\rho-1)^6]{\Bv}(x,y,z)\,\d{z}}{\int_{z_{\rm min}}^{z_{\rm max}}\exp[-2000 (\rho-1)^6]\,\d{z}},
\en
which we refer to as the `varying fine' average. The various helicity quantities calculated using this average will be labelled $H^{vf},\,L^{vf}\,{\cal H}^{vf}$ and ${\cal L}^{vf}$, for `varying fine'. As the name indicates, this measure will sample the field based around the evolving $\rho=1$ surface but average over a domain that is significantly smaller compared to the previous measure. This measure will be used to try to separate the effect on topological quantities of averaging the field over a finite domain from the effect of sampling the field on a moving surface.}
\begin{figure}
\begin{center}
\subfigure[]{\includegraphics[width=7.5cm]{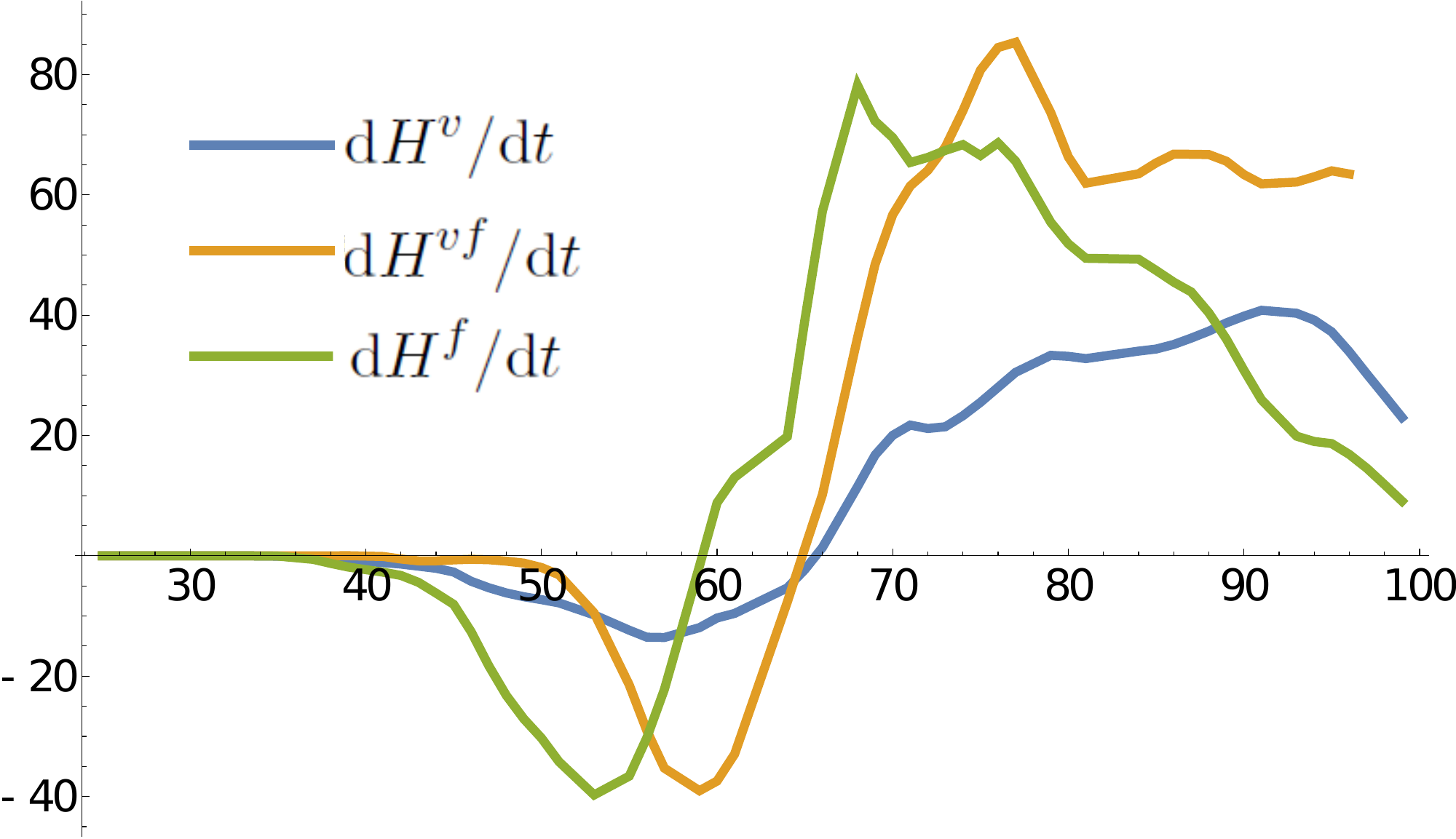}}\quad \subfigure[]{\includegraphics[width=7.5cm]{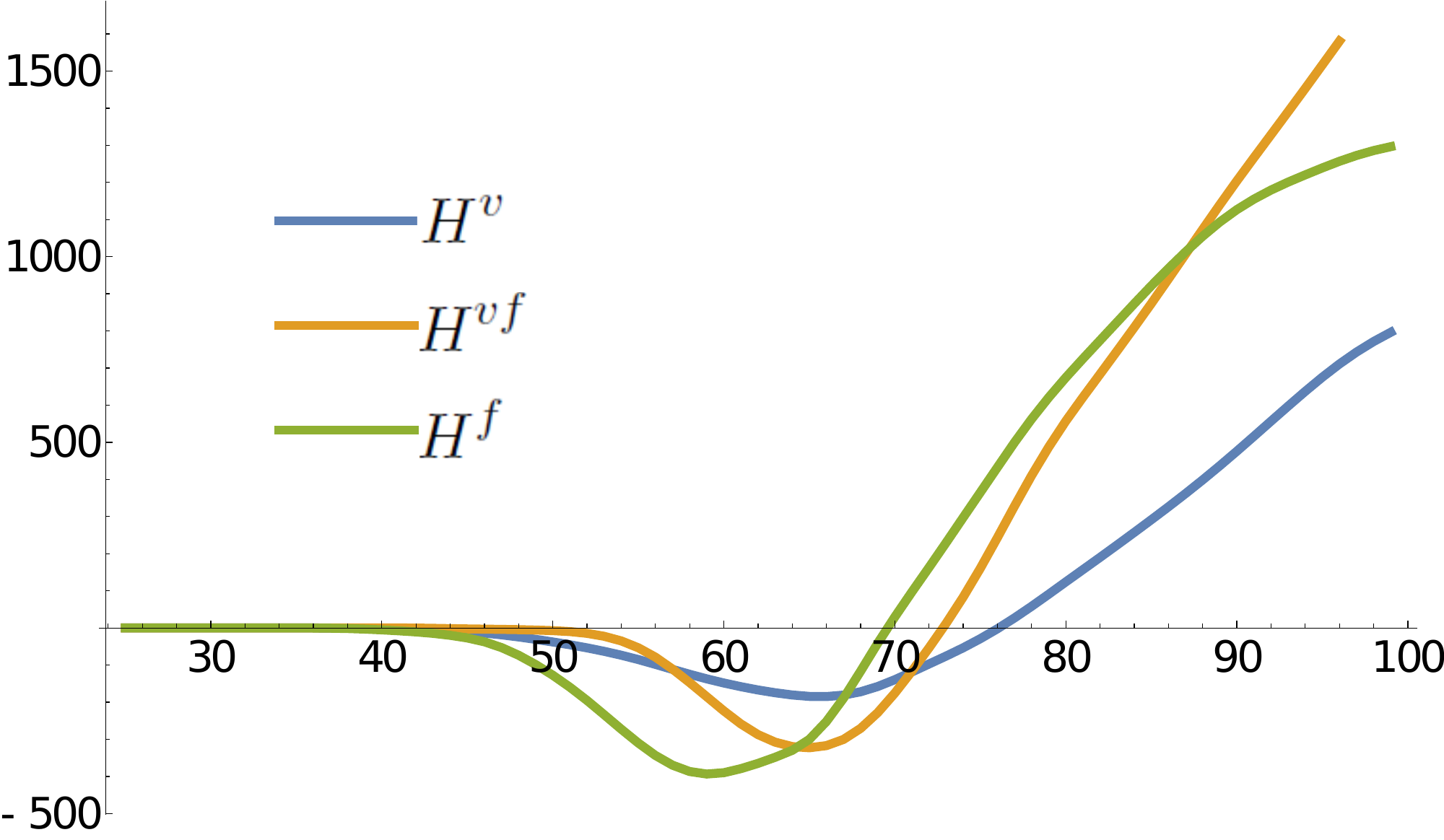}}
\caption{\label{helicityvaryb03}Helicity  time series for the $B_0=3$, $\alpha=-0.4$ case. Panel (a) shows the helicity input rates $\d{H}^v/\d{t}$, $\d{H}^f/\d{t}$ and $\d{H}^{vf}/\d{t}$. Panel (b) shows the integrated helicities $H^v(t),\,H^f(t)$ and $H^{fv}(t)$ over the same period. }
\end{center}
\end{figure}

{The helicity input rates $\d H^v/\d t$,  $\d H^f/\d t$ and $\d H^{vf}/\d t$  for the  $B_0=3$ case are shown in figure \ref{helicityvaryb03}(a). At first we focus on $\d H^v/\d t$ and $\d H^f/\d t$ in order to compare to the $B_0=5,7$ cases. They are qualitatively similar to the $B_0=5,7$ cases in that there is a negative helicity input rate associated with the initial emergence of the negatively twisted flux rope,  which later changes to positive input. However, in this case the positive input period is much more significant than the negative input stage, as indicated by a positive net helicity input over the simulation period  (figure \ref{helicityvaryb03}(b)). The varying fine input rate $\d H^{vf}/\d t$ has the intriguing property that its magnitudes are comparable to those of the flat input rate, but its peaks are shifted temporally. This second property is in line with the narrative that the core of the flux rope reaches this surface after passing through the $z=0$ plane. The fact that the magnitude of the averaged varying measure is significantly smaller than the two other measures is indicative of the averaging, in this case, weighting the helicity calculation with weaker magnetic field that is higher up in the atmosphere (and is not recorded by the flat and varying fine cases). This property will be confirmed shortly when we show that the winding input for the varying case is larger than the other two, hence it is recording more topological complexity but weighted with a weaker field strength.

\begin{figure}
\begin{center}
\subfigure[]{\includegraphics[width=7.5cm]{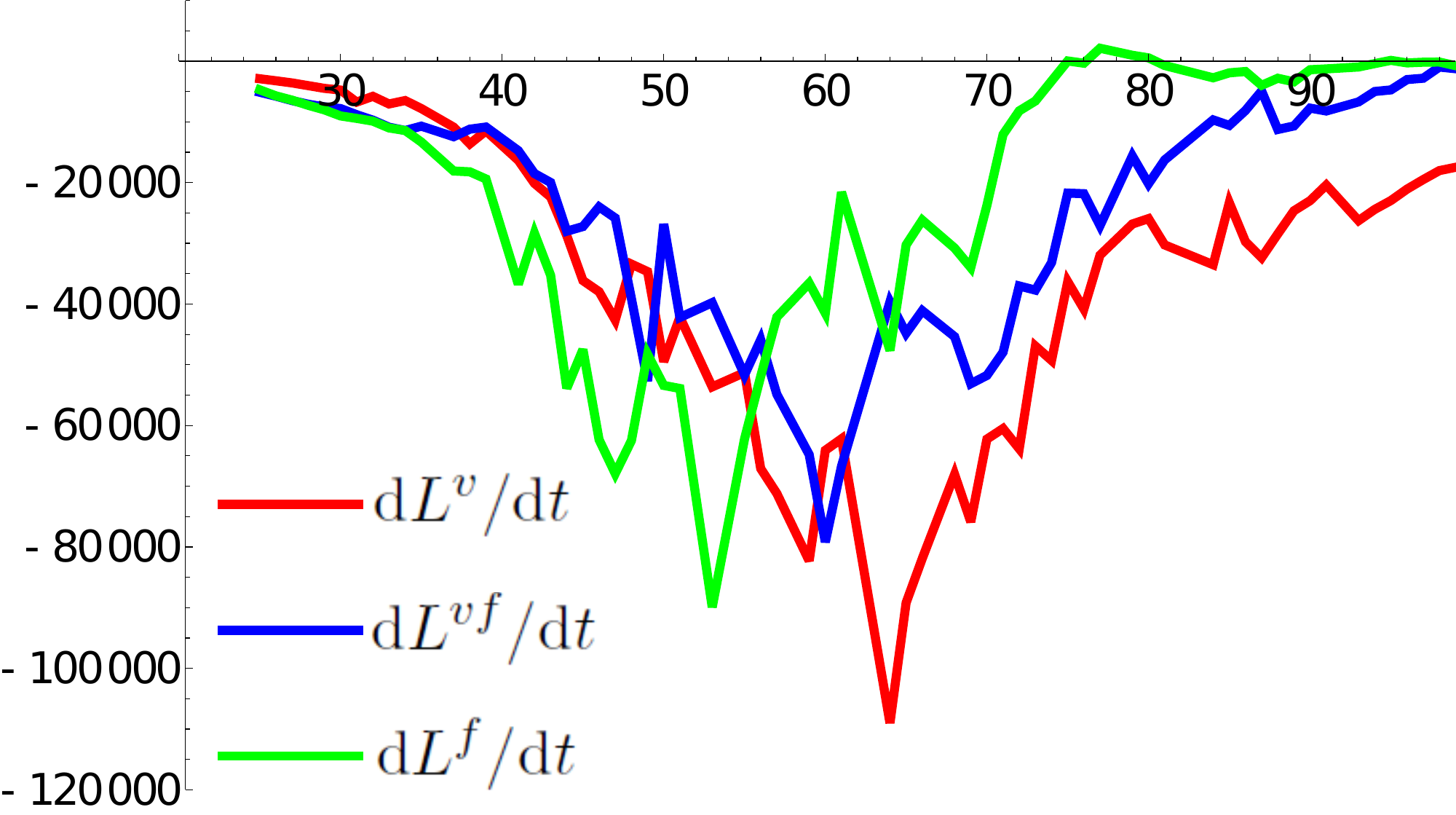}}\quad \subfigure[]{\includegraphics[width=7.5cm]{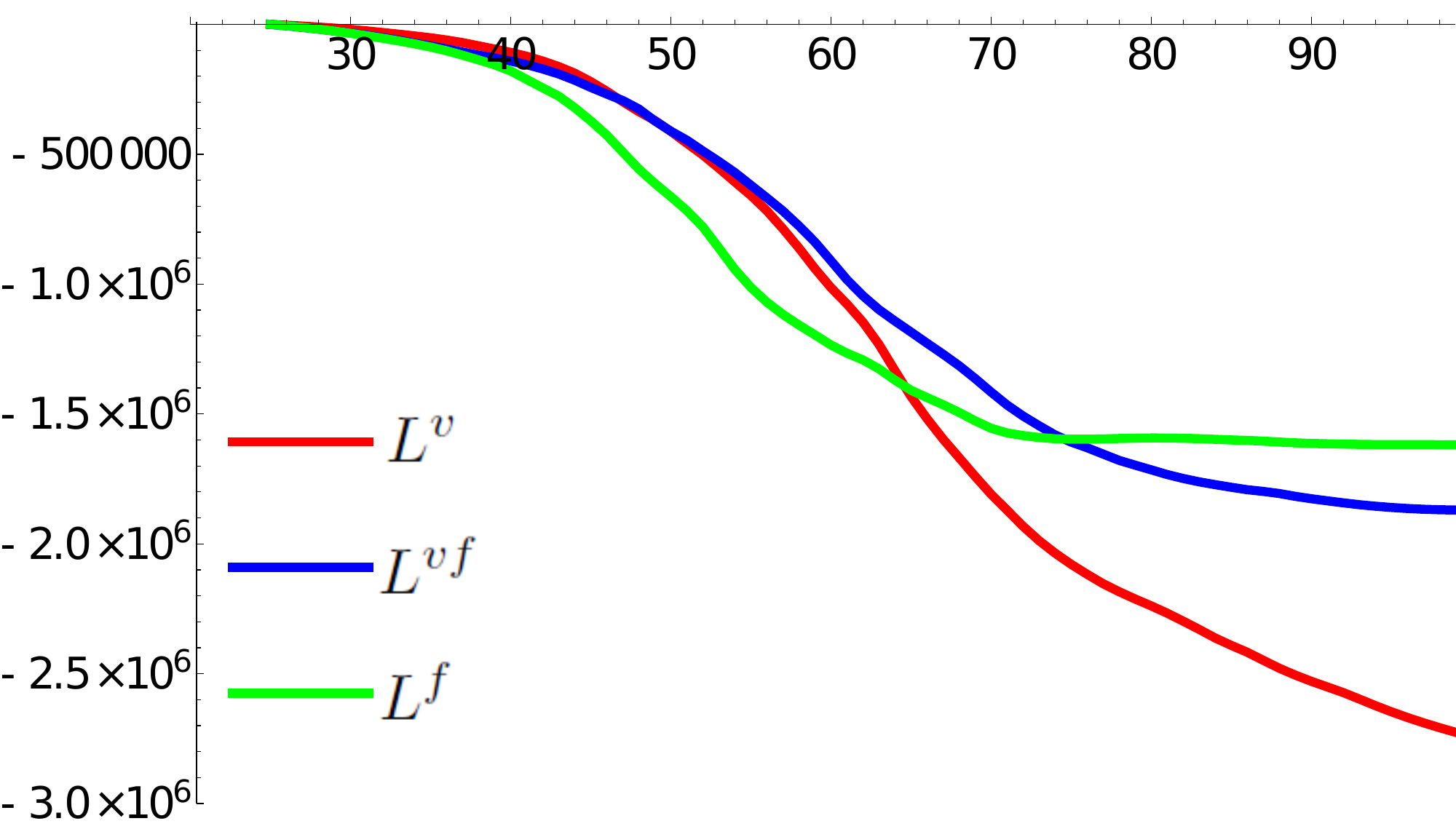}}
\caption{\label{windingvaryb031}Winding  time series for the $B_0=3$, $\alpha=-0.4$ case. Panel (a) shows the winding input rates $\d{L}^v/\d{t}$, $\d{L}^f/\d{t}$ and $\d{L}^{vf}/\d{t}$. Panel (b) shows the integrated winding $L^v(t),\,L^f(t)$ and $L^{fv}(t)$ over the same period. }
\end{center}
\end{figure}
\begin{figure}
\begin{center}
\subfigure[]{\includegraphics[width=7.5cm]{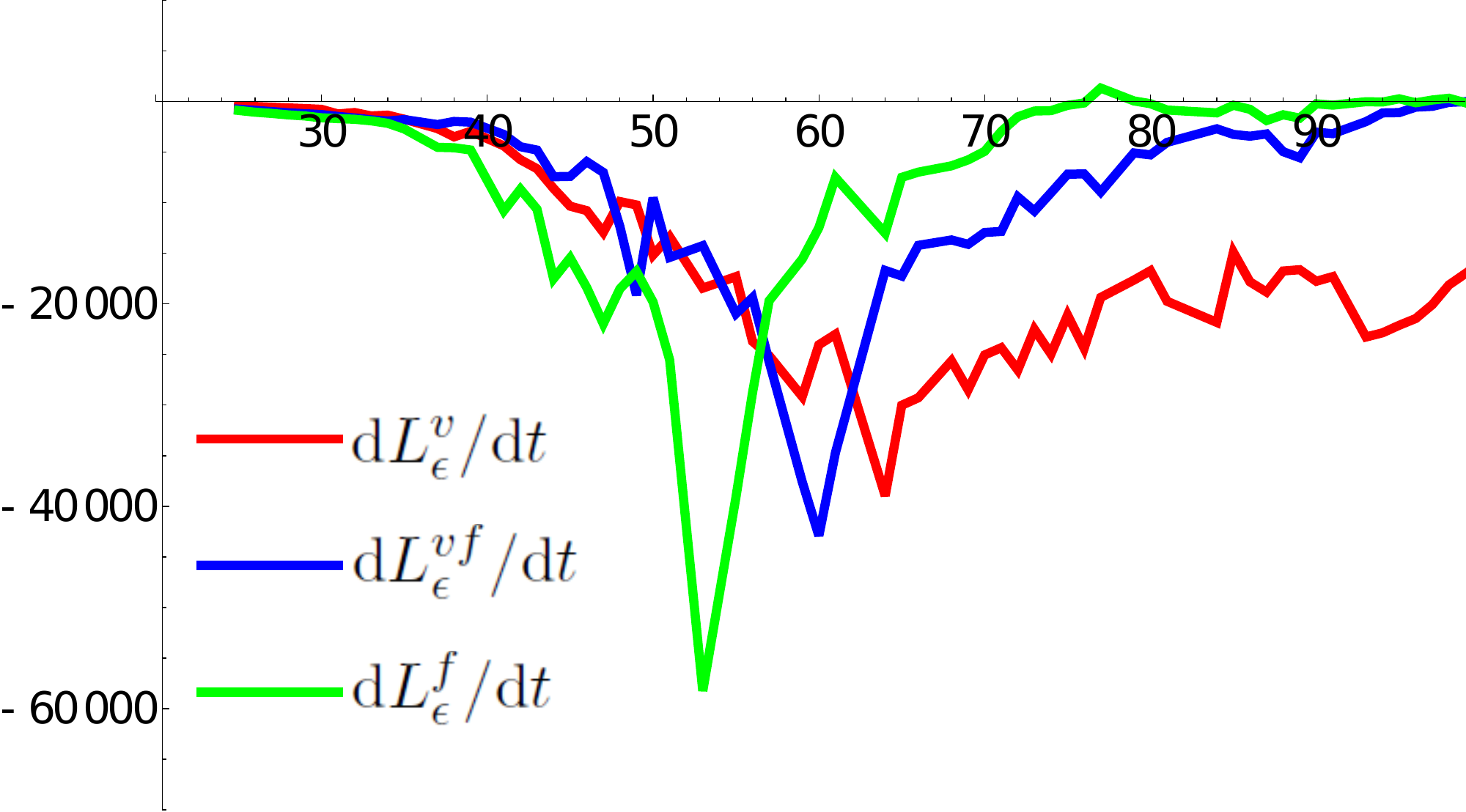}}\quad \subfigure[]{\includegraphics[width=7.5cm]{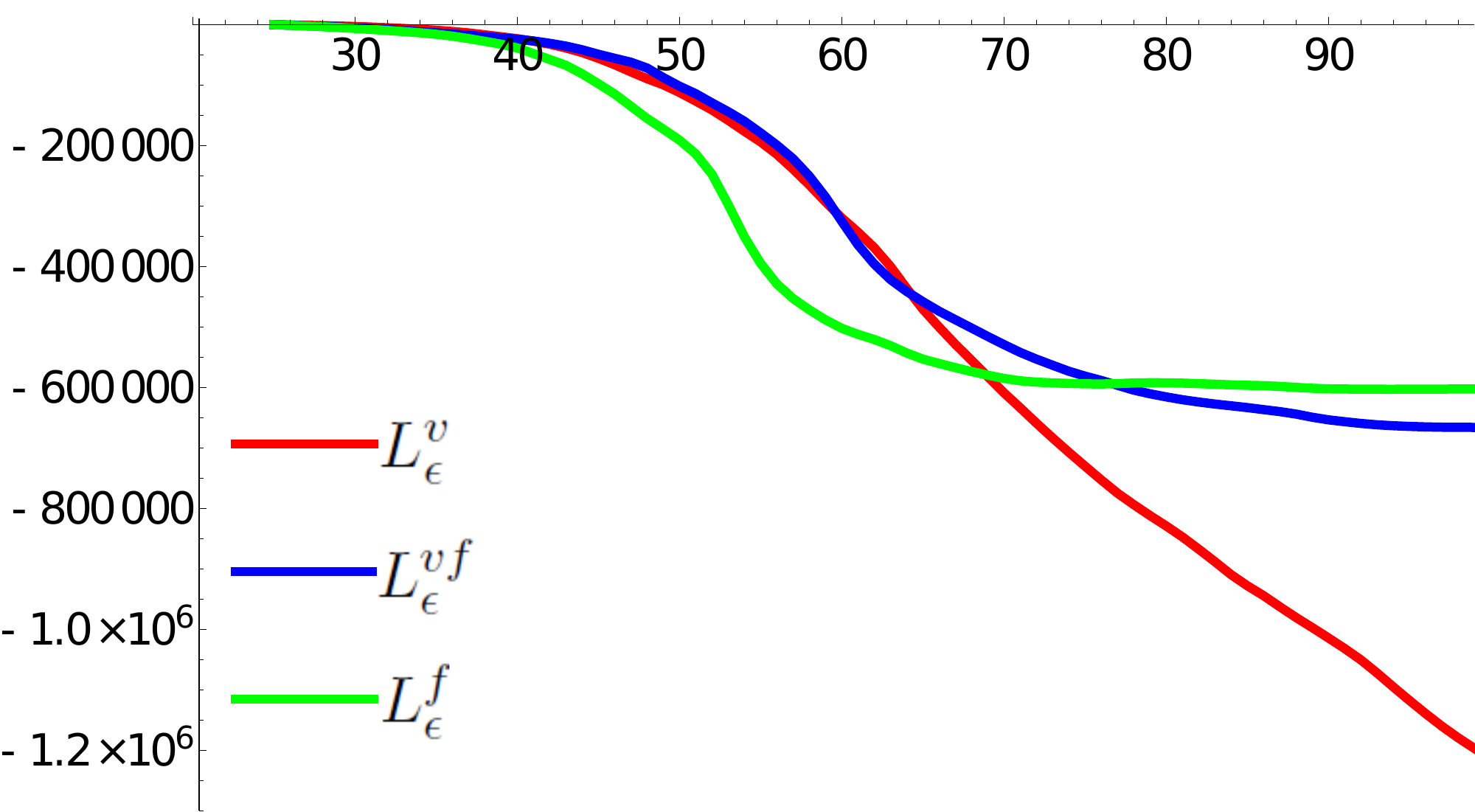}}
\caption{\label{windingvaryb032}Winding  time series with cut-off $\epsilon$ for the $B_0=3$ $\alpha=-0.4$ case. Panel (a) shows the winding input rates $\d{L}_{\epsilon}^v/\d{t}$, $\d{L}_{\epsilon}^f/\d{t}$ and $\d{L}_{\epsilon}^{vf}/\d{t}$. Panel (b) shows the integrated winding $L_{\epsilon}^v(t),\,L_{\epsilon}^f(t)$ and $L_{\epsilon}^{fv}(t)$ over the same period. }
\end{center}
\end{figure}
The winding rates $\d L^v/\d t$,  $\d L^f/\d t$ and $\d L^{vf}/\d t$ (no cut-off) are shown in figure \ref{windingvaryb031}, whilst the cut-off rates $\d L_{\epsilon}^v/\d t$,  $\d L_{\epsilon}^f/\d t$ and $\d L_{\epsilon}^{vf}/\d t$ are shown in figure \ref{windingvaryb032}. As expected, the magnitudes are larger for the non-cut-off cases, but the qualitative picture is basically the same. Therefore,  most of the following comments apply to both sets of plots. As with the $B_0=5,7$ cases the input is dominantly negative  leading to a net input of negative winding. Comparing the varying and flat measures we see that the peak in the flat case $\d L^f/\d t$ occurs before the peak in both the varying fine and varying cases. Interestingly, when the cut-off is applied the input rate is largest for the flat case, whilst without the cut-off the highest input rate is found in the varying case. This contrasts with the $B_0=5,7$ cases where the flat measure has a significantly higher rate \emph{with and without}  applying the cut-off (see figure \ref{windingvaryb031}(a), figure \ref{helicityvaryb05}(b) and figure \ref{helicityvaryb07}(b)). 

It is interesting to compare the winding rates $\d L^{v}/\d t$ and $\d L^{vf}/\d t$. Early in the simulation (up to about $t=58$) the rates are very close to each other. At $t=60$ the varying fine measure $\d L^{vf}/\d t$ peaks and then begins to decline whilst the varying measure  $\d L^{v}/\d t$ continues rising for a little while longer. After about $t=65$ both are in decline at a similar rate, however, the drop from the peak value is more pronounced in $\d L^{v}/\d t$. As a consequence $\d L^{vf}/\d t$ drops to nearly zero by the end of the simulation whilst $\d L^{v}/\d t$ does not.
}

\begin{figure}
\begin{center}
\subfigure[$t=55$]{\includegraphics[width=6cm]{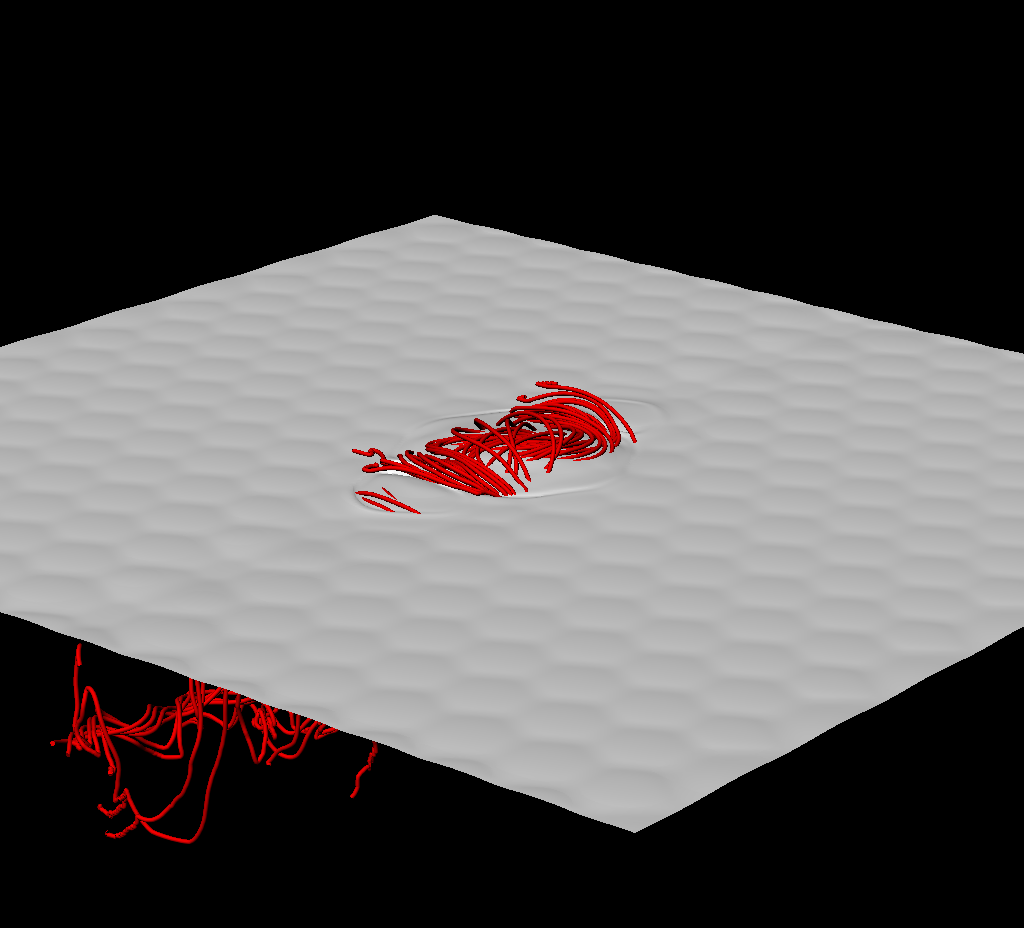}}\quad\subfigure[$t=55$]{\includegraphics[width=6cm]{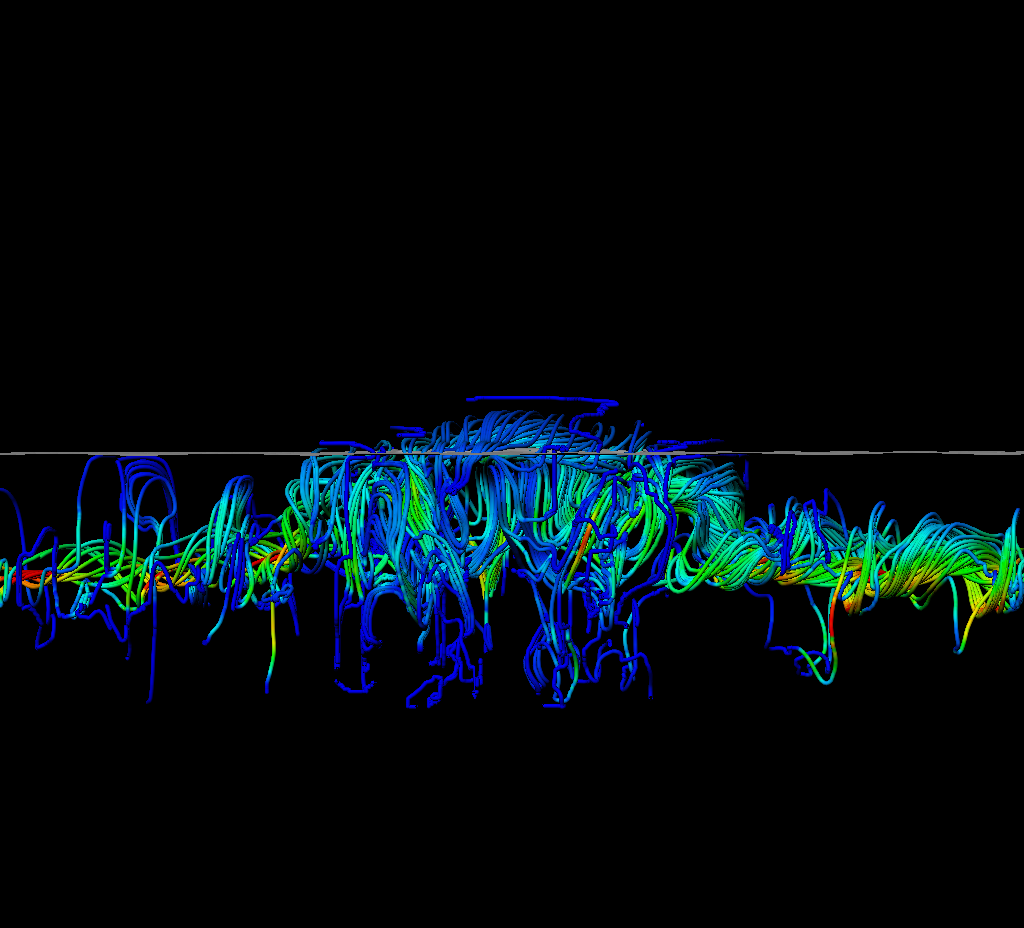}}\quad\subfigure[$t=75$]{\includegraphics[width=6cm]{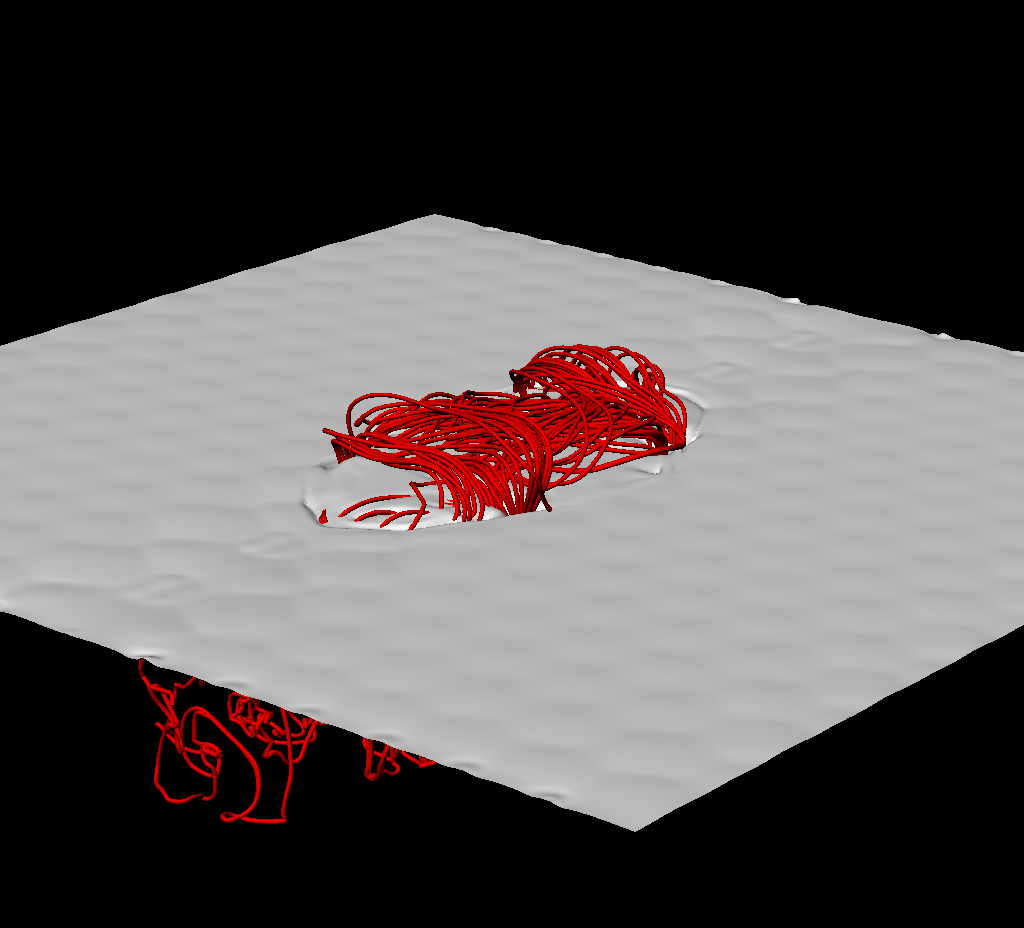}}\quad\subfigure[$t=75$]{\includegraphics[width=6cm]{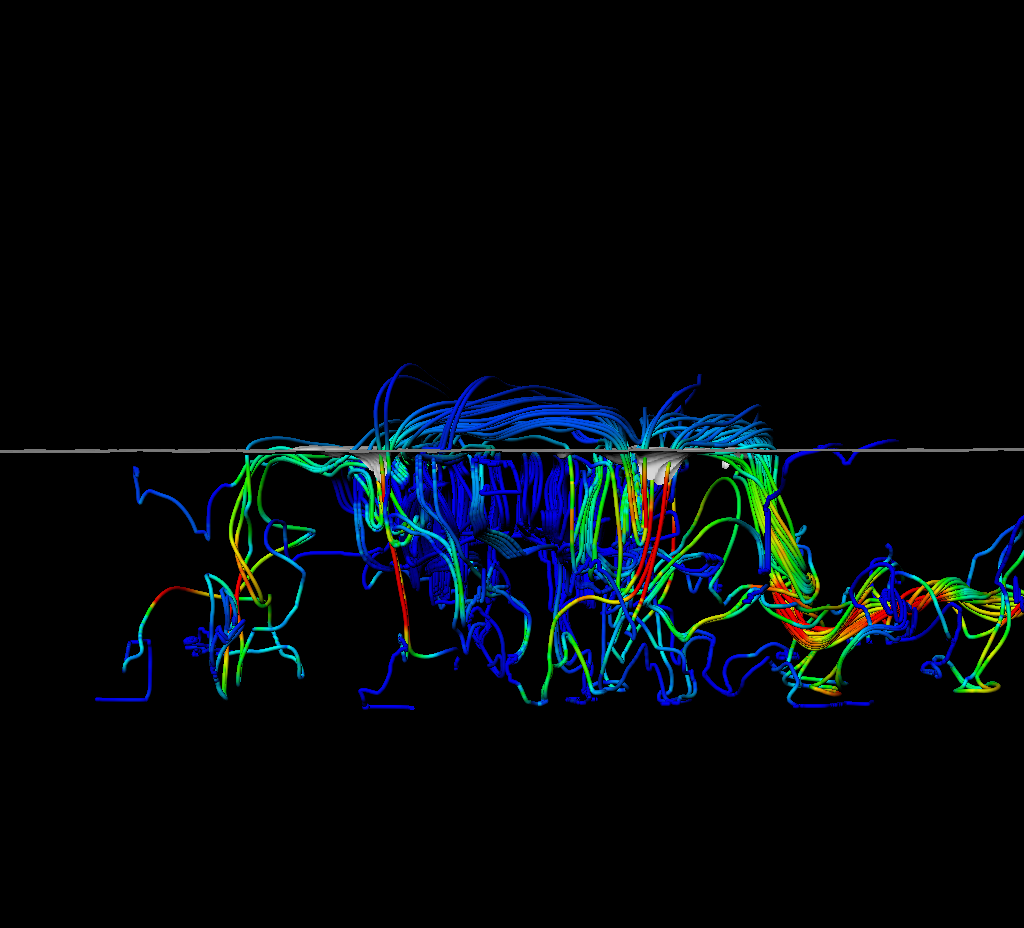}}\quad\subfigure[$t=98$]{\includegraphics[width=6cm]{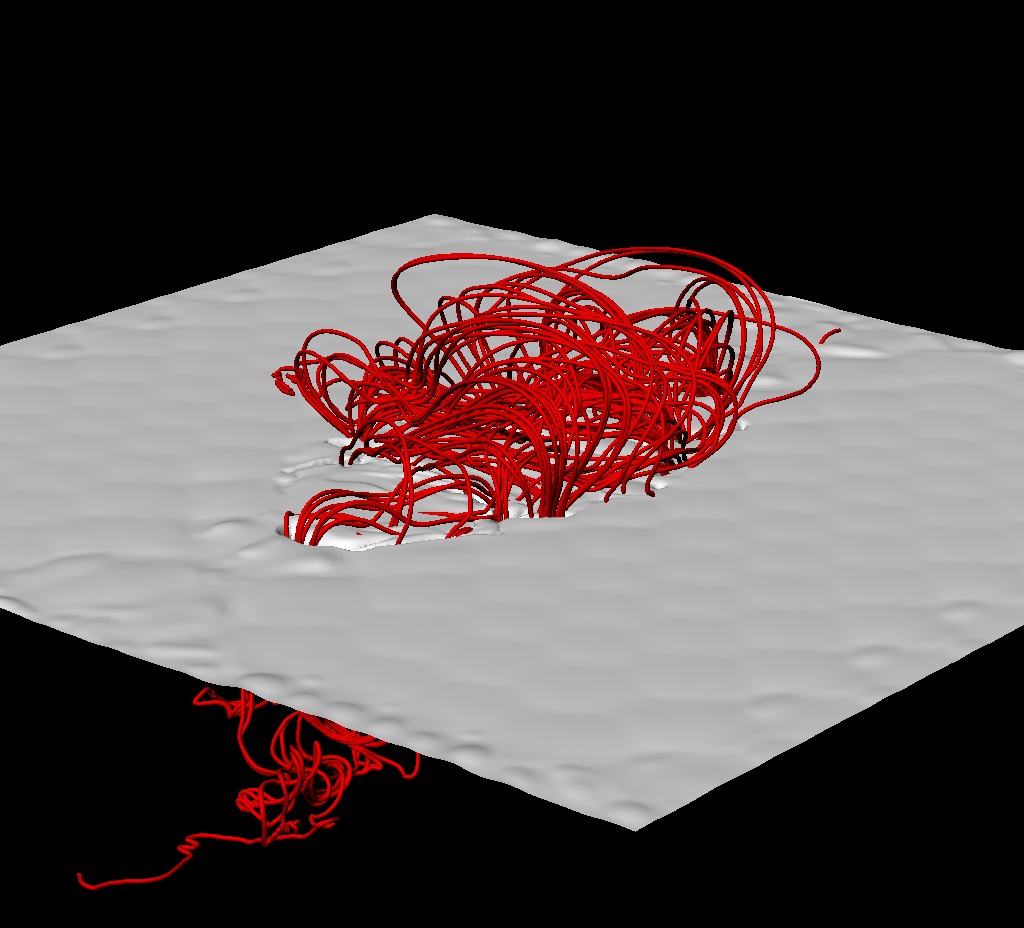}}\quad\subfigure[$t=98$]{\includegraphics[width=6cm]{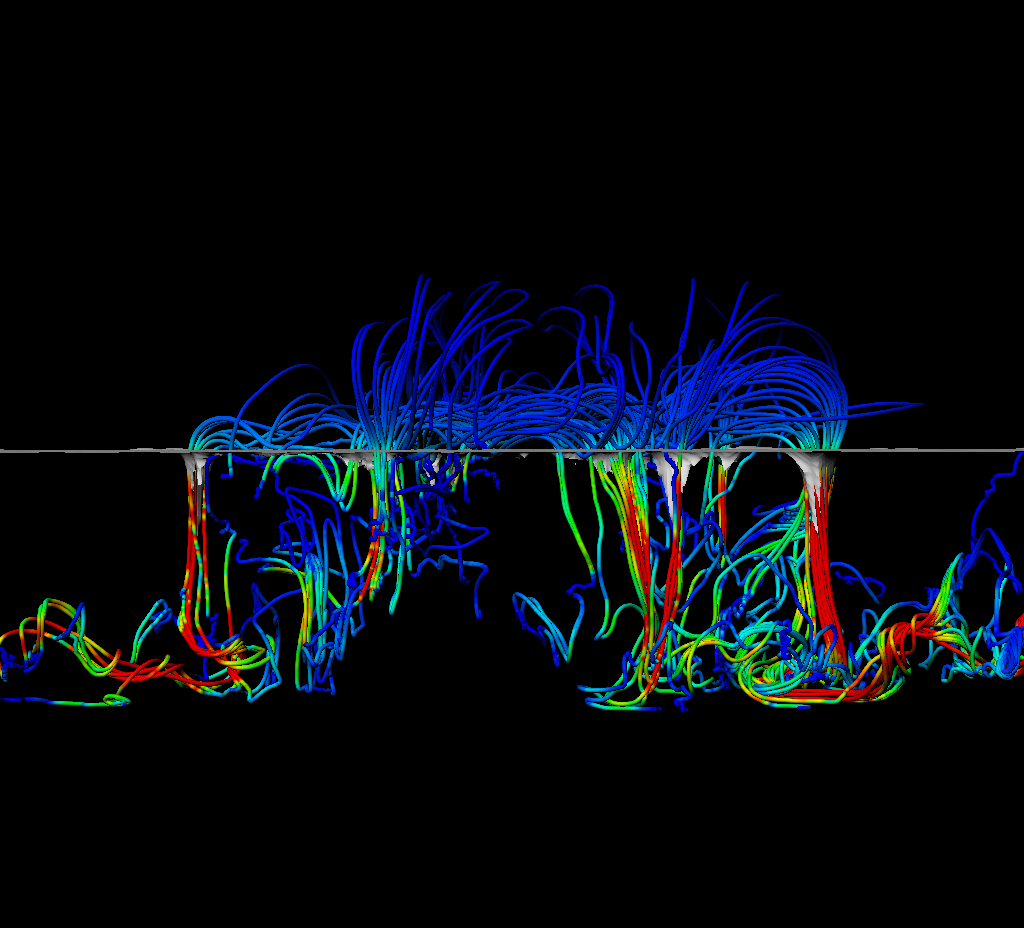}}
\caption{\label{b03fieldlines}Field line renderings for the $B_0=3$, $\alpha=-0.4$ case at various points in its evolution. Panels (a) and (b) depict the field at $t=55$. The surface shown is that of the plasma density $\rho=1$. Panels (c) and (d) depict the field at $t=75$. Panels (e) and (f) depict the emerged field at $t=98$. In panels (b), (d) and (f), the field lines are coloured by field strength with lighter colors indicating increased strength. }
\end{center}
\end{figure}
In figures  \ref{b03fieldlines}(a) and (b), corresponding to $t=55$, we see visualizations of field lines when the helicity rate has its most negative value. Here, the twisted core has reached the photosphere but the magnetic field does not emerge into the atmosphere as much as for the two stronger cases. This slower emergence has also been found in non-convective models \citep{murray2006}. 

 At $t=75$ (figures \ref{b03fieldlines}(c) and (d)) we see the formation of serpentine structures in the emerging field, similar to the $B_0=5,7$ cases.  By $t=98$ (figures \ref{b03fieldlines}(e) and (f)) the field has emerged much more extensively into the transition region and retains a significantly complex serpentine-like structure. In figure \ref{b03fieldlines}(f), regions of strong field strength can been seen clearly beneath dips in the photospheric surface. These features are clear signatures of how convection deforms the magnetic field, particularly for this weaker tube. {These field line plots indicate that the dominant net positive helicity signature observed in figure \ref{helicityvaryb03}(b) is a result of the fact that the magnetic field fails to  emerge fully (compared to the $B_0=5,7$ cases) before the serpentine field structures develop.}
 
{
\subsubsection{Helicity distributions and the effect of averaging}
\begin{figure}
\centering
\subfigure[]{\includegraphics[width=7cm]{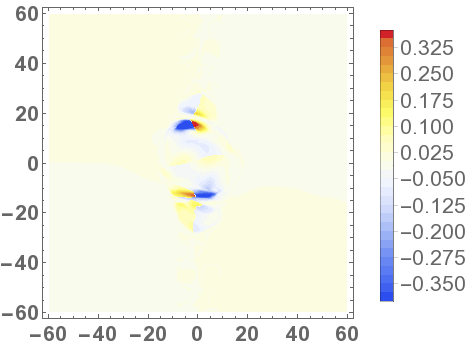}}\quad \subfigure[]{\includegraphics[width=7cm]{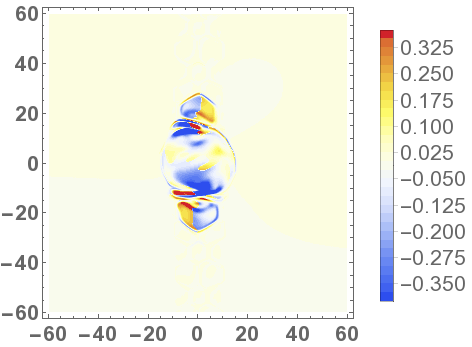}}\quad \subfigure[]{\includegraphics[width=7cm]{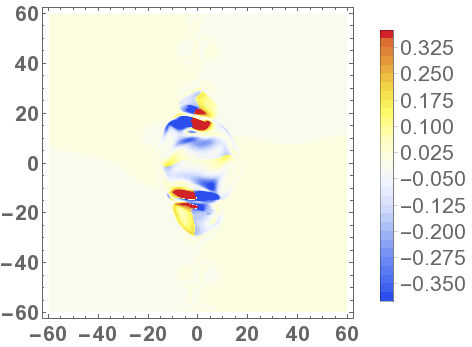}}
\caption{\label{helicitydistsb031} Helicity input distributions: (a) $\d {{\cal H}}^v/\d t$ at $t=56$, (b) $\d {{\cal H}}^f/\d t$ at $t=53$ and (c) $\d {{\cal H}}^{vf}/\d t$ at $t=59$. The times are when each respective spatial average is found to peak in figure  \ref{helicityvaryb03}(a). }
\end{figure}
\begin{figure}
\centering
\includegraphics[width=8cm]{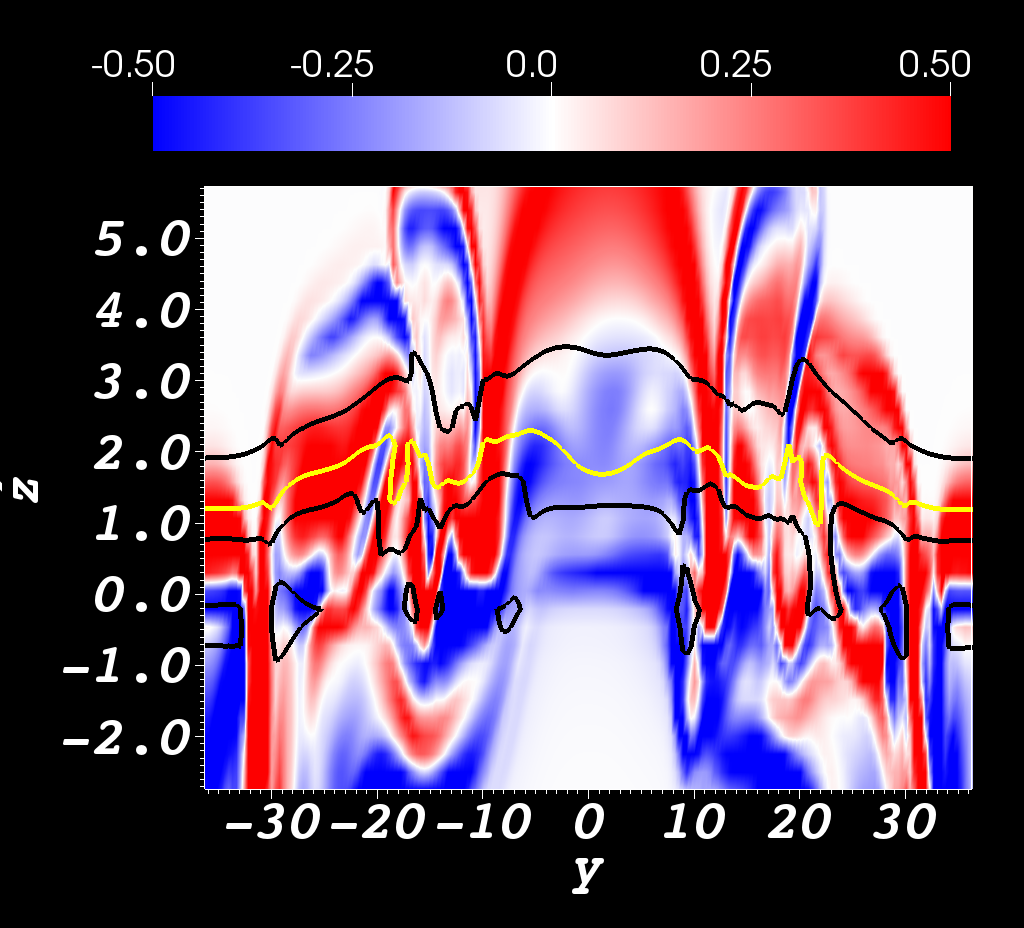}
\caption{\label{fieldcomp}Distribution of $\ev_y\cdot\nabla \times \Bv/\vert \Bv\vert$ in the $y$-$z$ plane at $x=0$. The yellow curve is the intersection of the surface $\rho=1$ with this plane and the black curves the surfaces of $\rho=0.5$ and $1.5$. Vertical averages at many points in this plane would  cross both negatively and positively rotating field structures. }
\end{figure}
In figure \ref{helicitydistsb031} we see the helicity distributions at the peak of each time series:   (a) $t=56$ for $\d {{\cal H}}^v/\d t$, (b) $t=59$ for $\d {H}^{vf}/\d t$ and (c) $t=52$ for $\d {H}^f/\d t$. In all cases it is notable that there is no clear sigmoidal structure or dominance of negative helicity input as there is in the $B_0=5,7$ cases (see, for example, \ref{helicitydistsb07}(b)). It is also notable that there is much less structure in the  $\d {H}^{v}/\d t$ distribution compared to the other two, with only the clear bipolar distribution at $x=0,y=\pm 18$ standing out. {In figure (\ref{fieldcomp}) is displayed the distribution of $\ev_y\cdot\nabla \times \Bv/\vert \Bv\vert$ in the $y$-$z$ plane at $x=0$ (a slice across the flux rope's length). This quantity is a measure of the local twisting of the field and, hence, indicative of the local magnetic field topology (a significant contributor to the helicity calculations). Three curves are indicated on this distribution, the yellow curve is the intersection of the $\rho=1$ surface and the black curves are the intersections of the $\rho=0.5$ and $\rho=1.5$ surfaces. The varying average gives significant weight to the magnetic field  between these surfaces. It is clear that there is often significant vertical variation in the sign of this local topology. This indicates why the vertical average associated with the $\d {H}^{vf}/\d t$ measure records less field topology compared to $\d {H}^{v}/\d t$.
}

\subsubsection{Winding spikes and topological events}\label{wind_spike1}
\begin{figure}
\centering
\subfigure[]{\includegraphics[width=7cm]{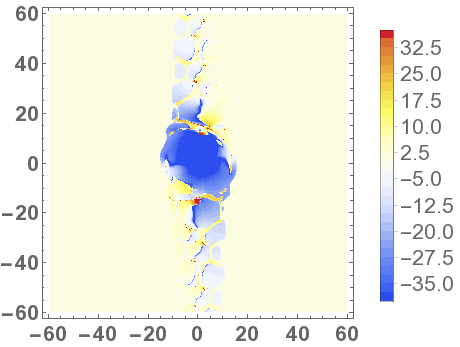}}\quad\subfigure[]{\includegraphics[width=7cm]{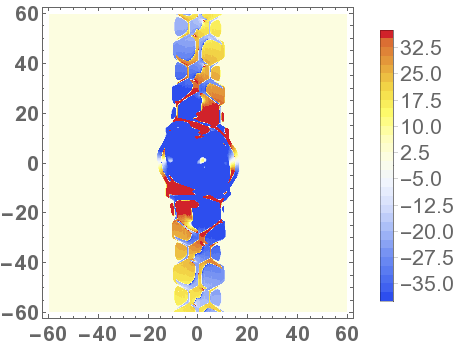}}\quad\subfigure[]{\includegraphics[width=7cm]{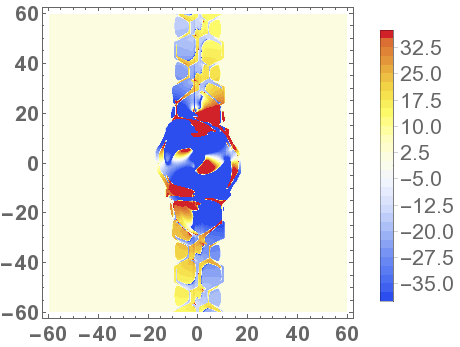}}
\caption{\label{flatwindingdists}Distributions of $\d\clf(\av_0)/\d t$ at (a) $t=51$, (b) $t=53$ and (c) $t=55$. }
\end{figure}
\begin{figure}
\centering
\subfigure[]{\includegraphics[width=8cm]{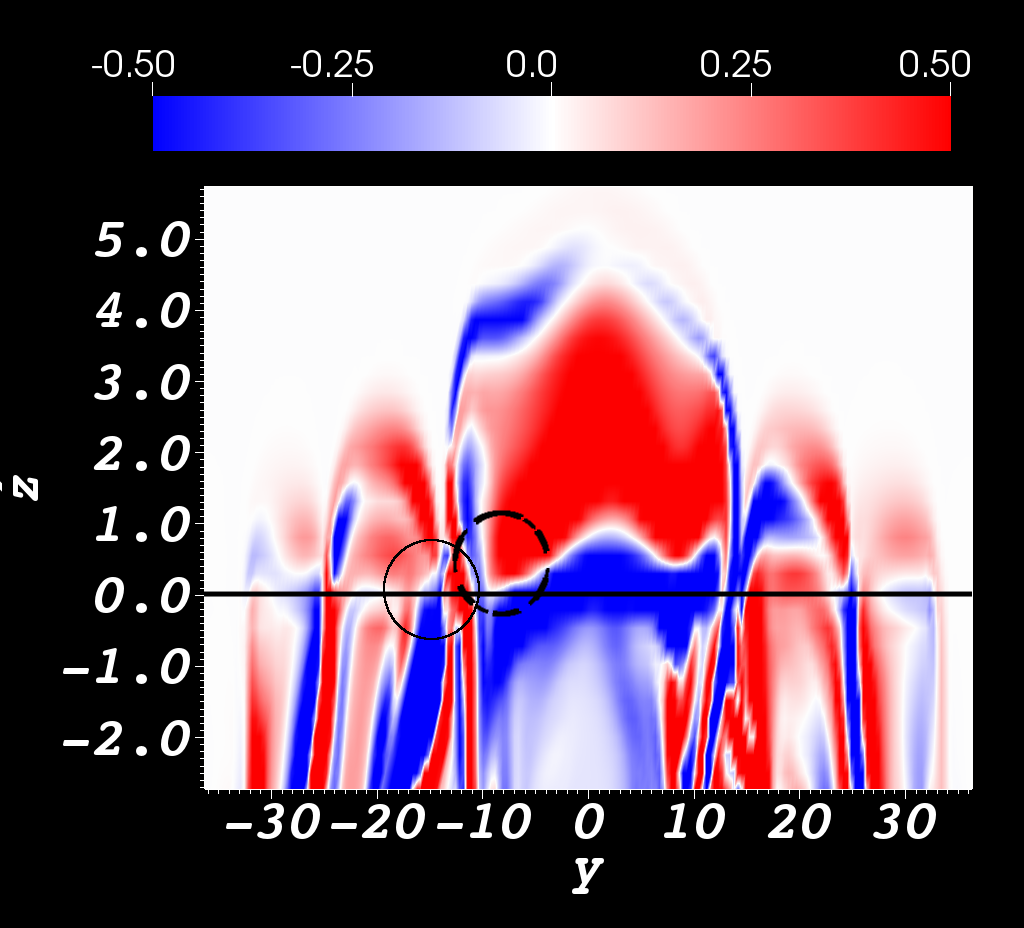}}\quad \subfigure[]{\includegraphics[width=8cm]{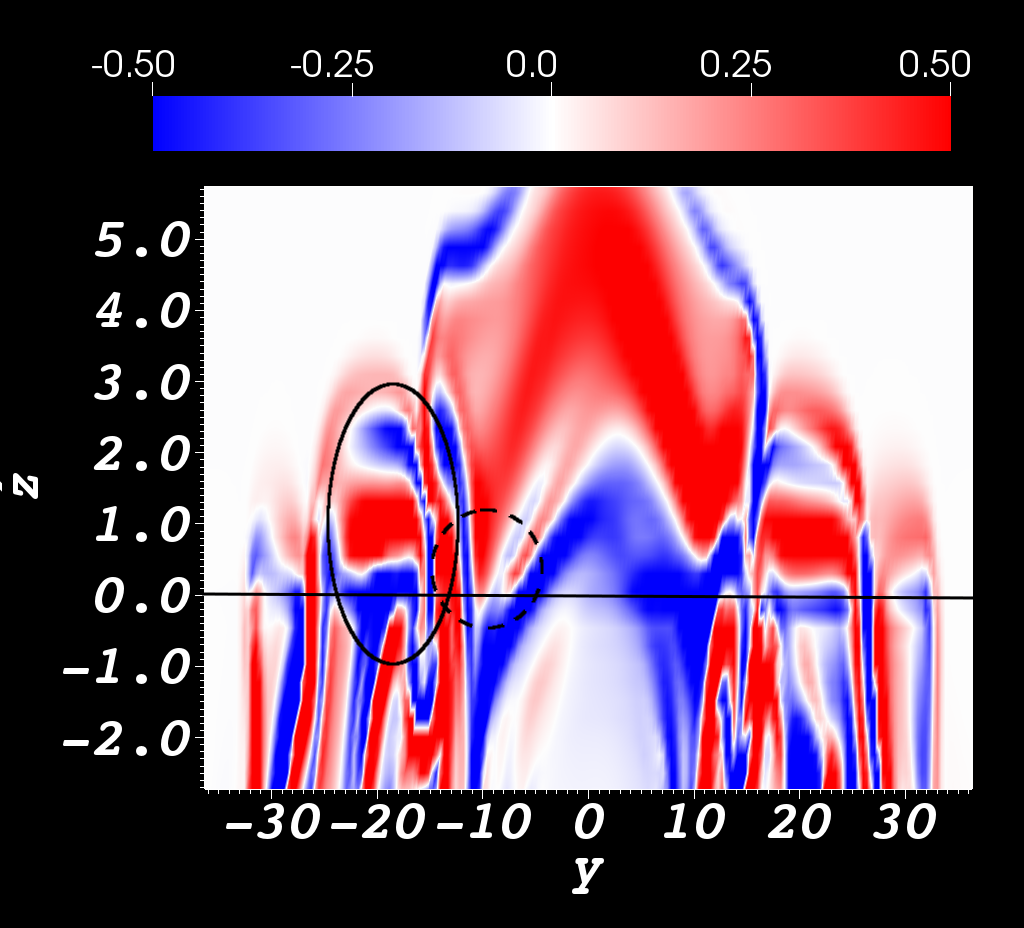}}
\caption{\label{flatwindingslices}Distributions of $\ev_y\cdot\nabla \times \Bv/\vert \Bv\vert$ in the $y$-$z$ plane at $x=5$ for times (a) $t=50$ and (b) $t=55$, before and after the peak in $\d L^f/\d t$ seen in figure \ref{windingvaryb032}(a). The line $z=0$ is shown. Two structures are circled in (a). The first (solid) is a small region of negative out of plane rotation. In (b) this is seen to have distorted and expanded both sideways and vertically into and above the $z=0$ line. In (a) the second circle (dashed) is at the edge of a region of positive  out of plane rotation. In (b) this is seen to have submerged below the $z=0$ line. }
\end{figure}
The noticeable peaks in all three winding input rate series (in both figure \ref{windingvaryb031}(a) and figure \ref{windingvaryb032}(a)) can be shown to coincide with significant emergence events. We begin by analysing the peak in the flat series which we find is the first indicator of a significant emergence event. In figure \ref{flatwindingdists} we see the flat winding distributions $\d\clf(\av_0)/\d t$ at (a) $t=51$, (b) $t=53$ and (c) $t=55$ covering the major peak in the net input rate $\d L^f/\d t$ seen in figure \ref{windingvaryb032}. From figure \ref{flatwindingdists}(a) to (b) there is a general increase in the magnitude of the distribution as a whole, which leads to the peak. Then, from figure \ref{flatwindingdists}(b) to (c), two significant patches of positive winding emerge at the distribution's centre, thus lowering the input rate. 

In figure \ref{flatwindingslices} we see slices of the distribution of $\ev_y\cdot\nabla \times \Bv/\vert \Bv\vert$ in the $y$-$z$ plane at $x=5$ for $t=50$ and $t=55$, i.e. before and after the peak in $\d L^f/\d t$ seen in figure \ref{windingvaryb032}(a). Two circled regions indicate respectively a region of significant negative twisting (solid circle) which emerges into and above the plane $z=0$, and a region of positive twisting (dashed circle) which submerges back down beneath the plane $z=0$ and which causes the positive winding inputs that appear in figure \ref{flatwindingdists}. Therefore, the winding rate spike is indicative of current-carrying field structure emerging/submerging through the surface at which the field is sampled. This is similar to a phenomenon found in `mixed helicity' flux emergence simulations in \cite{prior2019interpreting} and reinforces the conclusion in that paper that the winding series can be used for event detection in photospheric helicity input studies (the events being the emergence/submergence of helicity-carrying structures).

\begin{figure}
\centering
\subfigure[]{\includegraphics[width=7cm]{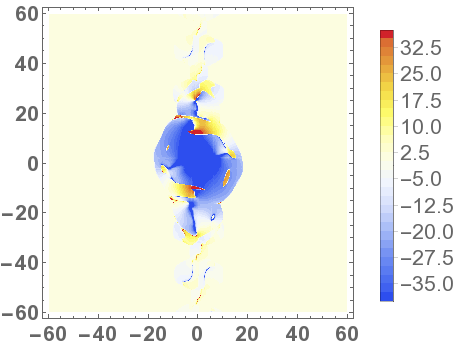}}\quad\subfigure[]{\includegraphics[width=7cm]{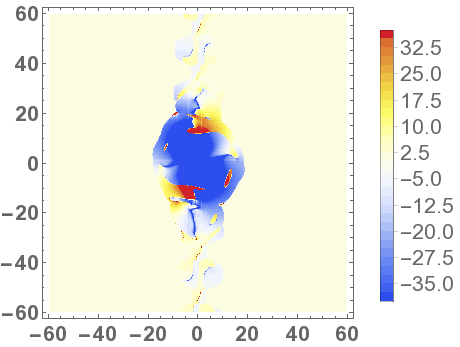}}\quad\subfigure[]{\includegraphics[width=7cm]{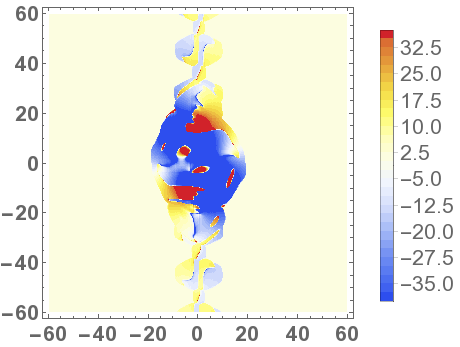}}
\caption{\label{varfinewindingdists}Distributions of $\d\clv(\av_0)/\d t$ at (a) $t=57$, (b) $t=59$ and  (c) $t=61$. }
\end{figure}
\begin{figure}
\centering
\subfigure[]{\includegraphics[width=7cm]{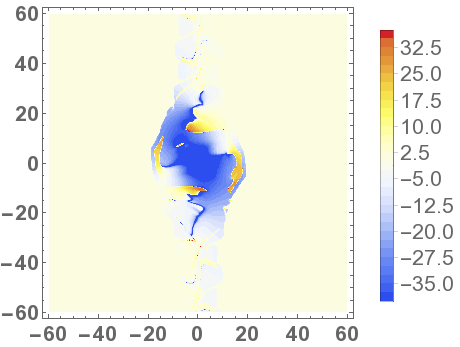}}\quad\subfigure[]{\includegraphics[width=7cm]{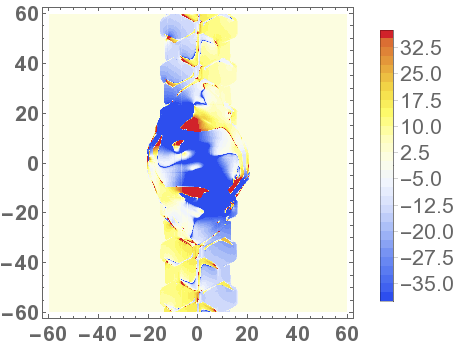}}\quad\subfigure[]{\includegraphics[width=7cm]{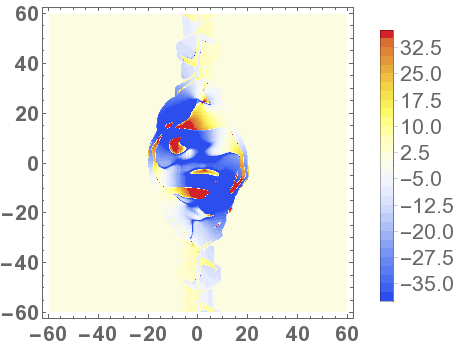}}
\caption{\label{finewindingdists}Distributions of $\d\cl(\av_0)/\d t$ at (a) $t=62$, (b) $t=64$ and  (c) $t=66$. }
\end{figure}
In figure \ref{varfinewindingdists} we see a set of changes in the varying fine winding rate $\d\clv(\av_0)/\d t$ in  a set of distributions at (a) $t=57$, (b) $t=59$ and (c) $t=61$ covering the major peak in the net input rate $\d L^{vf}/\d t$ (figure \ref{windingvaryb032}). Form figures \ref{varfinewindingdists}(a) to (b) there is a (relatively) large increase in the general magnitude of the distribution and then from figures \ref{varfinewindingdists}(b) to (c), the emergence of a set of positive winding rate islands towards the centre of the distribution. Again, it was checked that these islands develop further in the proceding evolution. It can be confirmed that these changes result from similar emergence/submergence events as indicated in figure \ref{flatwindingslices}. Finally, in figure \ref{finewindingdists},  a similar transition is found in  the varying winding rate $\d\cl(\av_0)/\d t$ in  a set of distributions at (a) $t=61$, (b) $t=63$ and (c) $t=65$ covering the major peak in the net input rate $\d L^{vf}/\d $ (figure \ref{windingvaryb032}). 

With all of the above results, we have been able to track the temporal development of a particular topological structure rising into the solar atmosphere. This emphasizes that whilst the differing means of field specification can affect the net input, they are consistent in their diagnosis of the large-scale topological development of the field.
}

\section{Summary and discussion}
In this work we have analyzed helicity and winding fluxes in simulations of twisted flux tube emergence with a range of different initial field strengths. { In particular, we have developed a model where a strongly-twisted flux tube reaches the photosphere in order to determine what the helicity and winding signatures would be for this commonly assumed scenario.} We have focused on the effect of convection on the topological quantities and have compared the results to those of other work that does not include convection. In the present study, we can select some general results:
{

\begin{enumerate}
\item[1.] {Convective flows act to pull down sections of the emerging field, leading to a more complex serpentine field line connectivity at the photospheric surface than typically observed for non-convective flux rope simulations. This effect also leads to a reversal in sign of the rate of helicity input $\d H/\d t$. In the case of the weaker magnetic field strengths, this effect further acts to lead to a change in sign of the \emph{net} helicity input $H(t)$. Thus mixed helicity input time series can be produced by single-sign helicity flux ropes if convection is accounted for. The strength of the field appears to dictate the degree of this serpentine structure and its effect on the helicity input series {(compare, for example, figures \ref{helicityvaryb07}, \ref{helicityvaryb05} and \ref{helicityvaryb03} and also the field line structures in figures \ref{b07fieldlines1}, \ref{b05fieldlines} and \ref{b03fieldlines}).} 

\item[2.] {As was found in \cite{prior2019interpreting}, the winding input rate time series tend to exhibit spikes when structures of significant topology pass into the photospheric region {(see the discussion in section \ref{wind_spike1})}. With the inclusion of convection in the model,  the winding signature still provides a clear signature for the arrival of { elements of helicity-carrying field structure through  photosphere.} A combination of winding time series and distributions can be used to establish {such signatures}. }

\item[3.] {The winding input rate $\d L/\d t$ and net input $L(t)$, which are independent of any magnetic field strength weighting, are not so influenced by the general downward pulling of the field by convective flows. Thus the winding series generally show a steady input of negative winding, concordant with the flux tube's initial structure.} 

\item[4.] {Analyzing helicity and winding signatures together provides a detailed picture of the initial stages of active region emergence into the solar atmosphere. }

\item[5.] Taking into account how the magnetic field readings are produced in real magnetgorams could be crucial for interpreting the amount of helicity injected into solar atmosphere. For example, for the $B_0=3$ field, which develops significantly complex serpentine structure, the effect of averaging along the line of sight (which likely occurs to some degree in actual magnetograms) results in a significant drop in the magnitude of helicity input. 

\item[6.] Despite the quantitative differences arising from averaging and varying the location where the magnetic field is measured, the qualitative results for both the helicity and the winding are consistent in all cases. The results of this work, therefore, provide clear signatures for the emergence of strongly-twisted flux tubes.}
\end{enumerate}

We plan to extend this work in three directions. The first is to analyze the helicity and winding rates in observations and, in particular, determine early-warning signatures for possible eruptions. The second is to study the topological signatures of the emergence of magnetic field structures different from twisted flux tubes, such as `mixed helicity' fields \citep{prior2019interpreting}. The third avenue that we will pursue is to examine how helicity and winding signatures can be used to interpret more advanced stages of emergence, including the onset of eruptions.

\section*{}
Results were obtained using the ARCHIE-WeSt High Performance Computer (www.archie-west.ac.uk) based at the University of Strathclyde.

\bibliographystyle{gGAF}
\bibliography{convection}

\end{document}